\documentclass[journal]{IEEEtran}

\usepackage{xcolor}
\usepackage{cite}
\usepackage{amsmath,amssymb,amsfonts}
\usepackage{algorithmic}
\usepackage{graphicx}
\usepackage{textcomp}

\usepackage{colortbl}
\DeclareRobustCommand{\rev}[1]{#1}
\newenvironment{revtable}{}{}
\newlength{\NAwidth}
\AtBeginDocument{\settowidth{\NAwidth}{$83\times83$}}

\newcommand{\DASH}{\makebox[\NAwidth][c]{--}}

\newcommand{\argmax}{\mathop{\rm arg~max}\limits}
\newcommand{\argmin}{\mathop{\rm arg~min}\limits}
\usepackage{tabularx}

\usepackage{bm}       
\usepackage{float}
\usepackage{makecell}
\usepackage{xurl}
\usepackage{placeins}

\newlength{\FigWidth}
\newcommand{\ignore}[1]{}  
\usepackage{comment}
\usepackage{booktabs}
\usepackage{multirow}

\usepackage{amsthm}

\newtheorem{remark}{Remark}

\newtheorem{definition}{Definition}

\newtheorem{assumption}{Assumption}
\newtheorem{example}{Example}

\usepackage{algorithm}

\usepackage{capt-of} 

\usepackage[caption=false,font=footnotesize]{subfig}

\renewcommand{\thetable}{\Roman{table}}

\def\BibTeX{{\rm B\kern-.05em{\sc i\kern-.025em b}\kern-.08em
    T\kern-.1667em\lower.7ex\hbox{E}\kern-.125emX}}

\usepackage[hidelinks,bookmarks=false]{hyperref}

\begin{document}
\title{System Design Framework for Distributed Space Antennas Using Electromagnetic Formation Flight}
\author{Seang Shim, Yuta Takahashi,~\IEEEmembership{Student Member,~IEEE},
Naoto Usami,~\IEEEmembership{Member,~IEEE}, and Shin-ichiro Sakai%
\thanks{Seang Shim is with The Graduate University for Advanced Studies, 3-1-1 Yoshinodai, Chuo-ku, Sagamihara, Kanagawa 252-5210, Japan (e-mail: shim.seang@ac.jaxa.jp).}%
\thanks{Yuta Takahashi is with the Institute of Science Tokyo, 2-12-1 Ookayama, Meguro-ku, Tokyo 152-8550, Japan, and Interstellar Technologies Inc., Hiroo, Hokkaido 089-2113, Japan.}%
\thanks{Naoto Usami and Shin-ichiro Sakai are with the Japan Aerospace Exploration Agency, 3-1-1 Yoshinodai, Chuo-ku, Sagamihara, Kanagawa 252-5210, Japan.}%
\thanks{Seang Shim and Yuta Takahashi contributed equally as co-first authors. This work was supported by JST SPRING, Japan Grant Number JPMJSP2104, and in part by The Graduate University for Advanced Studies, SOKENDAI.}%
\thanks{Corresponding author: Seang Shim.}}
\maketitle
\begin{abstract}
Distributed space antennas using electromagnetic formation flight (EMFF) are a promising architecture for large-aperture, long-life space communication systems. Their \rev{system design}, however, is governed by coupled constraints on antenna performance, satellite mass, power generation, coil geometry, and formation-keeping power. This paper proposes a system-level design framework for EMFF-based distributed space antennas. It \rev{links phased-array requirements with satellite and formation-system sizing for a static grid under a fixed total system mass.} \rev{Before system-level optimization, distributed-control simulations are used to derive a control index that quantifies the formation-maintenance requirement.} This index is integrated into an \rev{optimization problem with a user-defined objective and} sizing, power, coil, and sidelobe-envelope constraints. Parametric case studies examine \rev{the} margin magnetic moment, prescribed transmit power, and large inter-satellite spacing. Results show that increasing system mass improves footprint reduction or effective isotropic radiated power only while satellite-level design headroom remains. In direct-to-device cases with 0.15-m spacing, generated power and coil-geometry constraints dominate the feasible aperture. In the \rev{0.30-m} large-spacing case, the required coil burden can exceed satellite-level mass, size, and power capacities, making the design infeasible despite favorable communication performance. \rev{Higher-fidelity simulations of representative designs evaluate formation maintenance under nonlinear $J_2$ dynamics and antenna gain with mutual coupling, further supporting the framework's usefulness for preliminary design.} The proposed framework enables the design and evaluation of feasible static grid-based EMFF distributed antennas under coupled antenna, satellite, and control constraints.
\end{abstract}

\section{Introduction}
\label{introduction}
\begin{figure*}[!htbp]
    \centering
    \includegraphics[width=\linewidth]{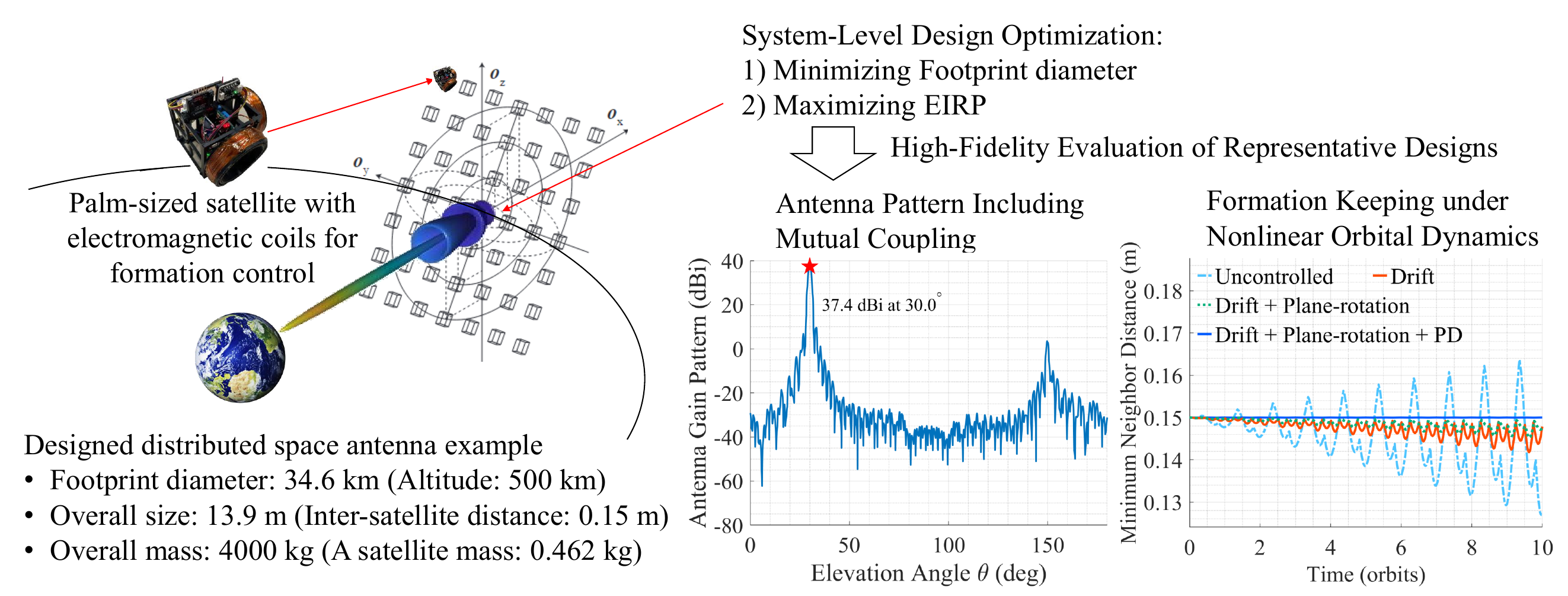}
    \caption{Overview of design optimization of distributed space antennas for the trade-off between formation control and antenna performance. \rev{Magnetic interactions of the satellite coils enable long-term formation maintenance, for which our previous studies provide control frameworks} \cite{takahashi2025distance,takahashi2026graph,takahashi2025noda_mmh}.
    }
\end{figure*}
\IEEEPARstart{D}{istributed} space antennas composed of palm-sized satellites provide a scalable architecture for high-performance communication systems under fuel-free formation maintenance, such as electromagnetic formation flight (EMFF). Previous space antennas typically required narrow beams to suppress interference with other beams \cite{tuzi2023multi} and high-gain transmission over long distances, which in turn required larger antenna structures through more elements or wider inter-element spacing \cite{balanis2005antenna,chen2016handbook}. Examples include deep-space antennas \cite{marco2019distributed}, ground terminals with tiny receivers \cite{tuzi2023satellite}, and space membrane antennas \cite{you2021ka}. Compared with monolithic antennas, distributed architectures alleviate single-point failures, payload-size limitations, and cost escalation while enabling large effective apertures \cite{shim2026integrated,she2019constructing,hadaegh2014development,marco2019distributed,kong2004electromagnetic,takahashi2022kinematics,takahashi2025distance,barnhart2007very}. \rev{Beyond antenna applications, large-scale distributed satellite systems have also been proposed for space-based AI computing \cite{beals2025towards}. Related distributed antenna concepts have also been investigated for unmanned aerial vehicle swarms, where multiple mobile platforms form distributed arrays for collaborative beamforming in wireless communications and radar sensing \cite{sun2022collaborative, yang2024beamforming}. In these systems, the spatial configuration of the platforms can be exploited as an additional design variable for improving array performance. For spaceborne distributed arrays, however,} practical realization requires active position control to preserve the formation against orbital disturbances such as $J_2$ effect \cite{schweighart2002high}. Without such control, antenna performance deteriorates over time: excessive intersatellite spacing induces grating lobes, whereas insufficient spacing increases mutual coupling \cite{craeye2011review,balanis2005antenna,chen2016handbook}. Because the propulsion required for formation maintenance directly affects mission lifetime, fuel-free control is of particular interest.

\rev{EMFF offers a fuel-free approach to controlling distributed satellite systems.} By exploiting magnetic torquers (MTQs), which are standard actuators on small satellites, EMFF enables simultaneous control of relative positions and absolute attitudes \cite{ahsun2006dynamics,takahashi2022kinematics,kong2004electromagnetic,takahashi2021simultaneous,takahashi2025distance}. Electromagnetic formation control using alternating current (AC) has also been studied with learning-based control \cite{takahashi2024neural,takahashi2025noda_mmh}. \rev{The underlying magnetic-actuation concept has also been examined experimentally using custom-built MTQs on air-bearing and linear-track ground testbeds, where controlled relative motion through magnetic interactions was demonstrated \cite{takahashi2025noda_mmh,takahashi2024neural,takahashi2026certified}. Such experiments complement numerical studies by providing hardware-level verification of key EMFF actuation and control principles.} Prior studies have established exact interaction models based on the Biot--Savart law \cite{schweighart2006electromagnetic}, dipole approximations for far-field applications \cite{ahsun2006dynamics}, and learning-based approximations \cite{takahashi2026certified}. In addition, the Earth's magnetic field can regulate the angular momentum \cite{takahashi2022kinematics} and \rev{multi-frequency coordination extends EMFF to large swarms \cite{takahashi2026decentralized}.} \rev{Other fuel-free methods face limitations for distributed space antennas: differential drag requires attitude maneuvers \cite{Varma2012} that conflict with antenna pointing during communication operations, while tethers limit scalability and reconfiguration \cite{Huang2018,wang2008grating}. These results support EMFF as a candidate for fuel-free formation control of distributed satellite systems.}

\rev{However, assessing the feasibility of large EMFF-based distributed antennas still requires a method that combines antenna performance, EMFF control demand, and component-level satellite sizing. Inter-satellite spacing links these requirements: increasing it can narrow the main beam, but the magnetic force available for a given magnetic moment decreases as the inverse fourth power of distance. Satellite design therefore balances the potential antenna benefit against the magnetic-moment and power required for formation maintenance. Existing studies address parts of this problem at different scales. A previous EMFF study examined a formation of a few satellites \cite{kong2004electromagnetic}, whereas distributed space antennas may require substantially larger arrays \cite{tuzi2023satellite}. A convex-optimization study evaluated formation-keeping power for large magnetically actuated swarms \cite{takahashi2026power}, but its conservative worst-case specifications limit detailed component-level satellite sizing. A system-level feasibility framework must therefore link antenna and EMFF requirements to satellite size, mass, and power at large array scales.}

\rev{To address this need, this paper presents an extended system-level design methodology that evaluates achievable antenna performance and satellite sizing under coupled EMFF, mass, size, and power constraints.} This paper builds on our previous work on the feasibility study of EMFF-based distributed space antennas \cite{shim2025feasibility}. In that study, the formation-maintenance requirement was derived using a bucket-brigade compensation model for a grid formation. Although this model provided an analytically tractable first estimate of the required magnetic moment and control power, it relied on an idealized compensation structure and did not directly reflect the behavior of distributed formation control. \rev{In contrast, this paper estimates the required control power from numerical simulations of distributed formation control. Consequently, the proposed framework assesses nominal antenna and satellite feasibility using control requirements derived from simulated distributed formation behavior.}

\rev{The proposed framework evaluates EMFF-based distributed antennas under coupled antenna, control, and satellite constraints. For a prescribed total system mass, it optimizes a user-defined evaluation function over satellite count, satellite size, and coil design variables. Formation-maintenance power and antenna requirements constrain these choices. The case studies consider beam footprint and EIRP as objectives. They vary the magnetic-moment margin, RF transmit power, and inter-satellite spacing to identify the limiting resources. At larger spacing, coil mass, size, and power demands can make a design infeasible despite favorable communication performance. These results describe a nominal feasible design region under simplified system models. 
The framework thus links different antenna objectives to the satellite and control resources needed to achieve them.}

The remainder of this paper is organized as follows. Section~\ref{section2} reviews the underlying orbital dynamics, magnetic interaction model, and planar phased-array antenna formulation. Our problem formulation is presented in Section~\ref {problem_formulation} to define the variables describing the distributed antenna. \rev{Section~\ref{sec:proposed_dsa_opt} develops a system-level design optimization framework under coupled antenna, formation-control, and satellite resource constraints.} Section~\ref{Result} presents the numerical case studies based on the integrated optimization framework. Section~\ref{conclusion} concludes the paper. 

\section{Preliminaries}
\label{section2}
This section summarizes the control of the relative distance from $J_2$ disturbance using EMFF and the derivation of the antenna performance, focusing on the EIRP and beamwidth. 
\subsection{Linearized/Averaged Relative Orbital Dynamics}
We introduce the approximated dynamics of relative orbital motion under $J_2$ disturbances \cite{schweighart2002high}. Let $P_j=[r_{\mathrm{ref}}+x_j;y_j;z_j]$ be the position vector of the $j$-th satellite from the center of the Earth in the local vertical, local horizontal (LVLH) coordinate frame, where $r_{\mathrm{ref}}$ is the constant orbital radius of the reference orbit. We obtain $\ddot{P}_j=\nabla U_{J_2}$ where $\nabla U_{J_2}\in\mathbb{R}^3$ is the gravity gradient associated with its orbital dynamics \cite{schweighart2002high}:
\begin{equation}
    \label{J2 orbital dynamics}
\nabla U_{J_2}(P_j,i,\theta)=
-
{\begin{bmatrix}
\frac{\mu_g}{\|P_j\|^2}\\0\\0
\end{bmatrix}}
-\frac{k_{J_2}}{\|P_j\|^4}
{\begin{bmatrix}
1-3 \sin ^2 i \sin ^2 \theta\\
 \sin ^2 i \sin 2 \theta\\
 \sin 2 i \sin \theta
\end{bmatrix}},
\end{equation}
where $\mu_g$ is the gravitational parameter, $i$ is the inclination, $\theta$ is the argument of latitude, and $k_{J_2}=2.63e^{10} \mathrm{~km}^5 / \mathrm{s}^2$ is the J2 coefficient.
A previous study defined a reference orbit to compensate $J_2$ effect:
\begin{equation}
    \label{J2_constant}
    \begin{aligned}
&\omega_{\mathrm{ref}}=
\begin{bmatrix}
    0\\0\\ c_+ \omega_{\mathrm{o}}
\end{bmatrix}\ 
\mathrm{s.t.}\ \omega_{\mathrm{ref}}\times \omega_{\mathrm{ref}}\times r_{\mathrm{ref}}=\int_0^{2\pi}\nabla U_{J_2(\theta)}\frac{\mathrm{d}\theta}{2\pi},\\
&\omega_{\mathrm{o}}=\sqrt{\frac{\mu_g}{r_{\mathrm{ref}}^3}},\ c_{\pm}=\sqrt{1\pm s_{J_2}},\ s_{J_2}=\frac{k_{J_2}(1+3 \cos 2 i_{\text {ref }})}{4 \mu_g r_{\mathrm{ref}}^2}.
    \end{aligned}
\end{equation}
We include an additional compensation term for the separation of the orbital planes owing to the time-varying longitudes of the ascending nodes $\Omega(t)$ such that $\theta(t)={\omega}_{z\mathrm{ref}}t$ where
\begin{equation}
\label{omega_zref}
    {\omega}_{z\mathrm{ref}}=c_+\omega -\dot{\Omega}_{\mathrm{avg}j} \cos i_j=
\omega_{\mathrm{o}} \left(c_{+} + \frac{k_{J_2}\cos^2i}{\mu r_{\mathrm{ref}}^{2}}\right),
\end{equation}
\rev{ where $\dot{\Omega}(\theta)=-{2 k_{J_2} \cos i \sin ^2 \theta}/({\|\mathbf{h}\| r_{\mathrm{ref}}^3})$, $|\dot{\Omega}_{\mathrm{avg}}|=|\int_0^{2\pi}\dot{\Omega}(\theta){\mathrm{d}\theta}/{2\pi}|\lesssim 2e^{-6}\cos i_{\mathrm{ref}}$ under $i_j\approx i_k$ \cite{takahashi2025distance},\cite{takahashi2025scalable}. }

Linearization based on the reference orbit is used to derive the relative motion between two satellites. Let ${r_{jk}}={r}_j-{r}_k=[x_{jk};y_{jk};z_{jk}]$ be the relative position from the $k$-th satellite to the $j$-th one. The dynamics of the linearized relative motion around the reference orbit in the LVLH frame are \cite{takahashi2025distance,takahashi2025scalable}
\begin{equation}
\label{Hill_dynamics}
\begin{aligned}
&\ddot{\overline{x}}-2\omega_{xy}\dot{\overline{y}}-3\omega_{xy}^2 \overline{x}-\frac{4 \omega_{xy}^2}{c_-^2/s_{J_2}}\left(2 \overline{x}+\frac{\dot{ \overline{y}}}{ \omega_{xy}}\right)\\
&\qquad{}=c_+(\rev{u_{c,x}}+\rev{d_{\mathrm{orb},x}}) \\
& \ddot{\overline{y}}+2\omega_{xy}\dot{\overline{x}}={c_-}(\rev{u_{c,y}}+\rev{d_{\mathrm{orb},y}})\\
&\ddot{z}+\omega_z^2  z=2 l \omega_z \cos (\omega_z t+\theta_z)+(\rev{u_{c,z}}+\rev{d_{\mathrm{orb},z}}) \\
&\left\{
\begin{aligned}
&\overline{x} =c_{+} x,\quad \overline{y} = c_- y,\quad{\omega}_{xy}=c_-\omega_{\mathrm{o}},\\
&\omega_z =\omega_{z\mathrm{ref}}+f_1(\delta \dot{\Omega}_{\mathrm{avg}})\approx\omega_{z\mathrm{ref}},\ r_z \sin \theta_z=z, \\
&l \sin \theta_z+\omega_z r_z \cos \theta_z=\dot{z},
\end{aligned}
\right.
\end{aligned}
\end{equation}
where the subscript $jk$ will be omitted, $l(\delta \dot{\Omega}_{\mathrm{avg}jk}) = -r_{\text{ref}} \sin i_j \sin i_k f_2(\delta \dot{\Omega}_{\mathrm{avg}jk})$, and the dynamics along the $z$-axis include a term for compensating for errors in the cross-track motion owing to time averaging. Assuming that the satellites have identical $i$, the amplitude of the relative orbital motion along the $z$-axis reaches a constant value, which means that $l(\delta \dot{\Omega}_{\mathrm{avg}}) \approx 0$ in (\ref{Hill_dynamics}).
\subsection{Averaged $J_2$ Relative Orbital Parameters}
We \rev{derive} the orbital indices, averaged $J_2$ relative orbital parameters, and this denotes desired stable trajectories $p_d(t)$. An analytical solution of (\ref{Hill_dynamics}) is 
\cite{takahashi2025distance,takahashi2025scalable}:
\begin{equation}
\label{CWsol}
\begin{aligned}
\begin{bmatrix}
    {x}(t)\\
    {y}(t)\\
    {z}(t)
\end{bmatrix}
&=
\begin{bmatrix}
r_{\mathrm{o}}(0,t)\\
0
\end{bmatrix}
+
\begin{bmatrix}
   r_{xy}\sin{(\omega_{xy} t + \theta_{xy})}/c_+\\
    2r_{xy}\cos{(\omega_{xy} t + \theta_{xy})}/c_-\\
    (r_z+l t) \sin{({\omega}_z t+\theta_z)}
\end{bmatrix},\quad \\
r_{\mathrm{o}}(0,t)&=
\begin{bmatrix}
    {2C_{1}}&{C_4-\epsilon_{2} C_1 t}
\end{bmatrix}^\top,\ \epsilon_2 = \frac{3+5s_{J_2}}{c_+c_-}\omega_{xy},
\end{aligned}
\end{equation} 
where $r_{\mathrm{o}}(0,t)\in\mathbb{R}^2$ represents the center position of the relative orbit at time $t$, $l(\delta \dot{\Omega}_{\mathrm{avg}jk})\approx 0$, where $f_{1,2} \in \mathbb{R}$ is a function of $\delta \Omega_{jk0}$. The orbital indices calculated at the estimation time ($t=0$) are specified as \cite{takahashi2025distance}
\begin{equation}
    \label{definition_C}
    \left\{
    \begin{aligned}
    &C_{1}={c_+}/{c_-^2}(2 \overline{x}+{\dot{\overline{y}}}/{ {\omega}_{xy}}),\ C_{4}= (\overline{y}-{2 \dot{ \overline{x}}}/{{\omega}_{xy}})/c_-\\
    &r_{xy}^2= {C_2^2+C_3^2},\ \theta_{xy}= \tan^{-1}(C_3,C_2)\\ 
    &C_{2}=( \overline{y}-c_-C_4)/2, \ C_{3}=\overline{x}-2c_+C_{1}\\
    &r_{z}^2= {C_6^2+C_5^2},\ \theta_{z}= \tan^{-1}(C_6,C_5)\\
    &C_5 = \dot{z}/\omega_z,\ C_6 = z.
    \end{aligned}
    \right.
\end{equation}
It is clear from (\ref{CWsol}) that the motions are governed by $r_{xy,z}$ and $\delta \theta=\theta_z-\theta_{xy}$, as defined in (\ref{definition_C}). A simple calculation yields the following relationship between $r_{xy,z}$, $\delta \theta$, and the satellite swarm angle $(\Theta_P,\Theta_{z-xy})$ for the aperture design:
\begin{equation}
\label{r_xy_and_r_z}
\begin{aligned}
r_z &= \frac{r_{xy}}{\tan\Theta_P}\frac{\cos\Theta_{z-xy}}{\cos(\theta_z - \theta_{xy})},\\
\theta_z &=  \theta_{xy}+\tan^{-1}(2\tan \Theta_{z-xy}).
\end{aligned}
\end{equation}
The desired trajectory along the $z$-axis is selected based on the $x$--$y$ motion: $\omega_{zd}=\omega_{xy}$ and $\theta_{zd}=\theta_{xy}+\tan^{-1}(2\tan \Theta_{z-xy})$. This derives the desired stable trajectories $p_d(t)$
\begin{equation}
\label{desired_position}
{p}_d(t)=
\begin{bmatrix}
   (1/c_+)r_{xyd}\sin{(\omega_{xy} t + \theta_{xy})}\\
    (1/c_-)2r_{xyd}\cos{(\omega_{xy} t + \theta_{xy})}\\
    \frac{r_{xyd}}{\tan\Theta_P}\frac{\cos(\Theta_{z-xy})}{\cos(\theta_z - \theta_{xy})} \sin{({\omega}_{xy} t+\theta_{zd}(\Theta_{z-xy},t))}
\end{bmatrix}.
\end{equation}
Note that $\omega_{zd} = \omega_{xy}$ is realized via the $z$-axis orbit control to maintain the $z$-axis orbital frequency $\omega_z$ in $\omega_{xy}$ or it works with the disturbance $d_{fz}=(\omega_{xy}^2-\omega_z^2) z$ on ${p}_d$ \cite{takahashi2025distance,takahashi2025scalable}:
\begin{equation}
\label{d_fz_disturbance}
\begin{aligned}
d_{fz(t)}&=r_{zd}(\omega_{xy}^2 \sin{(\omega_{xy} t + \theta_z)}-\omega_{z}^2 \sin{(\omega_z t + \theta_z)}).
\end{aligned}
\end{equation}
\subsection{Magnetic Interaction Approximation Model}
This subsection introduces the approximation of the magnetic field interaction model. We define the magnetic moment $\bm{\mu}$ and the resistance of a single-axis coil $R_{\mathrm{coil}}$ as
\begin{equation}
    \label{2_1::coil dipole moment}
        \bm{\mu} =\pi N_t a_{\mathrm{coil}}^2 c_{\mathrm{coil}} \rev{\bm{\hat{n}}_{\mathrm{coil}}},\quad 
        R_{\mathrm{coil}} = {2 a_{\mathrm{coil}} N_t p_c}/{r_{\mathrm{coil}}^2},
\end{equation}
where $N_t$ is the number of coil turns, $a_{\mathrm{coil}}$ is the coil radius, $c_{\mathrm{coil}}$ is the coil current, $\rev{\bm{\hat{n}}_{\mathrm{coil}}}$ is a vector normal to the coil plane, $p_c$ is the wire resistivity, and $r_{\mathrm{coil}}$ is the wire radius. A previous study \cite{takahashi2024neural} presented a dipole model approximation \cite{ahsun2006dynamics} as the bilinear polynomial formulation using the line-of-sight frame $\{\mathcal{LOS}_{j\leftarrow k}\}$ (see Definition~\ref{NODA_C_LOS_new}). Then, the input exerted on the $j$-th agent by the $k$-th agent is :
\begin{equation}
\label{Feasibility_bilinear_polynomial_model_1}
u_{j\leftarrow k}^{a}=
\begin{bmatrix}
    f_{j\leftarrow k}^{a}\\
    \tau_{j\leftarrow k}^{a}
\end{bmatrix}=
\frac{\mu_0}{4\pi}Q_{{j\leftarrow k}}(r_{j\leftarrow k})
\left (
{\mu_{k}^{a}}
\otimes
{\mu_{j}^{a}}
\right ),
\end{equation}
where $\mu_0$ is the permeability of free space, $\bm{\mu}_{j,k}$ are the magnetic moments of the $j,k$-th coils, $Q_{{j\leftarrow k}}\in \mathbb{R}^{6\times 9}$ is \cite{takahashi2024neural}
\begin{equation*}
Q_{{j\leftarrow k}}=(I_2\otimes C^{A/L_{j\leftarrow k}}) 
\begin{bmatrix}
\Psi_{f}(r_{j\leftarrow k})\\
\Psi_{\tau}(r_{j\leftarrow k})\end{bmatrix}
(C^{L_{j\leftarrow k}/A}\otimes C^{L_{j\leftarrow k}/A}),
\end{equation*}
and the vector from the $k$-th coil to the $j$-th coil ${r}_{j\leftarrow k}$ yields
\begin{equation*}
    \begin{aligned}
        &\left\{
        \begin{aligned}
            &\Psi_f=
            \frac{1}{{\|r_{j\leftarrow k}\|^4}}
            {\begin{bmatrix}
                -6&0&0&0&3&0&0&0&3\\
                0&3&0&3&0&0&0&0&0\\
                0&0&3&0&0&0&3&0&0
                \end{bmatrix}}\\
                &\Psi_\tau=
                \frac{1}{\|r_{j\leftarrow k}\|^3}
                {\begin{bmatrix}
                0&0&0&0&0&1&0&-1&0\\
                0&0&2&0&0&0&1&0&0\\
                0&-2&0&-1&0&0&0&0&0
            \end{bmatrix}}
        \end{aligned}
        \right.
    \end{aligned}.
\end{equation*}
\begin{definition}\label{NODA_C_LOS_new} 
    The line-of-sight frame $\{\mathcal{LOS}_{j\leftarrow k}\}$ \cite{takahashi2024neural} is attached to the $k$-th agent and oriented toward the $j$-th agent. It is defined in (\ref{NODA_new_frame}), with the associated coordinate transformation matrix $C^{O/L{j\leftarrow k}}\in\mathbb{R}^{3\times 3}$ expressed as $C^{A/L_{j\leftarrow k}}=\mathcal{C}(r_{j\leftarrow k}^a,\ f^a_{j}\times r^a_{j\leftarrow k})$ where we define the coordinate transformation matrix given the arbitrary vectors $v{^a},w{^a}\in\mathbb{R}^3$:
    \begin{equation}
        \label{NODA_new_frame}
        \mathcal{C}{(v{^a},w{^a})}=
              \begin{bmatrix}
                \mathsf{e}_x\triangleq\frac{v{^a}}{\|v{^a}\|},\mathsf{e}_y\triangleq
                \frac{{v{^a}}\times w{^a}}{\|{v{^a}}\times w{^a}\|},\mathsf{e}_x\times \mathsf{e}_y
             \end{bmatrix}.
    \end{equation}
\end{definition}

\subsection{Optimal Electromagnetic Swarm Control}
\label{Optimal_Electromagnetic_Swarm_Control}
We consider the position and attitude control of multiple satellites using the magnetic interaction model defined in the previous subsection. We assume that the $j$-th magnetic moment is driven by a sinusoidal wave \cite{takahashi2024neural}:
\begin{equation}
    \label{2_1::AC coil dipole moment}
    \bm{\mu}_{j}(t)= \overline{\bm{\mu}}_j \sin(\omega_{j}t+\bm{\theta}_j)=\bm{s}_{j} \sin(\omega_{{j}}t)+\bm{c}_{j} \cos(\omega_{{j}}t),
\end{equation}
where the amplitudes of the cosine and sine components are $\bm{s}_j\in\mathbb{R}^3$ and $\bm{c}_j\in\mathbb{R}^3$, respectively, and $\bm{\theta}_j\in \mathbb{R}^3$ are phases for the $j$-th MTQ. The first-order averaged input $\overline{u}_{j \leftarrow k}\in\mathbb{R}^6$ is\begin{equation}
\label{Feasibility_bilinear_polynomial_model}
\overline{u}_{j\leftarrow k}=
\frac{1}{2}\frac{\mu_0}{4\pi}Q_{{j\leftarrow k}}
\left (
{s_{k}^{a}}
\otimes
{s_{j}^{a}}
+
{c_{k}^{a}}
\otimes
{c_{j}^{a}}
\right )\ \mathrm{if}\ \omega_j=\omega_k.
\end{equation}
We formulate the power-optimal dipole allocation \cite{takahashi2024neural} as
\begin{equation}
    \label{2-2::opt1}
    \begin{aligned}
    \mathrm{min}\quad&J_{\mathrm{p}}=\left\|[{s}_j^a;{s}_k^a;{c}_j^a;{c}_k^a]\right\|^2/2\\
        \mathrm{s.t.}\quad& \frac{\mu_0}{8\pi}Q_{j\leftarrow k}
        \left (
        {s_{k}^a}
        \otimes
        {s_{j}^a}
        +
        {c_{k}^a}
        \otimes
        {c_{j}^a}
        \right )=u_{j\leftarrow k}^a
    \end{aligned}.
\end{equation}
and its Lagrange dual problem of the problem in (\ref{2-2::opt1}) is
\begin{equation}
    \label{2-2::opt2}
    \max_{\rev{\lambda_d} \in \mathbb{R}^{6}}\ J_{\mathrm{d}}=\frac{-\rev{\lambda_d^\top} u^{a}_{j\leftarrow k}}{\mu_0/(8\pi)}\quad\mathrm{s.t.}\quad P_{\rev{\lambda_d}}=
    \begin{bmatrix}
    E_3&R_{\rev{\lambda_d}}\\
    R_{\rev{\lambda_d}}^\top &E_3
    \end{bmatrix} \succeq 0,
\end{equation}
where the Lagrange multiplier vector $\lambda_d \in \mathbb{R}^6$ and $R_{\rev{\lambda_d}}\in\mathbb{R}^{3\times 3}$ satisfies $\mathrm{vec}(R_{\rev{\lambda_d}})=Q_{j\leftarrow k}^\top\rev{\lambda_d}$ \cite{takahashi2024neural}. 
Thus, the required magnetic dipole power magnitude for control can be derived using (\ref{2-2::opt2}) \cite{takahashi2024neural} because the lower bound is obtained as $J_d\leq J_p$ \cite{boyd2004convex} and $J_d= J_p$ for $n=2$ \cite{takahashi2026power}.
\subsection{Planar Phased-Array Antenna Formulation}
\label{Planar_Phased_Array_Antenna_Formulation}
This subsection introduces the effective isotropic radiated power (EIRP) for a planar phased-array antenna characterized by uniform element spacing, omnidirectional elements, and identical excitation amplitudes across all elements. The EIRP is the product of antenna gain $G_T$ and the transmitting power $P_T$ \cite{maral2009satellite}, where $P_T$ is the total of each element transmitting power $P_t$. Then, we obtain $EIRP = P_T G_T= ( P_t N_{lx}N_{ly})G_T$ where $N_{lx}$ and $N_{ly}$ are the number of satellites per grid side, respectively. 
Assuming that the antenna loss is negligible, \(G_T\) is equal to the maximum directivity \(\overline{D}\) \cite{balanis2005antenna}:
\begin{equation}
    \label{2-3:: antenna directivity}
    \overline{D}
    \triangleq
    \max_{\theta,\phi}D(\theta,\phi)
    =
    \max_{\theta,\phi}
    \frac{4\pi B(\theta,\phi)}
    {\int_0^{2\pi}\int_0^{\pi} B(\theta,\phi)\sin\theta\,\mathrm{d}\theta\mathrm{d}\phi},
\end{equation}
where \(\theta\) and \(\phi\) are the elevation and azimuth angles in the polar coordinate system, as shown in Fig.~\ref{fig:enter-label}. Here, \(B(\theta,\phi)\) denotes the dimensionless normalized gain pattern, which is defined from the normalized field array factor \(A(\theta,\phi)\) as
\begin{equation}
    \label{2-3:: normalized gain pattern}
    B(\theta,\phi)=|A(\theta,\phi)|^2.
\end{equation}
For a rectangular planar array with uniform excitation, the normalized field array factor is expressed as the product of two linear-array factors $A(\theta,\phi) = A_x(u_x)A_y(u_y)$ \cite{balanis2005antenna} where 
\begin{equation}
    \label{2-3:: plane array factor}
    \begin{gathered}
        A_x(u_x) = \frac{1}{N_{lx}} \frac{\sin\left(N_{lx}u_x\right)}{\sin\left(u_x\right)}, \quad
        A_y(u_y) = \frac{1}{N_{ly}} \frac{\sin\left(N_{ly}u_y\right)}{\sin\left(u_y\right)}.
    \end{gathered}
\end{equation}
Here, \(u_x\) and \(u_y\) are defined as
\begin{equation}
    \label{2-3:: array phase variables}
    \begin{aligned}
        u_x
        &=
        k d_x
        \left(
        \sin\theta\cos\phi
        -
        \sin\theta_0\cos\phi_0
        \right),\\
        u_y
        &=
        k d_y
        \left(
        \sin\theta\sin\phi
        -
        \sin\theta_0\sin\phi_0
        \right),
    \end{aligned}
\end{equation}
where \(k=\pi/\lambda\), \(\lambda\) is the wavelength, \(d_x\) and \(d_y\) are the element spacings along the two array axes, and \((\theta_0,\phi_0)\) denotes the main-beam direction.
When mutual coupling is neglected, averaging (\ref{2-3:: antenna directivity}) over all directions gives \cite{lo2013antenna, gilbert1955optimum}
\begin{equation}
    \label{2-3:: simple plane directivity}
    G_T=\overline{D}\approx N_l^2.
\end{equation}
Finally, $N_{l}=N_{lx}=N_{ly}$ derives EIRP as follows:
\begin{equation}
    \label{2-4:: EIRP}
        \mathrm{EIRP}_{\max}\triangleq P_TG_T=P_tN_l^4.
\end{equation}
\section{Problem Formulation}
\label{problem_formulation}
\subsubsection{Grid Structure Array Model}
\begin{figure}[!tb]
    \centering
    \begin{minipage}[b]{0.47\FigWidth}
        \centering
        \subfloat[Planar-array antenna.]{\includegraphics[width=1\linewidth]{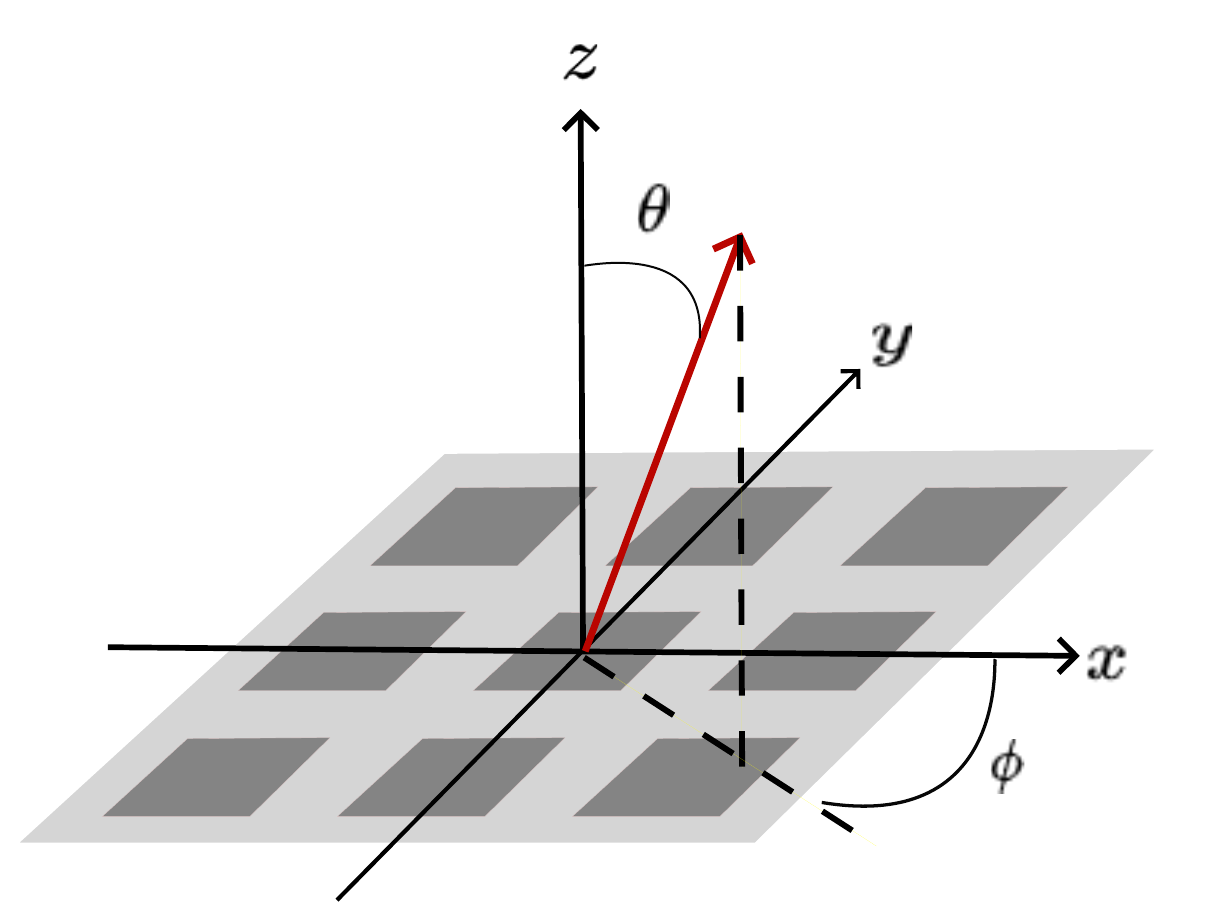}\label{fig:enter-label}}
    \end{minipage}
    \begin{minipage}[b]{0.5\FigWidth}
        \centering
        \subfloat[Distributed space antenna \cite{takahashi2025distance,takahashi2025scalable}.]{\includegraphics[width=1\columnwidth]{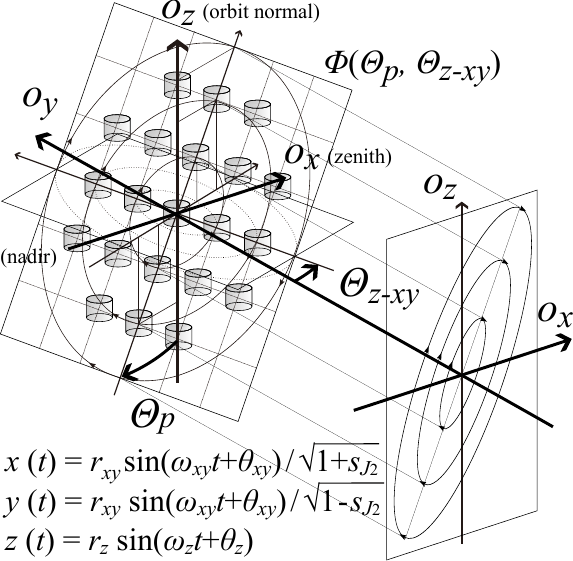}\label{grid_formation_ver2}}
    \end{minipage}\\
    \vspace{0.25cm}
        \begin{minipage}[b]{0.95\FigWidth}
        \centering
        \subfloat[Linear formation model of $2n+1$ satellites \cite{takahashi2026power}.]{\includegraphics[width=1\columnwidth]{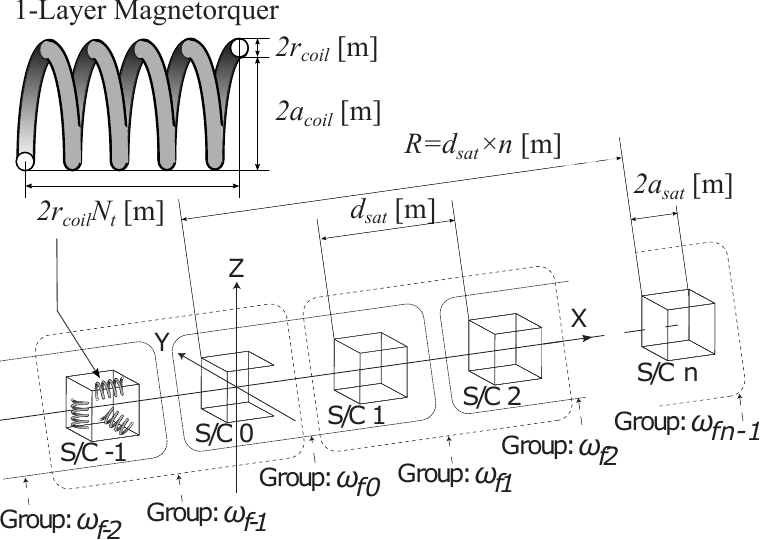}\label{PSA_figure_grid_linear_approximation}}
    \end{minipage}
    \caption{Grid-structured approximation for distributed space system design. This formation consists of equally spaced linear arrays.}\label{fig:Grid_structure_overview}
\end{figure}

We primarily use a simplified model of a square phased array in Fig.~(\ref{fig:Grid_structure_overview}) and the elements are equally spaced on a grid. 
\begin{assumption}
\label{mass_size_constant_mass_and_distance_assumption}
We consider a distributed space antenna of a square phased array as shown in Fig.~\ref{grid_formation_ver2}, i.e.,
\begin{equation}
\label{N_all_square_assumption}
N_{\mathrm{all}}\triangleq N_l^2\triangleq (2n+1)^2,
\end{equation}
where $N_l$ is used for antenna context in subsection~\ref{Planar_Phased_Array_Antenna_Formulation}. Its total system mass $\overline{m}_{\mathrm{sys}}$, their inter-element distance $d_{\mathrm{sat}}$, and associated total length $r_l$ are user-defined constant values:
$$
\overline{m}_{\mathrm{sys}}\triangleq \mathrm{const.},\quad d_{\mathrm{sat}}\triangleq \mathrm{const.},\quad r_l\triangleq(2n+1)d_{\mathrm{sat}},
$$
Its arbitrary row-wise linear formation illustrated in Fig.~\ref{PSA_figure_grid_linear_approximation} havs the distance vector $R_l(\tau)\in\mathbb{R}^3$ from the ($-n$)th satellite to the ($n$)th satellite as  
$$
R_l(\tau)\triangleq r_l(\tau)\hat{p}(\tau)\ \parallel\ {p}_d(\tau) ,\quad \tau\in[0,2\pi/\omega_{xy}).
$$
\end{assumption}
\begin{example}
\label{antenna_wavelength_distance}
The distance $d_{\mathrm{sat}}$ is set to be no greater than half a wavelength $\lambda$ to suppress grating lobes—undesired radiation with comparable strength to the main lobe peak. 
\begin{equation}
\label{antenna_wavelength_distance_equation}
d_{\mathrm{sat}}\triangleq \mathrm{const.}\leq \lambda/2 .   
\end{equation}
\end{example}
\begin{definition}
\label{component_assumption_}
The total system mass $m_{\mathrm{sys}}$ and its relaxed constraint are defined by the total number of satellites $N_{\mathrm{all}}=(2n+1)^2$ by (\ref{N_all_square_assumption}) in Assumption~\ref{mass_size_constant_mass_and_distance_assumption}:
\begin{equation}
    \label{3-1:: M_sys gamma}
       N_{\mathrm{all}}\ m_{\mathrm{sat}} \triangleq \overline{m}_{\mathrm{sys}},\quad |N_{\mathrm{all}}\ m_{\mathrm{sat}}-\overline{m}_{\mathrm{sys}} | \overset{\triangle}{\leq} \gamma_{\mathrm{sys}} ,
\end{equation}
where $\overline{m}_{\mathrm{sys}}$ is the user-defined total system mass and $\gamma_{\mathrm{sys}}$ is the tolerance of the total system mass for computation.
\end{definition}
\subsubsection{Decentralized Control Model}
To suppress unintended coupling among nonadjacent satellites in close proximity, we first assume the use of frequency allocation introduced in subsection~\ref{Optimal_Electromagnetic_Swarm_Control} to confine electromagnetic interactions to neighboring satellites.
\begin{assumption}
\label{AC_backet_condition}
Control pairs are defined by assigning distinct AC angular frequencies $\omega_{fk}$ for $k\in[-n,\ n]$ to adjacent satellite groups, as illustrated in Fig.~\ref{PSA_figure_grid_linear_approximation} (Please refer the detailed selection of angular frequencies \cite{takahashi2024neural}.). 
\end{assumption}
\section{Distributed Antenna Design Optimization}
\label{sec:proposed_dsa_opt}
This section presents a nonconvex optimization-based framework to evaluate the feasibility of distributed space antennas. In the next section, considering the practical constraints, the detailed design problem is solved to yield feasible design solutions and more refined scaling trends.
\subsection{Overview of Design Optimization Problem}
\label{Result_overview}
We provide an overview of the design framework for the distributed space antenna and clarify the constraints. We consider the design parameters $X$ of our framework for a given satellite distance $d_{\mathrm{sat}}$ in (\ref{antenna_wavelength_distance_equation}) of Example~\ref{antenna_wavelength_distance} and system mass $\overline{m}_{\mathrm{sys}}$ in Assumption~\ref{mass_size_constant_mass_and_distance_assumption}:
\begin{equation}
\label{design_parameters}
X=\left\{2a_{\mathrm{coil}},\ 2a_{\mathrm{sat}},\ q_{\mathrm{coil}},\ n,\ \rev{u_{\mathrm{psl}}}\right\},
\end{equation}
where the coil diameter $2a_{\mathrm{coil}}$, satellite size $2a_{\mathrm{sat}}$, coil parameter $q_{\mathrm{coil}}$, satellite number $n$, and peak sidelobe variable $u_{\mathrm{psl}}$. Then, the design optimization $\mathcal{P}_{\mathrm{DSA}}$ is formulated as follow:
\begin{equation}
    \label{problem_formulation_ver1}
    \begin{aligned}
            &\ \mathit{X^*}=\argmin_{X\in\chi_{\mathrm{feas}}} L_{\mathrm{bw}}=\argmax_{X\in\chi_{\mathrm{feas}}} N_{\mathrm{all}}\qquad 
            \text{subject to:}\\
            &\left\{
            \begin{aligned}
            &\text{{(A)\quad Satellite specification constraints}}\\        
            &\text{{(B)\quad Phased-array antenna constraints}}\\        
            &\text{{(C)\quad Magnetic formation-keeping constraints}}
            \end{aligned}
            \right.
    \end{aligned}
\end{equation}
where the constraints in (A)–(C) are described in Subsections~\ref{method}, \ref{subsec: antenna req}, and \ref{section:grid}, respectively. The optimal $X^{*}$ yields other dependent parameters, such as the generation and consumption power, EIRP, and component masses.
\subsection{Constraints 1/3: Satellite Specification}
\label{method}
This section defines the design parameters for each satellite and formulates the power consumption and satellite constraints. We define the satellite configuration as follows.
\begin{definition}
\label{component_assumption_1}
The satellite consists of 1) an identical three-axis coil, 2) solar panels, 3) a battery, 4) a satellite body, and 5) a bus that includes antennas, transmitters, and avionics:
$$
m_{\mathrm{sat}}\triangleq m_{\mathrm{3coil}} + \rev{m_{\mathrm{sap}}} + m_{\mathrm{bat}} + m_{\mathrm{str}} + m_{\mathrm{bus}},
$$
where each term corresponds to the mass of each subsystem.
\end{definition}
\subsubsection{Component Size Constraints}
This subsection summarizes the constraints on the satellite and coil sizes.
Size is a critical parameter that involves trade-offs between mass, power acquisition, and achieving of the magnetic moment. We formulate Assumption \ref{ass: size} considering the obvious physical constraints of satellites.
\begin{assumption}
\label{ass: size}
The geometric relationship among the satellite radius, coil radius, and inter-satellite distance is illustrated in Fig.~\ref{PSA_figure_grid_linear_approximation}.
The coil radius $a_{\mathrm{coil}}$ is smaller than the satellite one $a_{\mathrm{sat}}$ and $2a_{\mathrm{sat}}$ is smaller than the distance between the satellite centers $d_{\mathrm{sat}}$ with the coil size margin $r_{\mathrm{mar}}$
\begin{equation}
    \label{3-1:: coil and sat size}
        a_{\mathrm{coil}} \overset{\triangle}{\leq} a_{\mathrm{sat}} - r_{\mathrm{mar}},\quad 
        2a_{\mathrm{sat}} \overset{\triangle}{\leq} d_{\mathrm{sat}},\quad  k_Fa_{\mathrm{coil}} \overset{\triangle}{\leq} d_{\mathrm{sat}},
\end{equation}
where the final inequality is from the assumption that the distance between the coils is sufficiently large compared with the coil radius with defined coefficient $k_F$.
\end{assumption}
\subsubsection{Component Mass Definitions and Assumptions}
This subsection defines the masses of the satellite components. We implement Assumption~\ref{component_assumption_2} based on Definition~\ref{component_assumption_1} to define the component masses $m_{\mathrm{3coil}}$, $\rev{m_{\mathrm{sap}}}$, $m_{\mathrm{bat}}$, $m_{\mathrm{str}}$, and $m_{\mathrm{bus}}$ as dependent parameters of the other parameters.
\begin{assumption}
\label{component_assumption_2}
The satellite has a cubic shape and an identical three-axis coil, that is, with an equal number of turns $N_t$, wire radius $r_{\mathrm{coil}}$, and coil radius $a_{\mathrm{coil}}$. Solar panels are attached to the four sides of the satellite. The bus mass $m_{\mathrm{bus}}$ is treated as a user-defined constant, $m_{\mathrm{bus}0}$. The structural mass $m_{\mathrm{str}}$ is defined as a proportion of the total satellite mass. $P_{\mathrm{sap}}$ denotes the power generated during sunlit operation:
\begin{equation}
\label{eq:solar_power}
P_{\mathrm{sap}}(a_{\mathrm{sat}}) \triangleq k_{\mathrm{sap}}\, P_{\mathrm{W/m^2}} (2a_{\mathrm{sat}})^2,
\end{equation}
where $P_{\mathrm{W/m^2}}$ is the power generated per unit area of the solar panel and $k_{\mathrm{sap}}$ is the effective area coefficient.

The battery storage requirement is modeled as the solar energy that is harvested over a sunlit charging duration $h_{\mathrm{charge}}$ and a design coefficient that captures the charging efficiency and usable fraction $k_{\mathrm{bat}}$ using $P_{\mathrm{sap}}$ in \eqref{eq:solar_power}.
\end{assumption}
\begin{example}[Requirement for battery storage from sunlit charging]
\label{ex:bat_requirement}
As a concrete sizing example, consider a baseline case in which the satellite stores $10\%$ of the energy that is harvested during a $12$-h sunlit period. 
This corresponds to setting $k_{\mathrm{bat}}=0.1$ and $h_{\mathrm{charge}}=12$. 
Under this setting, the required battery storage is modeled as
\begin{equation}
\label{eq:ex_Ebat}
k_{\mathrm{bat}}\,h_{\mathrm{charge}}\,P_{\mathrm{sap}}(a_{\mathrm{sat}})
= 1.2\,P_{\mathrm{sap}}(a_{\mathrm{sat}}),
\end{equation}
where $P_{\mathrm{sap}}(a_{\mathrm{sat}})$ is determined by \eqref{eq:solar_power}.
\end{example}

\noindent
This Assumption~\ref{component_assumption_2} defines the mass of the three-axis coil, four solar panels, battery, structure, and bus system:
\begin{equation}
    \label{3-1::Mass of str bat bus}
    \left\{
    \begin{aligned}    
        m_{\mathrm{3coil}}(a_{\mathrm{coil}},\ q_{\mathrm{coil}})&\triangleq 
        3(2\pi^2 a_{\mathrm{coil}})q_{\mathrm{coil}}\rho_{c},\\
        m_{\mathrm{sap}}(a_{\mathrm{sat}})&\triangleq \nu_{\mathrm{sap}}\rho_{\mathrm{sap}}(2a_{\mathrm{sat}})^2, \\
        m_{\mathrm{bat}}(a_{\mathrm{sat}})&\triangleq\rho_{\mathrm{bat}}k_{\mathrm{bat}}\ h_{\mathrm{charge}}\ P_{\mathrm{sap}}(a_{\mathrm{sat}}),\\
        m_{\mathrm{str}}(m_{\mathrm{sat}})&\triangleq\eta_{\mathrm{str}} m_{\mathrm{sat}},\\ 
        m_{\mathrm{bus}}&\triangleq m_{\mathrm{bus}0} = \mathrm{const},
    \end{aligned}
    \right.
\end{equation}
\rev{where we define the coil parameter $q_{\mathrm{coil}}\triangleq N_t r_{\mathrm{coil}}^2$, leaving $N_t$ and $r_{\mathrm{coil}}$ free to fit the satellite geometry, $\rho_c$ denotes the mass density of the wire, $\rho_{\mathrm{sap}}$ is the mass per unit area density, $\nu_{\mathrm{sap}}$ is the number of solar panels that are mounted on the satellite, $\eta_{str}$ is a user-defined constant, and $\rho_{\mathrm{bat}}$ is the mass per unit power capacity of the battery.} \subsubsection{Component Mass Constraints}
\begin{figure}[t]
  \centering
\begin{minipage}[t]{1\linewidth}
\centering
\captionof{table}{Previous satellite designs}
\label{tab:previous_satellite_designs}
\begin{tabular}{c|c|c}\hline
&$V$ (cm$^3$)&$m_{\mathrm{sat}}$ (g)\\
\hline
\cite{barnhart2007very}&$2\times 2\times 0.075$&10\\
\cite{hadaegh2014development}&$4\times 4\times 4.25$&95.5\\
\cite{hu2019development,cao2015novel}&$3.3\times 3.3\times 0.5$&9.9\\
\hline
\end{tabular}
  \end{minipage}
  \begin{minipage}{1\linewidth}
    \centering
    \includegraphics[width=0.9\linewidth]{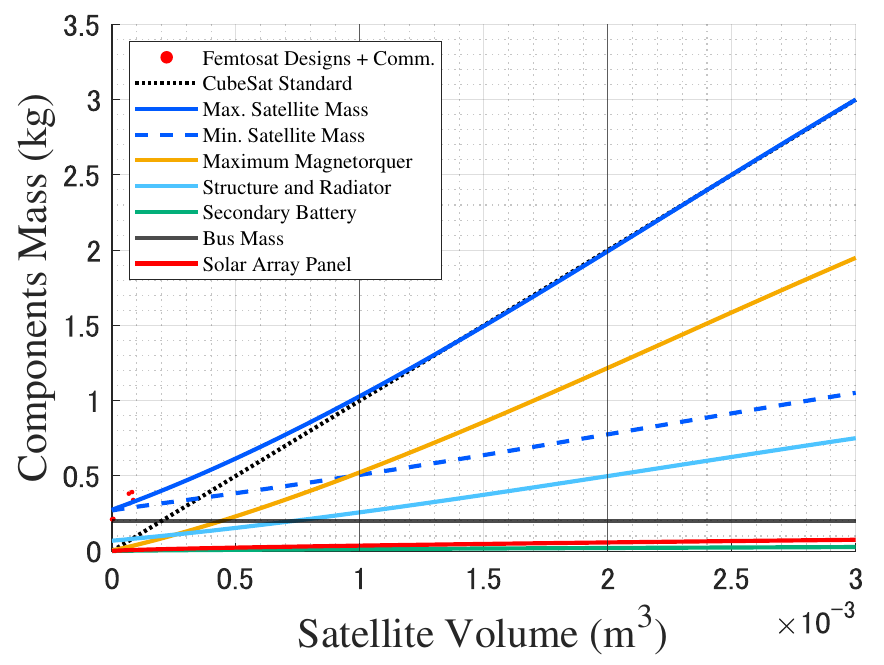}
    \caption{\rev{Relationship between satellite volume and component mass.}}
    \label{fig:existing sat}
  \end{minipage}
  \label{fig:fig_table_combined}
\end{figure}

Finally, we calculate the component mass constraints. As indicated in (\ref{3-1::Mass of str bat bus}), our assumption limits the masses of almost all components as dependent variables of $a_{\mathrm{sat}}$, except for the coil mass $m_{\mathrm{3coil}}(a_{\mathrm{coil}},q_{\mathrm{coil}})$. This $m_{\mathrm{3coil}}$ should be designed to satisfy the requirements of antenna performance and formation control. Subsequently, we derive the upper bound $\overline{m}_{\mathrm{sat}}$ to avoid impractical coil configurations by assuming that ${m}_{\mathrm{sat}} \propto a_{\mathrm{sat}}^{3}$.
Because the total mass should converge to a finite lower bound, we statistically derive this relationship from the existing satellite design parameters.
\begin{assumption}
\label{asp:: sat mass}
$\overline{m}_{\mathrm{sat}}$ is a polynomial function of $a_{\mathrm{sat}}$:
\begin{equation}
    \label{overline_mass}
\overline{m}_{\mathrm{sat}}(a_{\mathrm{sat}})\triangleq 
\left\{
\begin{aligned}
&k_{\overline{m}}\ a_{\mathrm{sat}}^3 &&\mathrm{if}\quad 2a_{\mathrm{sat}}\geq 0.1\\
&f_{\mathrm{emp}}(a_{\mathrm{sat}})&&\mathrm{otherwise},\ \\
\end{aligned}
\right.
\end{equation}
where $k_{\overline{m}}$ is an arbitrary parameter for which we mainly select $k_{\overline{m}}=1e^{3}$ for the standard CubeSat design, and $f_{\mathrm{emp}}(a_{\mathrm{sat}})$ denotes an empirically approximated function that is obtained by fitting the previously designed satellite mass, e.g., the parameters in Table~\ref{tab:previous_satellite_designs} and red markers in Fig.~\ref{fig:existing sat}.
\end{assumption}
\noindent
This methodology using $f_{\mathrm{emp}}$ yields design solutions that are consistent with current technology and is adaptable using updated data from next-generation satellites. Subsequently, we derive the upper and lower bounds ${m}_{\mathrm{sat}}(a_{\mathrm{sat}})$\rev{:} \begin{equation}
    \label{3-1::Mass of existing sat}
    \begin{aligned}
        &m_{\mathrm{sat}} \overset{\triangle}{\leq}\overline{m}_{\mathrm{sat}}(a_{\mathrm{sat}})\\
        &\rev{m_{\mathrm{sap}}(a_{\mathrm{sat}})+m_{\mathrm{bat}}(a_{\mathrm{sat}})+m_{\mathrm{str}}(m_{\mathrm{sat}})}+m_{\mathrm{bus}}\overset{\triangle}{\leq} m_{\mathrm{sat}},\\
    \end{aligned}
\end{equation}
where $\overline{m}_{\mathrm{sat}}(a_{\mathrm{sat}})$ is determined by (\ref{overline_mass}) and the lower bound is derived as $m_{\mathrm{3coil}}=0$.
\subsection{Constraints 2/3: Phased-Array Antenna}\label{subsec: antenna req}

The antenna performance constraints of the distributed space antenna are formulated in this section.
We adopt a simplified antenna model consisting of an $N_l \times N_l$ square grid to render the analysis tractable while preserving system-level trends. 
\subsubsection{Tractable Beam Footprint Model}
\label{3-3::coverage_beamwidth}
We first formulate a computationally tractable beam footprint model by approximating the beam edge with the first null of the main lobe. The footprint boundary is characterized by the beamwidth in the diagonal cut \((\phi=\pi/4)\), which reduces the two-dimensional beam projection to a uniaxial representation.

\begin{assumption}[Azimuth restriction for footprint evaluation]
\label{3-3::antenna_assumption_1}
The azimuth is fixed at \(\phi=\phi_0=\pi/4\), that is, the diagonal cut where \(u_x=u_y=u\), to reduce the dimensionality of the footprint evaluation.
\end{assumption}
\noindent
Under this restriction, the two-dimensional normalized field array factor of the square grid array is transformed into
\begin{equation}
\label{3-3:: plane grid array factor pi/4}
A_{\mathrm{diag}}(u)
=
\left(
\frac{\sin\left(N_lu\right)}
{N_l\sin\left(u\right)}
\right)^2,
\quad
u=
\frac{k d_{\mathrm{sat}}}{\sqrt{2}}
\left(\sin\theta-\sin\theta_0\right).
\end{equation}
Here, \(A_{\mathrm{diag}}(u)\) is the direction-restricted normalized field array factor. The corresponding normalized gain pattern is given by \(|A_{\mathrm{diag}}(u)|^2\), but only the null location of \(A_{\mathrm{diag}}(u)\) is required for the footprint approximation. The beam edge is therefore defined by the first null condition \(N_lu=\pi\), which yields the null angle \(\theta_{n1}\).
\begin{assumption}[Footprint Definition]
\label{3-3::antenna_assumption_2}
The beam footprint is defined by the first null of the main lobe evaluated along the specific azimuth direction defined in Assumption~\ref{3-3::antenna_assumption_1}. It is assumed that the orbital altitude $h$ is significantly larger than the element spacing $d_{\mathrm{sat}}$, i.e., $h \gg d_{\mathrm{sat}}$.
\end{assumption}
\noindent
Using the geometric relationship based on Assumption~\ref{3-3::antenna_assumption_2}, the footprint diameter $D_{fp}$ is derived as a function of $\theta_{n1}$ and the orbital altitude $h$:
\begin{equation}
\label{3-3:: null point}
\theta_{n1}\triangleq \arcsin\left(\frac{\sqrt{2}\lambda}{N_l d_{\mathrm{sat}}}+\sin\theta_0\right),\quad D_{fp} \approx 2|\theta_{n1}-\theta_0|h.
\end{equation}
\subsubsection{Peak-Sidelobe and Transmitter Power Derivation}
\label{3-3::psl_power}
The local maximizer of \(B_{\mathrm{env}}(u)\) within the first sidelobe region defines the peak-sidelobe envelope point \(u_{\mathrm{psl}}\):
\begin{equation}
    \label{3-3:: u_msl constraint}
    \left\{
    \begin{aligned}
        &N_l \sin(u_{\mathrm{psl}})\cos(N_lu_{\mathrm{psl}})-\cos(u_{\mathrm{psl}})\sin(N_lu_{\mathrm{psl}})\triangleq0,\\
        &\frac{\pi}{N_l}\overset{\triangle}{<}|u_{\mathrm{psl}}|\overset{\triangle}{<}\frac{2\pi}{N_l}.
    \end{aligned}
    \right.
\end{equation}
For peak-sidelobe-constrained transmit-power sizing, we use the one-dimensional sidelobe envelope of the separable square array. From the separable field array factor in (\ref{2-3:: plane array factor}) and the normalized gain definition in (\ref{2-3:: normalized gain pattern}), the two-dimensional response can be expressed as
\[
B_{\mathrm{2D}}(u_x,u_y)=B_x(u_x)B_y(u_y),
\]
where \(B_x\) and \(B_y\) are normalized by the main-beam peak. Since the maximum contribution from the other array axis is unity, i.e., \(\max_{u_y}B_y(u_y)=1\), the largest two-dimensional response for a given \(u_x=u\) is bounded by
\[
\max_{u_y}B_{\mathrm{2D}}(u,u_y)
=
B_x(u)\max_{u_y}B_y(u_y)
=
B_x(u).
\]
Therefore, we define the sidelobe-envelope measure as
\begin{equation}
    \label{3-3:: sidelobe envelope}
    B_{\mathrm{env}}(u)
    =
    B_x(u)
    =
    \left|
    \frac{\sin\left(N_lu\right)}
    {N_l\sin\left(u\right)}
    \right|^2 .
\end{equation}
This quantity is a dimensionless normalized gain pattern used for system-level sizing. It is distinguished from the direction-restricted field array factor \(A_{\mathrm{diag}}(u)\) used for the footprint approximation. The sidelobe EIRP evaluated by the envelope model must be smaller than the received power indicator \(I_R\):
\begin{equation}
    \label{3-3:: maximum sidelobe constraint}
    \mathrm{EIRP}_{\max} B_{\mathrm{env}}(u_{\mathrm{psl}})
    \overset{\triangle}{\leq}
    I_R .
\end{equation}
The RF transmit power satisfying the equality in (\ref{3-3:: maximum sidelobe constraint}) is obtained from \(\mathrm{EIRP}_{\max}=P_tN_l^4\) and (\ref{3-3:: sidelobe envelope}) as follows:
\begin{equation}
    \label{3-3:: P_T maximum sidelobe constraint_original}
    P_t
    \triangleq
    \frac{I_R}{N_l^4 B_{\mathrm{env}}(u_{\mathrm{psl}})}
    =
    \frac{I_R}{N_l^2}
    \left(
    \frac{\sin(u_{\mathrm{psl}})}
    {\sin(N_lu_{\mathrm{psl}})}
    \right)^2 .
\end{equation}

\begin{remark}
Assuming a direct-to-device (D2D) communication scenario, $I_R$ is expressed as an example as follows:
\begin{equation}
    I_R \triangleq  \zeta P_R L_f / G_R, \quad L_f = \left( \frac{4\pi h}{\lambda} \right)^2,
\end{equation}
where $P_R$ is the required received power, $G_R$ is the antenna gain, $\zeta \in (0, 1]$ is the attenuation factor, and $L_f$ is the free-space path loss. Then, the transmit power is
\begin{equation}
    \label{3-3:: P_T maximum sidelobe constraint_d2d}
    P_t\triangleq \frac{\zeta P_R L_f}{N^2_l G_R} \left(\frac{\sin(u_{\mathrm{psl}})}{\sin(N_lu_{\mathrm{psl}})}\right)^2.
\end{equation}
\end{remark}

\begin{remark}
The variable \(u_{\mathrm{psl}}\) is introduced to specify the finite-array peak point of the sidelobe envelope used in transmit-power sizing. It is not intended to represent a full two-dimensional radiation-pattern search.
\end{remark}

\subsection{Constraints 3/3: Magnetic Formation-Keeping}
\label{section:grid}
We describe the power requirements for a distributed space antenna to execute its communication mission while maintaining relative inter-satellite positions against disturbances. \subsubsection{Power Budget for Steady-Formation Maintenance}
\label{sec:steady_state_power_budget}

A steady-state power-budget constraint for long-term magnetic formation maintenance is formulated in this subsection. 
\begin{definition}
\label{def:power_breakdown}
The steady-state power consumption includes a control power $P_{\mathrm{cont}}$, mission power $P_{\mathrm{mis}}$, bus power $P_{\mathrm{bus}}$, and margin power $P_{\mathrm{mar}}$, where $P_{\mathrm{bus}}$ is a constant:
\begin{equation}
\label{eq:power_total}
    P_{\mathrm{tot}}\triangleq P_{\mathrm{cont}} + P_{\mathrm{mis}} + P_{\mathrm{bus}} + P_{\mathrm{mar}}.
\end{equation}
\end{definition}

\begin{assumption}
\label{asp:steady_power_balance}
As a steady-state condition, the power is supplied only by solar panels, and the time-averaged consumption remains below the generated power: $P_{\mathrm{tot}} \leq P_{\mathrm{sap}}$.
\end{assumption}
\noindent
\begin{assumption}
\label{asp:control_power_square_grid}
The required power for each axis is evaluated from the peak coil output. The control power for one axis is defined as
\begin{equation}
\label{eq:control_power_1axis_def}
P_{\mathrm{1axis}} = R_{\mathrm{coil}}\, c_{\mathrm{coil}}^2
= \frac{2 p_c \overline{\mu}_{\mathrm{2dir}}^2}{\pi^2 q_{\mathrm{coil}} a_{\mathrm{coil}}^3},
\end{equation}
where $R_{\mathrm{coil}}$ is the coil resistance defined in (\ref{2_1::coil dipole moment}), $c_{\mathrm{coil}}$ is the coil current, and $\overline{\mu}_{\mathrm{2dir}}$ is the upper bound of magnetic moment required for bidirectional compensation along one axis.
\end{assumption}
When AC actuation is used to alternate the moment command between two opposite directions along one axis, the synthesized magnetic moment $\mu_{\mathrm{2dir}}(t)$ is written as
\begin{equation}
    \label{3_3::coil alt dipole moment synthesis}
    \begin{aligned}
        \mu_{\mathrm{2dir}}(t)
        &= \mu_{\mathrm{1dir}} \cos(\omega_{f_n} t)
        + \mu_{\mathrm{1dir}} \cos\left(\omega_{f_n} t + \frac{\pi}{2}\right) \\
        &= \sqrt{2}\,\mu_{\mathrm{1dir}} \rev{\cos}\left(\omega_{f_n} t + \frac{\pi}{4}\right),
    \end{aligned}
\end{equation}
where $\omega_{f_n}$ is the AC frequency assigned to the $n$-th satellite and the upper bound of $\overline{\mu}_{\mathrm{2dir}}$ is $\sqrt{2}\mu_{\mathrm{1dir}}$.
Accordingly, the total control power at the central satellite is given by
\begin{equation}
    \label{eq:control_power_2axis}
    P_{\mathrm{cont}} = P_{\mathrm{2axis}} = 2 P_{\mathrm{1axis}}.
\end{equation}

We also model the mission power $P_{\mathrm{mis}}$ as the transmitter power consumption and the margin power $P_{\mathrm{mar}}$ for another purpose, e.g., deployment, are defined as 
\begin{equation}
    \label{eq:control_power_total}
    P_{\mathrm{mis}}=P_t  /  \eta_{\mathrm{tra}}
    ,\quad
    P_{\mathrm{mar}} = \frac{2p_c \overline{\mu}_{\mathrm{mar}}^2}{\pi^2 q_{\mathrm{coil}} a_{\mathrm{coil}}^3},
\end{equation}
where $\eta_{\mathrm{tra}}$ is the transmitter efficiency and $P_t$ is the per-satellite transmit power and $\overline{\mu}_{\mathrm{mar}}$ denotes the upper bound of margin magnetic moment. 
\subsubsection{Magnetic Moment Estimation by Numerical Integration}
We numerically calculate the power consumption during steady-state operations for a square-grid formation. We compute the maximum control power at each time step through actual numerical simulations in order to evaluate the largest power consumption within the formation. This avoids the conservatism of the convex optimization-based power analysis in \cite{takahashi2026power}, which assumes perfect disturbance cancellation. Although the optimal values $J_d^*$ in (\ref{2-2::opt2}) are equal to the total squared magnetic-dipole requirement over the three axes \cite{takahashi2026power}, the problem in (\ref{2-2::opt2}) does not specify its axis-wise allocation. Then, we adopt the simplified coil-sizing assumption. 
\begin{assumption}
A single-axis coil is sized to meet the worst-case axis power demand, and the same coil specification is applied to all three axes. By removing the time-averaging factor and considering the peak requirement, the one-direction magnetic moment $\mu_{\mathrm{1d}}$ is defined as
\begin{equation}
    \label{3-3:: magnetic moment 1d 0 and 1}
    \mu^2_{\mathrm{1d}}\triangleq J_d^*(n).
\end{equation}
\end{assumption}

Instead of strictly stabilizing the formation to an exact target trajectory, we relax the control objective. As an example, this study considers stabilizing the satellites within a tolerable position-error bound sufficient for antenna functionality to improve performance. The drift term in~(\ref{CWsol}) increases the relative distance between satellites over time. Then, we apply the previous controller \cite{takahashi2025distance,takahashi2025scalable} to reduce this term, and the closed-loop system related to this drift term \cite{takahashi2025distance,takahashi2025scalable} is
\begin{equation}
    \label{closed_loop_ro}
    \begin{aligned}
        \begin{bmatrix}
            \dot{\mathsf{e}}_{-2{C}_1}\\
            \dot{\mathsf{e}}_{{C}_4}
        \end{bmatrix}&=
        \begin{bmatrix}
           - \frac{k_A}{2}  L_e&O\\
             \frac{\epsilon_2}{2}\left(I    
             -\frac{\gamma k_\gamma}{2}L_e \right )&-\frac{\gamma k_A}{2}  L_e
        \end{bmatrix}\begin{bmatrix}
            {\mathsf{e}}_{-2{C}_1}\\
            {\mathsf{e}}_{{C}_4}
        \end{bmatrix}\\
&\quad{}-k_0
        \begin{bmatrix}
            {\mathsf{e}}_{d_y}\\
            {\mathsf{e}}_{d_x}
        \end{bmatrix}
    \end{aligned}
\end{equation}
where ${\mathsf{e}}_{-2{C}_1}=E^\top [-2{C}_1]$, ${\mathsf{e}}_{{C}_4}=E^\top [{C}_4]$, and ${\mathsf{e}}_{d_{x,y}}=E^\top[{d}_{x,y}]$. We construct approximate function $J_d^*(n)$ by exhaustively solving (\ref{2-2::opt2}) for each $n$ in Section~\ref{Analysis_representative_case_studies}. We emphasize that we can replace it with different controllers and control objectives.
\begin{remark}
Our goal is to evaluate the steady states of the formation keeping for $N_{\mathrm{all}}\gg1$ using scalable computations. Since theoretical bounds for the steady state can be conservative for $N_{\mathrm{all}}\gg 1$ \cite{takahashi2026graph}, we use numerical estimation to obtain exact values. One drawback of the numerical method is its high computational cost as the number of satellites grows. For example, we consider the linear time-invariant system $\dot{x} = A x(t) + d(t)$. The previous study \cite{takahashi2026power} avoids conducting a straightforward numerical integration and deriving analytical solutions for the system under $x(0)=0$:
$$
x(T)
=
\int_{0}^{T}
e^{A(T-\tau)} d(\tau)\, d\tau
=
\sum_{k=1}^{K}
\left[
\int_{t_k}^{t_{k+1}}
e^{A(T-\tau)} d\tau
\right]
d_k,
$$
where $d(t)=d_k$ for $t\in[t_k,t_{k+1})$, $A\in\mathbb{R}^{N_{\mathrm{all}}\times N_{\mathrm{all}}}$ is a stable matrix, and $d(t)\in\mathbb{R}^{N_{\mathrm{all}}}$ is a time-varying external input. Previous study \cite{takahashi2026power} shows its computational costs can be reduced to $\mathcal{O}(4N_{\mathrm{all}}^2K)$ or $\mathcal{O}(N_{\mathrm{all}}^3)+\mathcal{O}(N_{\mathrm{all}}^2K/P)$ with the parallelism $P$, which takes a high computational cost. In Section~\ref{Analysis_representative_case_studies}, we use Krylov subspace methods \cite{liesen2013krylov} to reduce the total costs for $N_{\mathrm{all}}\gg1$. This method approximately derives $e^{A(T-\tau_j)} d(\tau_j)$ with a cost of $\mathcal{O}(N_{\mathrm{all}}\log N_{\mathrm{all}})$ and the final cost is $\mathcal{O}(N_{\mathrm{all}}\log N_{\mathrm{all}} K/P)$, making the computation more tractable than direct numerical integration.
\end{remark}
\subsection{Overall Framework and Feasibility Region}\label{sec:problem statement}
\begin{algorithm}[bt!]
\caption{Distributed space antenna design framework}\label{alg:distributed_space_antennas_design}
\begin{algorithmic}[1]
\STATE \textbf{Input:} 1) satellite distance $d_{\mathrm{sat}}$, system mass $\overline{m}_{\mathrm{sys}}$, and investigation parameters $\chi_{inv}$: $\{$e.g., margin magnetic moment $\bar{\mu}_{mar}$ or transmit power $P_t$ $\}$ 2) feasible region $\chi_{\mathrm{feas}}$ for design parameters $\{$coil diameter $2a_{\mathrm{coil}}$, satellite size $2a_{\mathrm{sat}}$, coil parameter $q_{\mathrm{coil}}$, satellite number $n$, and maximum sidelobe variable $u_{\mathrm{psl}}$$\}$ in Table \ref{table:feasible_region_and_constants_for_DSA}, 3) multi-start trials $N_{\mathrm{GS}}$, 4) constants in Table~\ref{table:feasible_region_and_constants_for_DSA}, 5) user-defined evaluation function $\mathcal{J}(X)$
\STATE \textbf{Output:} Optimal $X^{*}$ and other parameters $f(X^{*})$: generation and consumption power, EIRP, satellite and component masses $m_{\mathrm{sat}}=m_{\mathrm{3coil}} + \rev{m_{\mathrm{sap}}} + m_{\mathrm{bat}} + m_{\mathrm{str}} + m_{\mathrm{bus}}$
\FOR{$s=1,\ldots,N_{\mathrm{GS}}$}
    \STATE Generate random $X_0^{(s)}\in \chi_{\mathrm{feas}}$ and solve $\mathcal{P}_{\mathrm{DSA}}$\ENDFOR

\STATE Select the best solution $X^{*}$ that maximizes $\mathcal{J}(X)$
\STATE \hrulefill \\
\STATE Distributed array design optimization $\mathcal{P}_{\mathrm{DSA}}(\mathit{d}_\text{sat}, \mathit{m}_{\text{sys}})$
$$
\begin{aligned}
        &\ \mathit{X^*}=\argmax_{X\in\chi_{\mathrm{feas}}} \quad
        \mathcal{J}(X)\quad 
        \text{subject to:}\\
        &\hspace{-0.5cm}\left\{
        \begin{aligned}
        &\text{{(A)\quad Satellite specification constraints}}\\
        &\quad a_{\mathrm{coil}} \leq a_{\mathrm{sat}} - r_{\mathrm{mar}},\quad k_F a_{\mathrm{coil}} \leq d_{\mathrm{sat}}\\
        &\quad 2a_{\mathrm{sat}} \leq d_{\mathrm{sat}},\quad \left|\frac{m_{\mathrm{sys}}}{N_{\mathrm{all}}} -m_{\mathrm{sat}} \right|\leq \gamma_{\mathrm{mass}} \\
        &\quad \overline{m}_{\mathrm{sat}}(a_{\mathrm{sat}})  - m_{3\mathrm{coil}} - \rev{m_{\mathrm{sap}}} \leq m_{\mathrm{sat}} \leq \overline{m}_{\mathrm{sat}}(a_{\mathrm{sat}})\\
        &\quad |\overline{m}_{\mathrm{sys}} - m_{\mathrm{sys}}| \leq \gamma_{\mathrm{sys}} \\
        &\quad P_{\mathrm{cont}} + P_{\mathrm{mis}} + P_{\mathrm{bus}} + P_{\mathrm{mar}} \leq P_{\mathrm{sap}} \\
        &\text{{(B)\quad Phased-array antenna constraints}}\\
        &\quad |N_l \sin_{(u_{\mathrm{psl}})} \cos_{(N_lu_{\mathrm{psl}})} - \cos_{(u_{\mathrm{psl}})} \sin_{(N_lu_{\mathrm{psl}})} |\leq \gamma_{\mathrm{psl}} \\
        &\quad \frac{\pi}{N_l} < |u_{\mathrm{psl}}| < \frac{2\pi}{N_l},\quad P_{\mathrm{mis}} =P_t / \eta_{\mathrm{tra}} \\
        &\quad G_T=\overline{D}\approx N_l^2, \quad P_t\triangleq \frac{I_R}{N_l^2} \left(\frac{\sin(u_{\mathrm{psl}})}{\sin(N_lu_{\mathrm{psl}})}\right)^2\\
        &\text{{(C)\quad Magnetic formation-keeping constraints}}\\
        &\quad 4P_{\mathrm{1d-cont}}=4\frac{2p_c \mu_{\mathrm{1d-cont}}^2}{\pi^2 q_{\mathrm{coil}} a_{\mathrm{coil}}^3},\quad \mu^2_{\mathrm{1d-cont}} = J_d^*(n)\\
        \end{aligned}
        \right.
    \end{aligned}
$$
\end{algorithmic}
\end{algorithm}

\rev{We integrate the objectives and constraints introduced in the previous sections into an optimization-based design framework, $\mathcal{P}_{\mathrm{DSA}}$, defined in \eqref{problem_formulation_ver1}. Algorithm~\ref{alg:distributed_space_antennas_design} summarizes the overall procedure using a user-defined evaluation function $\mathcal{J}(X)$. Constraints (A)–(C) are described in Subsections~\ref{method}, \ref{subsec: antenna req}, and \ref{section:grid}, respectively.}

To avoid an infeasible start of our nonconvex optimization problem, this subsection also denotes a warm-start strategy to specify the feasibility region $\chi_{\mathrm{feas}}$ of design parameters $X=\left\{2a_{\mathrm{coil}},\ 2a_{\mathrm{sat}},\ q_{\mathrm{coil}},\ n,\ \rev{u_{\mathrm{psl}}}\right\}$ in (\ref{design_parameters}). Here, $\underline{(\cdot)}$ and $\overline{(\cdot)}$ denote the lower and upper bounds, subscript 0 denotes the initial value, and $\xi_i\sim\mathcal{U}(0,1)$ are independent samples. 
The primary geometric variables are drawn uniformly within their bounds and then capped by simple geometric limits:
\begin{equation}
\begin{aligned}
a_{\mathrm{coil}0}
&= \min\Bigl(\frac{d_{\mathrm{sat}}}{k_F},\;
\underline{a}_{\mathrm{coil}} + \xi_{2}\bigl(\overline{a}_{\mathrm{coil}}-\underline{a}_{\mathrm{coil}}\bigr)\Bigr), \\
a_{\mathrm{sat}0}
&= \min\Bigl(\frac{d_{\mathrm{sat}}}{2},\;
\underline{a}_{\mathrm{sat}} + \xi_{1}\bigl(\overline{a}_{\mathrm{sat}}-\underline{a}_{\mathrm{sat}}\bigr)\Bigr), \\
m_{\mathrm{sys}0}
&= \underline{m}_{\mathrm{sys}} + \xi_{3}\bigl(\overline{m}_{\mathrm{sys}}-\underline{m}_{\mathrm{sys}}\bigr). \end{aligned}
\end{equation}
Then, we compute the remaining mass budget for the coil wiring, denoted $\overline{m}_{\mathrm{coil}0}$, and translate it into $\overline{q}_{\mathrm{coil}}$ using the coil-mass model (Definition~\ref{component_assumption_1} and Assumption~\ref{component_assumption_2}):
\begin{equation}
\label{eq:init_qcoil_bound}
\begin{aligned}
q_{\mathrm{coil}} &\leq \overline{q}_{\mathrm{coil}0} \triangleq 
{\overline{m}_{\mathrm{coil}0}}/({6\pi^2 \rho_c\, a_{\mathrm{coil}0}}).\\
q_{\mathrm{coil}0}
&= \underline{q}_{\mathrm{coil}} + \xi_{6}\Bigl(\min(\overline{q}_{\mathrm{coil}0},\overline{q}_{\mathrm{coil}})-\underline{q}_{\mathrm{coil}}\Bigr).
\end{aligned}
\end{equation}
Note that we bound $q_{\mathrm{coil}}$ in (\ref{3-1::Mass of str bat bus}) 
\begin{equation}
\label{eq:qcoil_range}
q_{\mathrm{coil}} \in 
\left[\underline{N}_{t}\,\underline{r}_{\mathrm{coil}}^2,\; \overline{N}_{t}\overline{r}_{\mathrm{coil}}^2\right]
\end{equation}
where the $\underline{N}_{t}$ and $\overline{N}_{t}$ define
\begin{equation}
\label{eq:qcoil_bounds_Nt_rc}
\left\{
\begin{aligned}
\underline{N}_{t} &= 1,\quad\overline{N}_{t} = \frac{\overline{a}_{\mathrm{sat}}}{\underline{r}_{\mathrm{coil}}}.
\end{aligned}
\right.
\end{equation}
Next, $n$ is sampled using an instance-dependent upper bound obtained from the empirical mass model in Assumption~\ref{asp:: sat mass}. We enforce ${m_{\mathrm{sys}0}}/{(2n_0+1)^2} \geq \underline{m}_{\mathrm{sat}}$ and solving this yields
\begin{equation}
\label{eq:init_n_max}
n_0 \leq \overline{n} \triangleq \left(-1+\sqrt{m_{\mathrm{sys}0}/\underline{m}_{\mathrm{sat}}}\right)/{2},
\end{equation}
to avoid unrealistically small satellite masses, and we use
\begin{equation}
\label{eq:init_n_sample}
n_0 = \underline{n} + \xi_{4}\Bigl(\overline{n}-\underline{n}\Bigr).
\end{equation}
The initial value $u_{\mathrm{psl}0}$ is randomly generated within a narrowed subset of its theoretical peak-sidelobe neighborhood. By introducing a small positive tolerance $\gamma_u$ to the interval boundaries, the warm-start point is determined as follows:
\begin{equation}
\label{eq:init_umsl}
u_{\mathrm{psl}0} = \left( \frac{(1 - 2\gamma_u)\pi}{2n_0 + 1} \right) \xi_5 - \frac{(2 - \gamma_u)\pi}{2n_0 + 1}.
\end{equation}
This formulation ensures that $u_{\mathrm{psl}0}$ is constrained within the range $[ \frac{(-2+\gamma_u)\pi}{2n_0+1}, \frac{(-1-\gamma_u)\pi}{2n_0+1} ]$, providing a well-conditioned starting point for the subsequent local optimization.

\section{Array Design Examples and Discussion}
\label{Result}
This section presents design examples obtained through the proposed framework and highlights the design trends of EMFF-based distributed space antennas.

\subsection{\rev{Setup} of Representative Case Studies}
\label{Analysis_representative_case_studies}

\begin{table*}[!tb]
\centering
\vspace{-0.3cm}
\captionof{table}{Calculation constants of design parameters for Algorithm~\ref{alg:distributed_space_antennas_design}: Distributed space antennas design.}
    \label{table:feasible_region_and_constants_for_DSA}
\begin{minipage}[t]{\textwidth}
    \centering
    \vspace{-0.65cm}
    \begin{minipage}[t]{0.49\linewidth}
        \vspace{0pt}
        \centering
        \resizebox{0.95\textwidth}{!}{
        \begin{tabular}{c|c|c}
            \hline
            \multicolumn{3}{c}{Satellite Design Parameters} \\ \hline
            \textbf{Description} & \textbf{Constant} & \textbf{Value} \\ \hline
            \rev{Nominal copper-wire resistivity} & $p_c$ (\ref{2_1::coil dipole moment}) & 1.68e$^{-8}$$\Omega \mathrm{m}$\cite{liu2022copperfilms}\\ \hline
            \rev{Nominal copper-wire mass density} & $\rho_c$ (\ref{3-1::Mass of str bat bus}) & $8960$ $\mathrm{kg/m^3}$\cite{crc2014} \\ \hline
            \rev{\makecell{Battery mass per unit energy}} & $\rho_{\mathrm{bat}}$ (\ref{3-1::Mass of str bat bus}) & \makecell{0.005 $\mathrm{kg/Wh}$ \rev{\cite{Weston2025SmallSpacecraftSOA}}} \\ \hline
            \rev{\makecell{Solar-cell-sheet areal mass}} & $\rho_{\mathrm{sap}}$ (\ref{3-1::Mass of str bat bus}) & $0.6$~$\mathrm{kg/m^2}$ \rev{\cite{kanayaDevelopmentOnorbitOperation2025}} \\ \hline
            \rev{\makecell{Number of body-mounted\\solar-cell sheets}} & $\nu_{\mathrm{sap}}$ (\ref{3-1::Mass of str bat bus}) & $4$ \\ \hline
            \makecell{Power generation per \\ unit area of solar panel} & $P_{\mathrm{W/m^2}}$ (\ref{eq:solar_power}) & \makecell{$1367 \times 0.3\ \mathrm{W/m^2}$} \rev{\cite{Weston2025SmallSpacecraftSOA}} \\ \hline
            \rev{\makecell{Effective illuminated-area\\coefficient}} & $k_{\mathrm{sap}}$ (\ref{eq:solar_power}) & $1$ \\ \hline
            \rev{Far-field separation coefficient} & \rev{$k_F$ (\ref{3-1:: coil and sat size})} & \rev{$4$ \cite{kwon2010micro}} \\ \hline
            \makecell{Coefficient of structure mass} & $\eta_{str}$ (\ref{3-1::Mass of str bat bus}) & $0.25$ \rev{\cite{franzese2025poquito, palmroth2019foresail}} \\ \hline
            Margin of coil size & $r_{\mathrm{mar}}$ (\ref{3-1:: coil and sat size}) & $0.005 ~ \mathrm{m}$ \\ \hline
            Bus power & $P_{\mathrm{bus}}$ (\ref{eq:power_total}) & $0.200~\mathrm{W}$~\rev{\cite{franzese2025poquito,nieto2019cubesat}}\\ \hline
            Bus mass & $m_{\mathrm{bus}}$ (\ref{3-1::Mass of str bat bus}) & $0.200~\mathrm{kg}$~\rev{\cite{nieto2019cubesat,palmroth2019foresail}}\\ \hline
            \makecell{Battery storage coefficient} & $k_{\mathrm{bat}}$ (\ref{eq:ex_Ebat}) & $0.1$ \\ \hline
            \makecell{Satellite/system-mass tolerance} & $\gamma_{\mathrm{mass}}/\gamma_{\mathrm{sys}}$ & $10^{-2}/10^{-2}$ \\ \hline
            \makecell{Peak-sidelobe/$u_{\mathrm{psl0}}$ tolerance} & $\gamma_{\mathrm{psl}}/\gamma_{\mathrm{u}}$ & $10^{-5}/0.1$ \\ \hline
        \end{tabular}}
    \end{minipage}
    \begin{minipage}[t]{0.49\linewidth}
        \vspace{0pt}
        \centering
        \resizebox{0.95\textwidth}{!}{
        \begin{tabular}{c|c|c}
            \hline
            \multicolumn{3}{c}{Orbital and Communication Constants} \\ \hline
            \textbf{Description} & \textbf{Constant} & \textbf{Value} \\ \hline
            Orbital height & $h$ (\ref{3-3:: null point}) & $500$~$\mathrm{km}$ \\ \hline
            Orbital inclination & $i_o$ & ${45\pi}/{180}~\mathrm{rad}$ \\ \hline
            Receiving power & $P_R$ (\ref{3-3:: maximum sidelobe constraint}) & $-87.2~\mathrm{dBm}$\cite{3gpp2022} \\ \hline
            Receiving antenna gain & $G_R$ (\ref{3-3:: maximum sidelobe constraint}) & $0$~$\mathrm{dBi}$ \\ \hline
            Attenuation rate & $\zeta$ (\ref{3-3:: maximum sidelobe constraint}) & $0.5$ \\ \hline
            Transmitter efficiency & $\eta_{\mathrm{tra}}$ (\ref{eq:control_power_total}) & $0.3$ \rev{\cite{nieto2019cubesat}} \\ \hline
            Direction of main beam & $\theta_0$ (\ref{2-3:: simple plane directivity}), (\ref{3-3:: null point}) & $\pi/6~\mathrm{rad}$ \\ \hline
            \makecell{Battery charging duration} & $h_{\mathrm{charge}}$ (\ref{eq:ex_Ebat}) & $12$ \rev{h} \\ \hline
            \multicolumn{3}{c}{Feasible region $\chi_{\mathrm{feas}}$ of optimization variable} \\ 
            \hline
            \textbf{Variable} & \textbf{Min} & \textbf{Max} \\\hline
            System mass $m_{\mathrm{sys}}$ & $(1-\gamma_{\mathrm{sys}})\overline{m}_{\mathrm{sys}}$ & $(1+\gamma_{\mathrm{sys}})\overline{m}_{\mathrm{sys}}$\\ \hline
            Satellite radius $a_{\text{sat}}$ & 0.015 & $d_{\mathrm{sat}}/2$ \\
            \hline
            Coil radius $a_{\text{coil}}$& 0.005 & $d_{\mathrm{sat}}/2 - r_{\mathrm{mar}}$ \\
            \hline
            Coil parameter $q_{\mathrm{coil}}$& $\underline{N}_{t} \underline{r}_{\mathrm{coil}}^2$ & $\overline{N}_{t}\overline{r}_{\mathrm{coil}}^2$ \\ \hline
            Satellite number $n$ & $3$ & $\overline{n}$ \\ \hline
            Peak sidelobe variable $u_{\mathrm{psl}}$ & $-2\pi /3$ & 0 \\ \hline
        \end{tabular}
        }
    \end{minipage}
    \vspace{1mm}
    \footnotesize
    \raggedright
    Note: The lower and upper bounds of the coil-wire radius used in the bounds of $q_{\mathrm{coil}}$ are 
    $\underline{r}_{\mathrm{coil}} = 0.03937 \times 10^{-3}~\mathrm{m}$ and 
    $\overline{r}_{\mathrm{coil}} = 0.00105~\mathrm{m}$, respectively.
    \vspace{-0.25cm}
\end{minipage}
\vspace{-0.2cm}
\end{table*}
To demonstrate the applicability of the proposed framework, this subsection considers three design scenarios. 
For Case 1, the design objective is to achieve a narrow beam by minimizing the footprint as follows: 
\begin{equation}
\argmin_{\mathcal{X}} \arcsin\left(\frac{\sqrt{2}\lambda}{N_l d_{\mathrm{sat}}} + \sin\theta_0\right)
=
\argmax_{\mathcal{X}} N_{\mathrm{all}}.
\end{equation}
For Cases 2 and 3, the design objective is to maximize the communication or sensing capability, and the EIRP maximization problem is reduced to the maximization of \(N_{\mathrm{all}}\):
\begin{equation}
\argmax_{\mathcal{X}} \mathrm{EIRP}
\propto
\argmax_{\mathcal{X}} \left(P_t N_l^2 \cdot G_T\right)
=
\argmax_{\mathcal{X}} N_{\mathrm{all}}.
\end{equation}

\rev{For all cases, the component-level parameters in Table~\ref{table:feasible_region_and_constants_for_DSA} were selected as rounded representative values based on published component data and spacecraft design examples rather than as specifications of a particular spacecraft. The supporting references are provided directly in Table II. The remaining orbital, geometric, and numerical parameters define common baseline conditions for consistent comparison among the case studies. The principal variables and parameters used in the proposed design framework are summarized in the Appendix.}

\begin{figure}[!tb]
    \centering
    \begin{minipage}[b]{1\FigWidth}
        \centering
        \subfloat[Residual drift vector\label{drift_vector}]{\includegraphics[width=\linewidth]{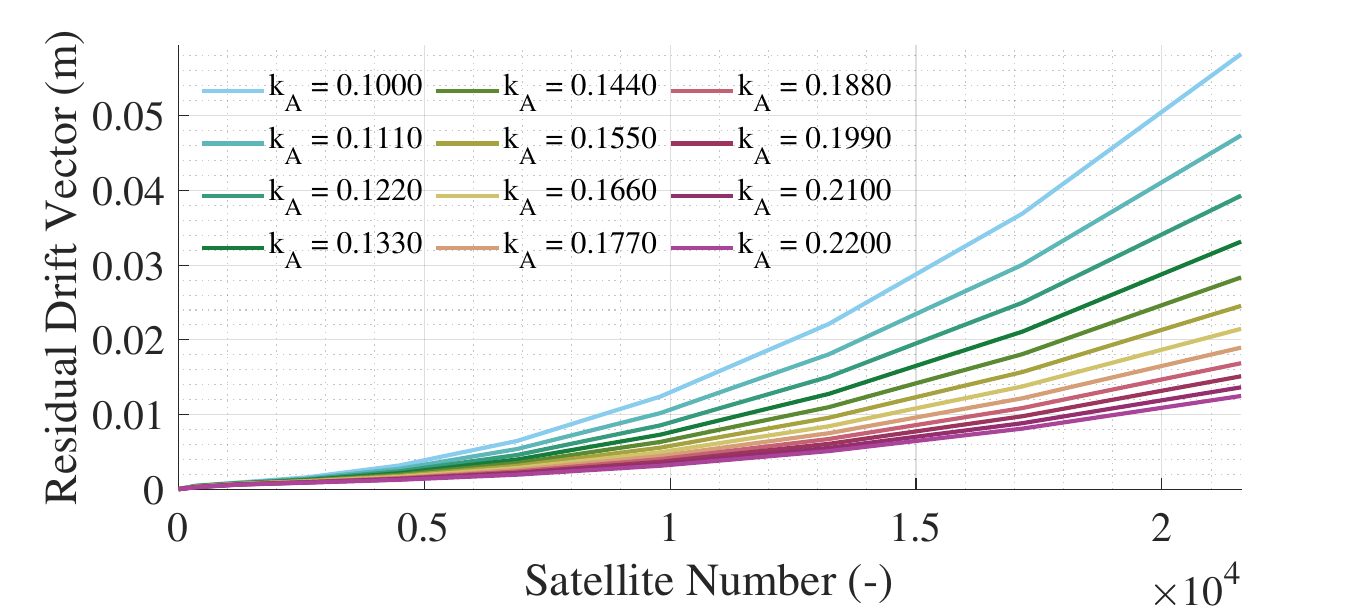}} 
    \end{minipage}
    \begin{minipage}[b]{1\FigWidth}
        \centering
        \subfloat[Maximum control input]{\includegraphics[width=\linewidth]{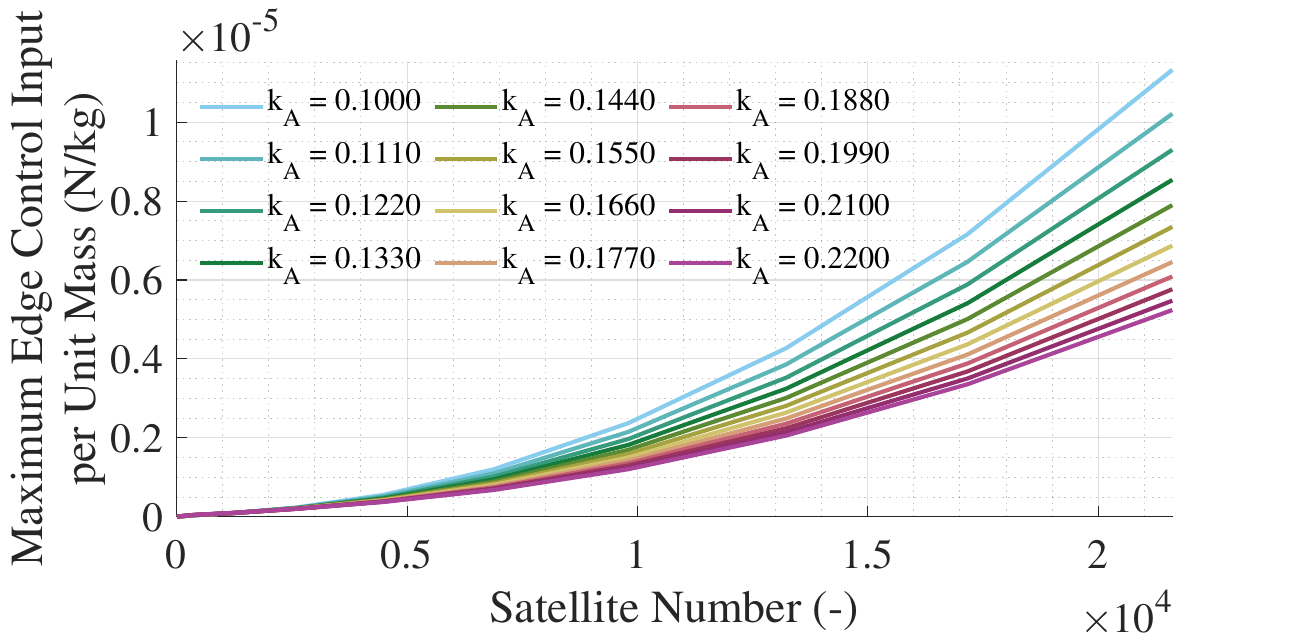}\label{control_input}}
    \end{minipage}
    \begin{minipage}[b]{0.49\FigWidth}
        \centering
        \subfloat[\rev{Mass-normalized control index \(j_d^*\)} at \(d_{\mathrm{sat}}=0.15~\mathrm{m}\).]{\includegraphics[width=\linewidth]{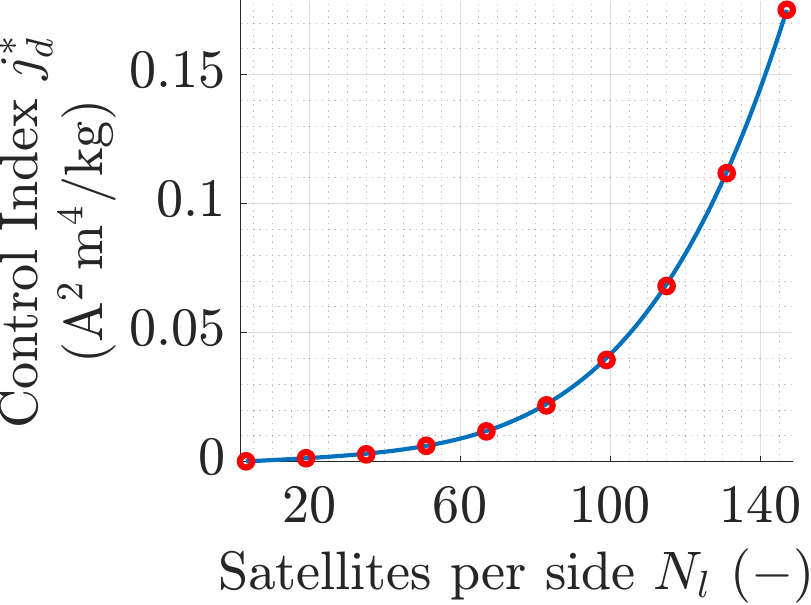}\label{fig:results_Jd_015}} 
    \end{minipage}
    \hfill
    \begin{minipage}[b]{0.49\FigWidth}
        \centering
        \subfloat[\rev{Mass-normalized control index \(j_d^*\)} at \(d_{\mathrm{sat}}=0.30~\mathrm{m}\).]{\includegraphics[width=\linewidth]{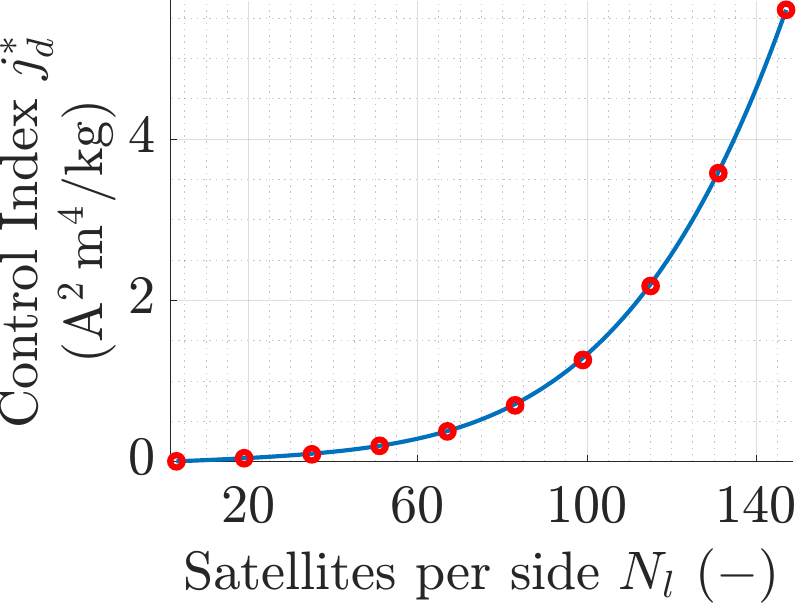}\label{fig:results_Jd_030}} 
    \end{minipage}
    \caption{Numerical estimation of the distributed-control requirement used in the EMFF power constraint. Panels (a) and (b) show the residual drift vector and maximum control input obtained from the distributed formation-control simulation for different feedback gains \(k_A\). \rev{The unit of \(k_A\) is \(\mathrm{s}^{-1}\), which is omitted from the figure labels.} Based on these results, \(k_A=\rev{0.1990~\mathrm{s}^{-1}}\) is selected as the nominal gain used to construct the control requirement. Panels (c) and (d) show the resulting \rev{mass-normalized control index \(j_d^*(N_l)\)} for \(d_{\mathrm{sat}}=0.15~\mathrm{m}\) and \(d_{\mathrm{sat}}=\rev{0.30}~\mathrm{m}\), respectively. The red circles denote the numerical values obtained by solving the magnetic-dipole allocation problem in (\ref{2-2::opt2}), and the blue curves denote the fourth-order polynomial fits used in the design optimization.}
    \label{fig:for_Jd_calculation}
\end{figure}

\rev{The disturbance-compensation requirement was precomputed per unit satellite mass before solving the system-level design problem. For \(d_{\mathrm{sat}}=0.15\) and \(0.30~\mathrm{m}\), the closed-loop drift dynamics in \eqref{closed_loop_ro} were numerically evaluated for square-grid formations with different side counts and feedback gains. Based on the residual drift and maximum single-edge specific control input shown in Figs.~\ref{drift_vector} and \ref{control_input}, \(k_A=0.199~\mathrm{s^{-1}}\) was selected. For each formation size, the maximum input was applied in the worst-case force direction, and the corresponding torque command was defined as \(\boldsymbol{\tau}^{\mathrm{cmd}}_{j\leftarrow k}=\mathbf r_{j\leftarrow k}\times\mathbf f^{\mathrm{cmd}}_{j\leftarrow k}\).}

\rev{The force and torque commands were evaluated per unit satellite mass using the dual allocation problem in \eqref{2-2::opt2}, yielding the mass-normalized control index \(j_d^*(N_l)\) in \(\mathrm{A^2\,m^4\,kg^{-1}}\). For \(N_l=2n+1\), the satellite-level control index is \(J_d^*(n)=m_{\mathrm{sat}}j_d^*(N_l)\) in \(\mathrm{A^2\,m^4}\), giving the requirement \(\mu_{1d}^{2}=J_d^*(n)\). The mass-normalized values were fitted using relative-error-weighted fourth-order polynomials with MATLAB R2022a \textnormal{\texttt{lsqlin}}, with the curves constrained to lie above all samples. Figures~\ref{fig:results_Jd_015} and \ref{fig:results_Jd_030} show the upper-envelope approximations for \(d_{\mathrm{sat}}=0.15\) and \(0.30~\mathrm{m}\), respectively, while Table~\ref{tab:Jd_fitting_accuracy_kA0199} summarizes their coefficients and fitting accuracy.}

\rev{All design solutions were obtained using Algorithm~\ref{alg:distributed_space_antennas_design} and MATLAB's GlobalSearch solver~\cite{MATLAB_GlobalSearch}, with the total system mass varied from $500$ to $6000~\mathrm{kg}$. 
All computations were performed in MATLAB R2022a on a workstation equipped with a 13th Gen Intel Core i9-13900 processor and 64~GB of RAM. GlobalSearch with fmincon was employed. For GlobalSearch, \textnormal{\texttt{NumTrialPoints}} and \textnormal{\texttt{NumStageOnePoints}} were both set to 1{,}000, while \textnormal{\texttt{MinNumStartPoints}} was set to 10{,}000. For fmincon, \textnormal{\texttt{MaxFunctionEvaluations}} and \textnormal{\texttt{MaxIterations}} were both set to 1{,}000, with the remaining options left at their MATLAB default values. The total system mass was evaluated at 12 points from 500 to 6{,}000~kg in increments of 500~kg. These mass-point problems were executed in parallel using \textnormal{\texttt{parfor}} with 12 workers.}

\begin{table}[!tb]
    \centering
    \begin{revtable}
    \caption{\rev{Coefficients and accuracy of the polynomial approximation for \(k_A=0.199~\mathrm{s^{-1}}\), defined as \(\widehat{j}_d^*(N_l)=a_4N_l^4+a_3N_l^3+a_2N_l^2+a_1N_l+a_0\), where \(a_0,\ldots,a_4\) denote the fitting coefficients. The coefficients and RMSE have units of \(\mathrm{A^2\,m^4\,kg^{-1}}\).}}
    \vspace{-0.3cm}
    \label{tab:Jd_fitting_accuracy_kA0199}
    \small
    \begin{tabular}{l|c|c}
        \hline
        Quantity & $d=0.15$ m & $d=0.30$ m \\
        \hline
        $a_4$ & $ 4.180\times10^{-10}$ & $ 1.338\times10^{-8}$ \\
        $a_3$ & $-1.564\times10^{-8}$  & $-5.004\times10^{-7}$ \\
        $a_2$ & $ 9.905\times10^{-7}$  & $ 3.170\times10^{-5}$ \\
        $a_1$ & $ 5.589\times10^{-5}$  & $ 1.788\times10^{-3}$ \\
        $a_0$ & $-1.496\times10^{-5}$  & $-4.787\times10^{-4}$ \\
        \hline
        $R^2$       & 0.9999 & 0.9999 \\
        RMSE         & $4.070\times10^{-4}$ & $1.344\times10^{-2}$ \\
        NRMSE (\%)   & 0.2326 & 0.2401 \\
        MAPE (\%)    & 1.589 & 1.607 \\
        \hline
    \end{tabular}
    \end{revtable}
\end{table}
\subsection{\rev{Results of Representative Case Studies}}
This subsection presents the results obtained using the proposed framework. The results for each case are described below.

\subsubsection{Case 1: Margin Magnetic Moment Analysis}
Case~1 evaluates how the margin magnetic moment affects footprint-oriented antenna sizing in a D2D-oriented 1-GHz case with \(\lambda=0.30~\mathrm{m}\) and \(d_{\mathrm{sat}}=0.15~\mathrm{m}\). This spacing corresponds to half a wavelength. The per-element RF transmit power \(P_t\) is calculated from the peak-sidelobe-envelope-limited model in (\ref{3-3:: P_T maximum sidelobe constraint_d2d}) using the D2D received-power requirement, and \(\bar{\mu}_{\mathrm{mar}}\) is treated as the investigation parameter representing the control capability reserved beyond disturbance compensation. The numerical setup, representative design values, optimized trends, and constraint margins are shown in Table~\ref{table:case1}, Table~\ref{tab:sat_spec_case1}, Fig.~\ref{fig:optimization_results_case_1}, and Figs.~\ref{fig:sat_mass_constraint_case1}--\ref{fig:power_constraint_case1}, respectively.

\subsubsection{Case 2: Transmit-Power Analysis}
\begin{figure*}[!tb]
\centering
\begin{minipage}[t]{\textwidth}
    \centering
    \vspace{-0.25cm}
    
    \captionof{table}{Numerical setup and input parameters for Case 2.}
    \vspace{-0.25cm}
    \label{table:case2}
    \resizebox{\textwidth}{!}{ \begin{tabular}{c|c|c|c|c}
            \hline
            Frequency [GHz] & Wavelength $\lambda$ [m] & Inter-satellite distance $d_{\mathrm{sat}}$ [m] & Surplus magnetic moment $\bar{\mu}_{\mathrm{mar}}$ [Am$^2$] & Transmitting power from an element $P_t$ [W] \\ \hline
            1 & 0.30 & 0.15 & 0.25 & [0.1, 0.2, 0.3, 0.4, 0.5]\\ \hline
        \end{tabular}
    }

    \vspace{0.15cm} 
    \captionof{table}{Case 2: Breakdown of size, mass, and power with overall system masses from 500 kg to 6000 kg. In terms of power, only the solar panel generates power, while all other components consume power. The bus mass and bus power are set as constants at 200 g and 200 mW, respectively. The SLL denotes the normalized peak sidelobe-envelope level relative to the main-beam peak.
    }
    \vspace{-0.25cm}
    \label{tab:sat_spec_case2}
    \begin{revtable}
    \resizebox{\textwidth}{!}{
    \begin{tabular}{@{}ll|c|c|c|c|c|c|c|c|c|c|c|c|c|c|c@{}}
        \hline
        \multicolumn{2}{c|}{$P_t$ (W)} & \multicolumn{3}{c|}{\textbf{0.1}} & \multicolumn{3}{c|}{\textbf{0.2}} & \multicolumn{3}{c|}{\textbf{0.3}} & \multicolumn{3}{c|}{\textbf{0.4}} & \multicolumn{3}{c}{\textbf{0.5}} \\ \hline
        \multicolumn{2}{l|}{\textbf{Overall system mass $\overline{m}_{\mathrm{sys}}$ (kg)}}  & \textbf{500} & \textbf{4000} & \textbf{6000} & \textbf{500} & \textbf{4000} & \textbf{6000} & \textbf{500} & \textbf{4000} & \textbf{6000} & \textbf{500} & \textbf{4000} & \textbf{6000} & \textbf{500} & \textbf{4000} & \textbf{6000} \\ \hline
        \multirow{3}{*}{Size} & Satellite $2a_{\mathrm{sat}}$ (mm) & 88.9 & 100.0 & 100.0 & 93.3 & 100.0 & 100.0 & 97.6 & 100.0 & 100.0 & 100.0 & 100.0 & 100.0 & 100.0 & 100.0 & 100.0 \\
         & Coil $2a_{\mathrm{coil}}$ (mm) & 75.0 & 75.0 & 75.0 & 75.0 & 75.0 & 75.0 & 75.0 & 75.0 & 75.0 & 75.0 & 75.0 & 75.0 & 75.0 & 75.0 & 75.0 \\
         & Coil parameter $q_{\mathrm{coil}}$ (mm$^2$) & 1.61 & 2.38 & 3.59 & 1.61 & 2.61 & 3.91 & 1.61 & 2.89 & 4.30 & 1.70 & 3.24 & 4.78 & 1.95 & 3.68 & 5.39 \\ \hline
        \multirow{5}{*}{Mass (g)} & Satellite & 356 & 389 & 421 & 361 & 395 & 429 & 366 & 403 & 440 & 372 & 412 & 452 & 378 & 423 & 468 \\
         & 3-axis coil & 32.1 & 47.4 & 71.3 & 32.1 & 52.0 & 77.8 & 32.1 & 57.5 & 85.5 & 33.8 & 64.4 & 95.1 & 38.8 & 73.3 & 107 \\
         & Battery & 19.4 & 24.6 & 24.6 & 21.4 & 24.6 & 24.6 & 23.4 & 24.6 & 24.6 & 24.6 & 24.6 & 24.6 & 24.6 & 24.6 & 24.6 \\
         & Body & 88.9 & 97.4 & 105 & 90.2 & 98.9 & 107 & 91.5 & 101 & 110 & 92.9 & 103 & 113 & 94.5 & 106 & 117 \\
         & 4 Solar panels & 18.9 & 24.0 & 24.0 & 20.9 & 24.0 & 24.0 & 22.9 & 24.0 & 24.0 & 24.0 & 24.0 & 24.0 & 24.0 & 24.0 & 24.0 \\ \hline
        \multirow{4}{*}{Power (mW)} & 4 Solar panels ($\times 10^{3}$) & 3.24 & 4.10 & 4.10 & 3.57 & 4.10 & 4.10 & 3.90 & 4.10 & 4.10 & 4.10 & 4.10 & 4.10 & 4.10 & 4.10 & 4.10 \\
         & Disturbance & 200 & 1870 & 2440 & 200 & 1690 & 2200 & 201 & 1510 & 1960 & 191 & 1320 & 1720 & 167 & 1140 & 1490 \\
         & Transmitter & 333 & 333 & 333 & 667 & 667 & 667 & 1000 & 1000 & 1000 & 1330 & 1330 & 1330 & 1670 & 1670 & 1670 \\
         & Margin ($\times 10^{3}$) & 2.50 & 1.69 & 1.13 & 2.50 & 1.54 & 1.03 & 2.50 & 1.40 & 0.939 & 2.38 & 1.25 & 0.844 & 2.07 & 1.10 & 0.749 \\ \hline
        \multirow{5}{*}{Performance} & Main-beam EIRP (dBW) & 53.0 & 70.3 & 73.2 & 55.9 & 73.2 & 76.0 & 57.6 & 74.8 & 77.6 & 58.7 & 75.9 & 78.6 & 59.5 & 76.6 & 79.2 \\
         & Peak SLL envelope (dB) & -13.2 & -13.3 & -13.3 & -13.2 & -13.3 & -13.3 & -13.2 & -13.3 & -13.3 & -13.2 & -13.3 & -13.3 & -13.2 & -13.3 & -13.3 \\
         & Maximum antenna gain (dBi) & 31.5 & 40.2 & 41.6 & 31.5 & 40.1 & 41.5 & 31.4 & 40.0 & 41.4 & 31.3 & 39.9 & 41.3 & 31.3 & 39.8 & 41.1 \\
         & Footprint diameter (km) & 84.7 & 31.8 & 27.0 & 85.3 & 32.0 & 27.3 & 85.9 & 32.3 & 27.6 & 86.5 & 32.7 & 28.0 & 87.3 & 33.1 & 28.5 \\
         & Satellite number $N_l\times N_l$ & $37\times37$ & $101\times101$ & $119\times119$ & $37\times37$ & $101\times101$ & $119\times119$ & $37\times37$ & $101\times101$ & $117\times117$ & $37\times37$ & $99\times99$ & $115\times115$ & $37\times37$ & $97\times97$ & $113\times113$ \\ \hline
    \end{tabular}
    }
    \end{revtable}
\end{minipage}
\begin{minipage}[t]{\textwidth}
    \centering
    \vspace{0.1cm}
    \begin{minipage}[t]{0.30\textwidth}
        \centering
        \subfloat[Antenna diameter ($P_t=0.2$ W)]{\includegraphics[width=\linewidth]{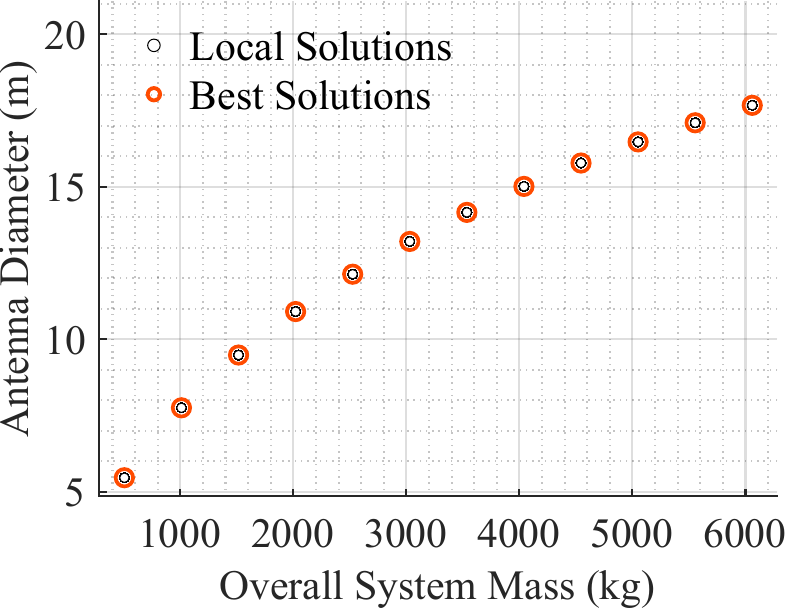}\label{fig:R_g_Pt200mW_case2}}
    \end{minipage}
    \begin{minipage}[t]{0.30\textwidth}
        \centering
        \subfloat[Antenna diameter]{\includegraphics[width=\linewidth]{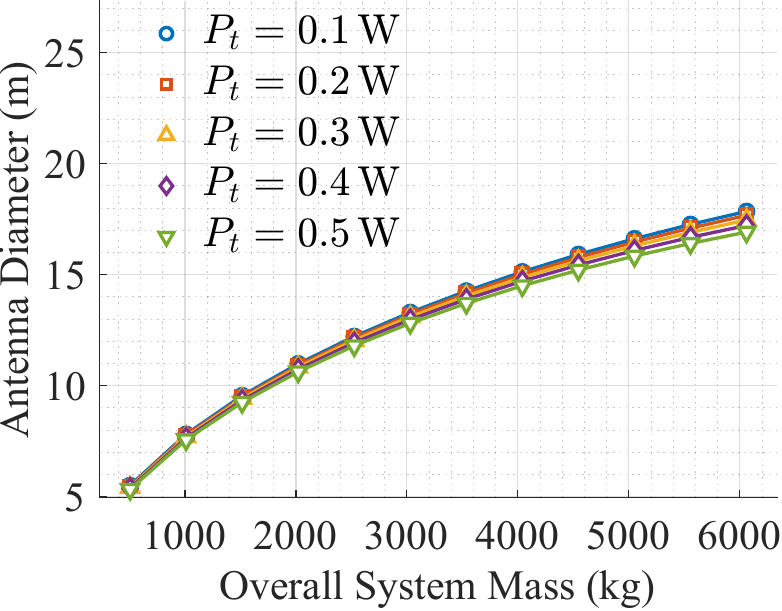}\label{fig:case2_Pt_R}}
    \end{minipage}
    \begin{minipage}[t]{0.30\textwidth}
        \centering
        \subfloat[EIRP]{\includegraphics[width=\linewidth]{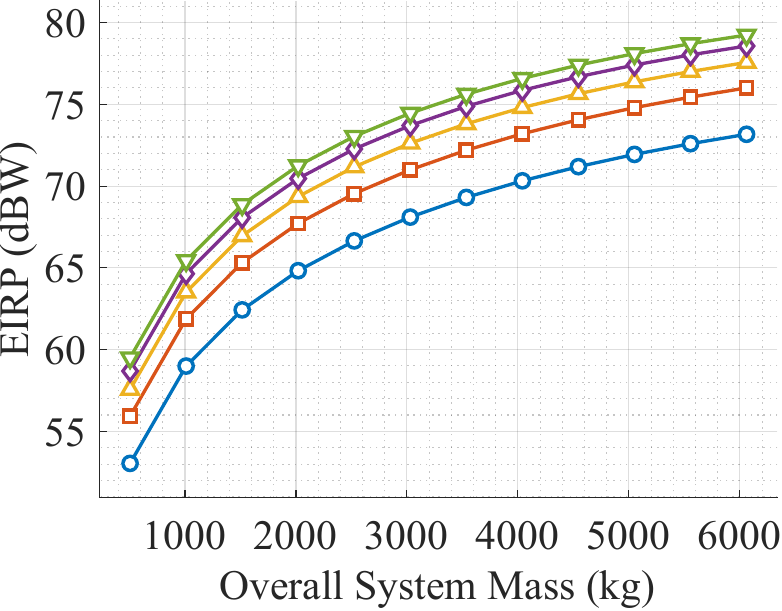}\label{fig:case2_Pt_eirp}}
    \end{minipage}\\
    \begin{minipage}[t]{0.30\textwidth}
        \centering
        \subfloat[Satellite number]{\includegraphics[width=\linewidth]{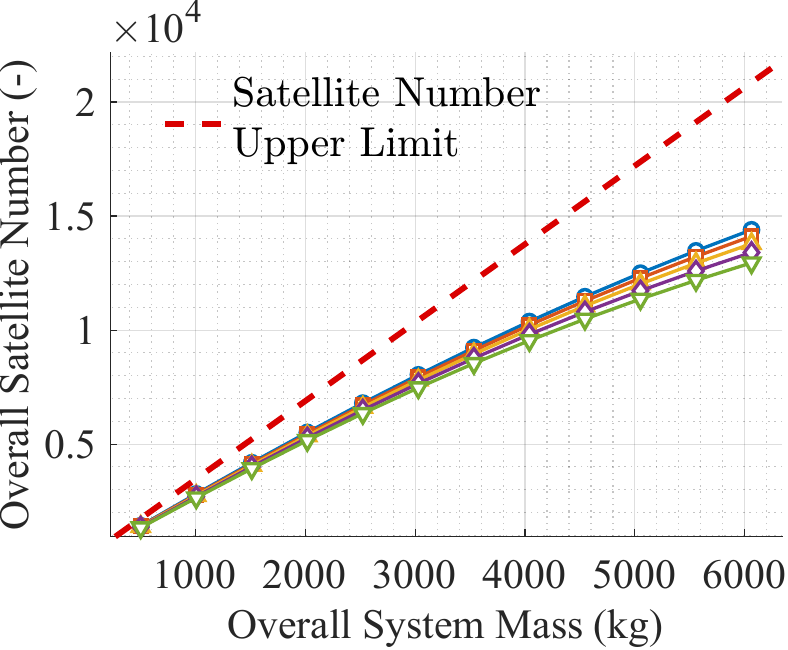}\label{fig:case2_Pt_Nall}}
    \end{minipage}
    \begin{minipage}[t]{0.30\textwidth}
        \centering
        \subfloat[Satellite mass]{\includegraphics[width=\linewidth]{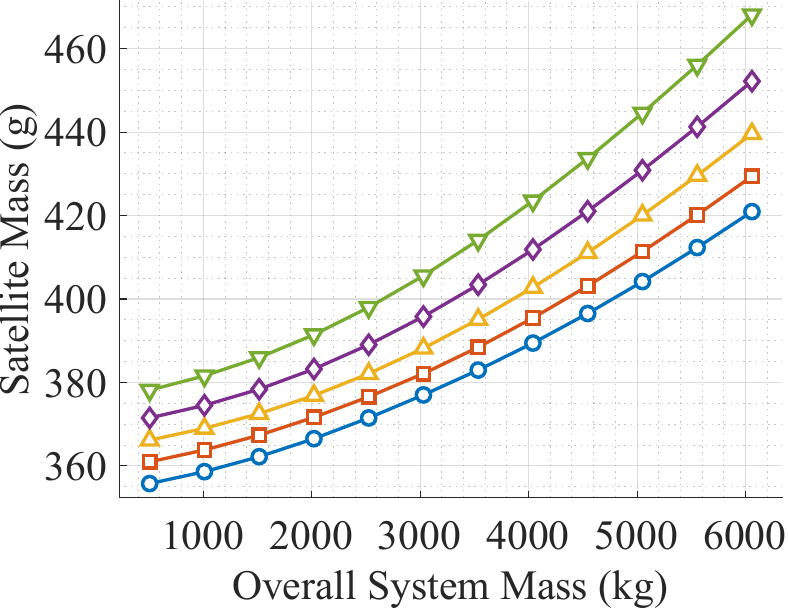}\label{fig:sat_mass_case2}}
    \end{minipage}
    \begin{minipage}[t]{0.30\textwidth}
        \centering
        \subfloat[Coil mass]{\includegraphics[width=\linewidth]{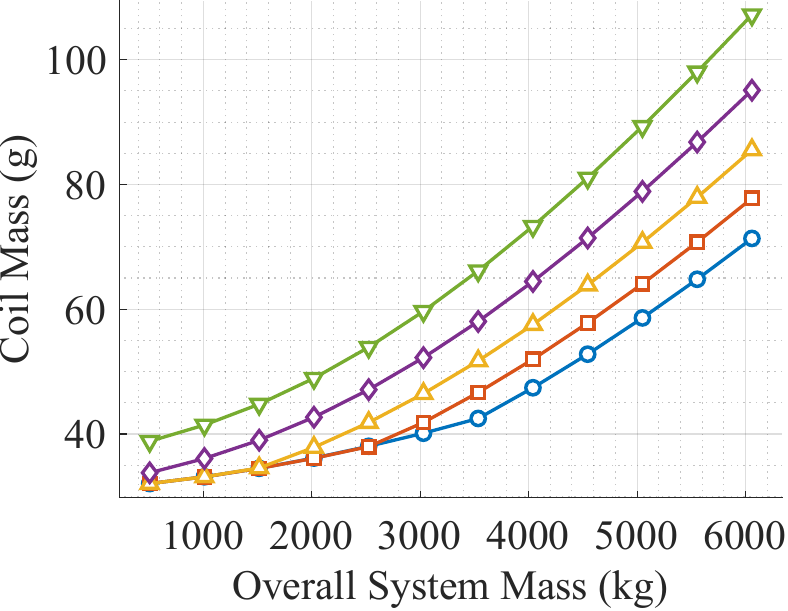}\label{fig:coil_mass_case2}}
    \end{minipage}\\
    \begin{minipage}[t]{0.30\textwidth}
        \centering
        \subfloat[Satellite size]{\includegraphics[width=\linewidth]{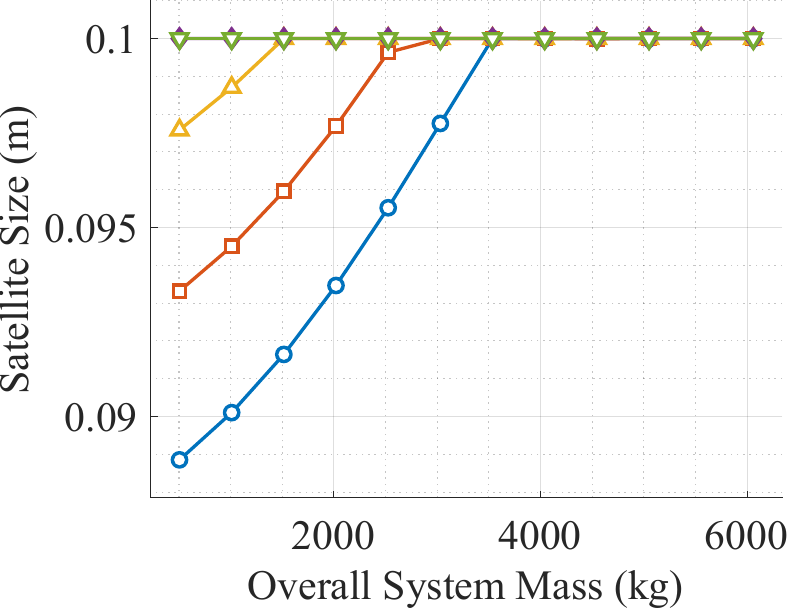}\label{fig:satellite_size_case2}}
    \end{minipage}
    \begin{minipage}[t]{0.30\textwidth}
        \centering
        \subfloat[Coil diameter]{\includegraphics[width=\linewidth]{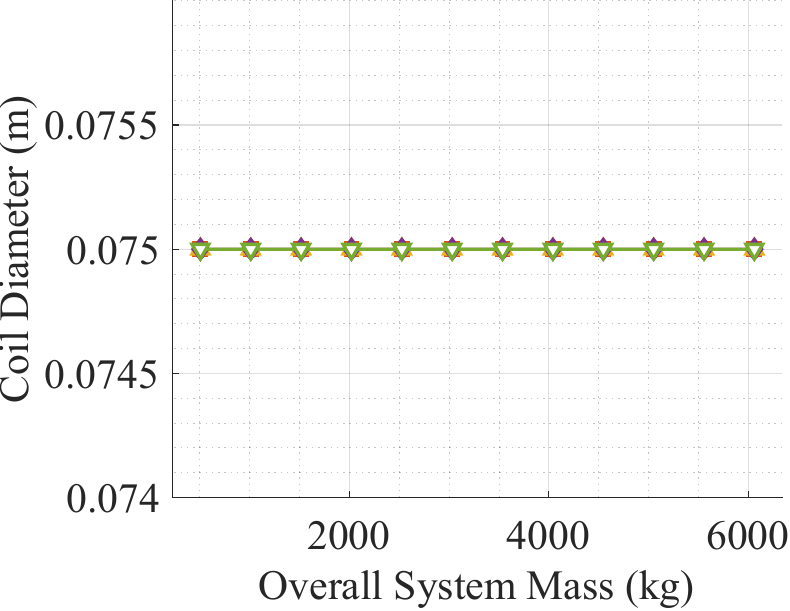}\label{fig:coil_diameter_case2}}
    \end{minipage}
    \begin{minipage}[t]{0.30\textwidth}
        \centering
        \subfloat[Coil parameter]{\includegraphics[width=\linewidth]{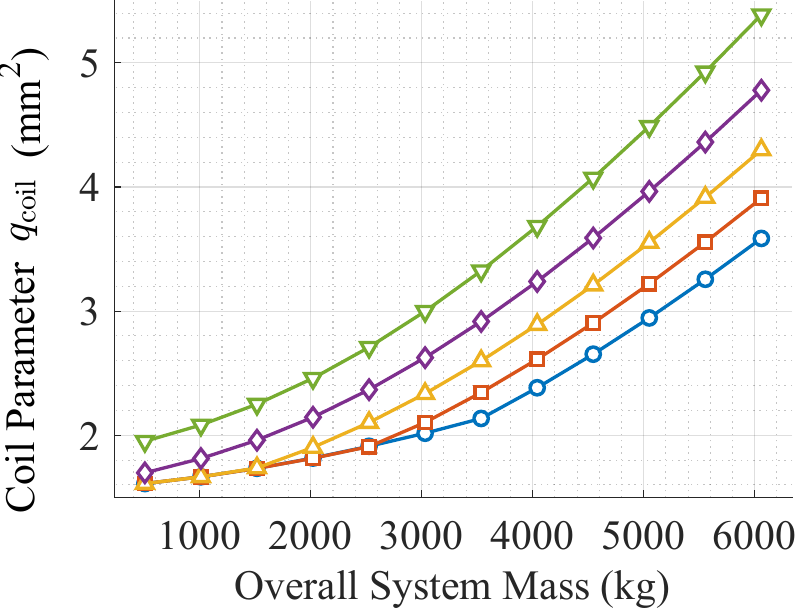}\label{fig:coil_parameter_case2}}
    \end{minipage}
\caption{Representative design solutions for Case~2, where the transmit power is swept at $\lambda=0.30~\mathrm{m}$, $d_{\mathrm{sat}}=0.15~\mathrm{m}$, and $\bar{\mu}_{\mathrm{mar}}=0.25~\mathrm{Am^2}$. 
Panel~(a) shows the concentrated local-solution trend for the fixed-$P_t$ formulation at $P_t=0.2~\mathrm{W}$. 
Panels~(b)--(i) show that increasing $P_t$ improves EIRP but changes the optimized aperture only moderately; the main design response is an increase in satellite size for solar-power generation, while the coil mass, coil diameter, and coil parameter remain comparatively stable because the EMFF control requirement is fixed.}
\label{fig:optimization_results_case_2}
\end{minipage}

\end{figure*}

Case~2 evaluates how prescribed transmit power affects EIRP and satellite sizing at $\lambda=0.30~\mathrm{m}$ and $d_{\mathrm{sat}}=0.15~\mathrm{m}$. 
The margin magnetic moment is fixed at $\bar{\mu}_{\mathrm{mar}}=0.25~\mathrm{Am^2}$, and $P_t$ is treated as the investigation parameter; unlike Case~1, the peak sidelobe level is evaluated as an output rather than used to determine $P_t$ inside the optimization. 
The numerical setup, representative design values, optimized trends, and constraint margins are shown in Table~\ref{table:case2}, Table~\ref{tab:sat_spec_case2}, Fig.~\ref{fig:optimization_results_case_2}, and Figs.~\ref{fig:sat_mass_constraint_case2}--\ref{fig:power_constraint_case2}, respectively.

\subsubsection{Case 3: Large-Spacing Feasibility Analysis}
\begin{figure*}[!tb]
\centering
\begin{minipage}[t]{\textwidth}
    \centering
    \vspace{-0.25cm}
    
    \captionof{table}{Numerical setup and input parameters for Case 3.}
    \vspace{-0.3cm}
    \label{table:case3}
    \resizebox{\textwidth}{!}{ \begin{tabular}{c|c|c|c|c}
            \hline
            Frequency [MHz] & Wavelength $\lambda$ [m] & Inter-satellite distance $d_{\mathrm{sat}}$ [m] & Surplus magnetic moment $\bar{\mu}_{\mathrm{mar}}$ [Am$^2$] & Transmitting power from an element $P_t$ [W] \\ \hline
            \rev{500} & \rev{0.60} & \rev{0.30} & 0.25 & [0.1, 0.2, 0.3, 0.4, 0.5]\\ \hline
        \end{tabular}
    }

    \vspace{0.15cm} 
    \captionof{table}{Case 3: Breakdown of size, mass, and power with overall system masses from 500 kg to 6000 kg. In terms of power, only the solar panel generates power, while all other components consume power. The bus mass and bus power are set as constants at 200 g and 200 mW, respectively. The SLL denotes the normalized peak sidelobe-envelope level relative to the main-beam peak.
    }
    \vspace{-0.3cm}
    \label{tab:sat_spec_case3}
    \begin{revtable}
    \resizebox{\textwidth}{!}{
    \begin{tabular}{@{}ll|c|c|c|c|c|c|c|c|c|c|c|c|c|c|c@{}}
        \hline
        \multicolumn{2}{c|}{$P_t$ (W)} & \multicolumn{3}{c|}{\textbf{0.1}} & \multicolumn{3}{c|}{\textbf{0.2}} & \multicolumn{3}{c|}{\textbf{0.3}} & \multicolumn{3}{c|}{\textbf{0.4}} & \multicolumn{3}{c}{\textbf{0.5}} \\ \hline
        \multicolumn{2}{l|}{\textbf{Overall system mass $\overline{m}_{\mathrm{sys}}$ (kg)}}  & \textbf{500} & \textbf{4000} & \textbf{5500} & \textbf{500} & \textbf{4000} & \textbf{5500} & \textbf{500} & \textbf{4000} & \textbf{5500} & \textbf{500} & \textbf{4000} & \textbf{5500} & \textbf{500} & \textbf{4000} & \textbf{5500} \\ \hline
        \multirow{3}{*}{Size} & Satellite $2a_{\mathrm{sat}}$ (mm) & 100.0 & 100.0 & 100.0 & 100.0 & 100.0 & \DASH & 100.0 & 100.0 & \DASH & 100.0 & 100.0 & \DASH & 100.0 & 100.0 & \DASH \\
         & Coil $2a_{\mathrm{coil}}$ (mm) & 90.0 & 90.0 & 90.0 & 90.0 & 90.0 & \DASH & 90.0 & 90.0 & \DASH & 90.0 & 90.0 & \DASH & 90.0 & 90.0 & \DASH \\
         & Coil parameter $q_{\mathrm{coil}}$ (mm$^2$) & 2.36 & 15.6 & 22.0 & 2.62 & 16.8 & \DASH & 2.93 & 18.4 & \DASH & 3.32 & 20.4 & \DASH & 3.84 & 22.5 & \DASH \\ \hline
        \multirow{5}{*}{Mass (g)} & Satellite & 401 & 816 & 1020 & 409 & 856 & \DASH & 419 & 906 & \DASH & 431 & 967 & \DASH & 448 & 1030 & \DASH \\
         & 3-axis coil & 56.4 & 371 & 526 & 62.4 & 402 & \DASH & 69.9 & 440 & \DASH & 79.3 & 487 & \DASH & 91.7 & 538 & \DASH \\
         & Battery & 24.6 & 24.6 & 24.6 & 24.6 & 24.6 & \DASH & 24.6 & 24.6 & \DASH & 24.6 & 24.6 & \DASH & 24.6 & 24.6 & \DASH \\
         & Body & 100 & 204 & 255 & 102 & 214 & \DASH & 105 & 226 & \DASH & 108 & 242 & \DASH & 112 & 259 & \DASH \\
         & 4 Solar panels & 24.0 & 24.0 & 24.0 & 24.0 & 24.0 & \DASH & 24.0 & 24.0 & \DASH & 24.0 & 24.0 & \DASH & 24.0 & 24.0 & \DASH \\ \hline
        \multirow{4}{*}{Power (mW)} & 4 Solar panels ($\times 10^{3}$) & 4.10 & 4.10 & 4.10 & 4.10 & 4.10 & \DASH & 4.10 & 4.10 & \DASH & 4.10 & 4.10 & \DASH & 4.10 & 4.10 & \DASH \\
         & Disturbance & 2580 & 3420 & 3460 & 2340 & 3100 & \DASH & 2100 & 2770 & \DASH & 1860 & 2450 & \DASH & 1630 & 2130 & \DASH \\
         & Transmitter & 333 & 333 & 333 & 667 & 667 & \DASH & 1000 & 1000 & \DASH & 1330 & 1330 & \DASH & 1670 & 1670 & \DASH \\
         & Margin ($\times 10^{3}$) & 0.988 & 0.150 & 0.106 & 0.893 & 0.139 & \DASH & 0.798 & 0.127 & \DASH & 0.703 & 0.115 & \DASH & 0.608 & 0.104 & \DASH \\ \hline
        \multirow{5}{*}{Performance} & Main-beam EIRP (dBW) & 52.0 & 63.9 & 64.7 & 54.8 & 66.5 & \DASH & 56.4 & 67.8 & \DASH & 57.4 & 68.4 & \DASH & 58.0 & 68.7 & \DASH \\
         & Peak SLL envelope (dB) & -13.2 & -13.3 & -13.3 & -13.2 & -13.3 & \DASH & -13.2 & -13.3 & \DASH & -13.2 & -13.3 & \DASH & -13.2 & -13.3 & \DASH \\
         & Maximum antenna gain (dBi) & 31.0 & 36.9 & 37.4 & 30.9 & 36.7 & \DASH & 30.8 & 36.5 & \DASH & 30.7 & 36.2 & \DASH & 30.5 & 35.8 & \DASH \\
         & Footprint diameter (km) & 89.9 & 45.8 & 43.7 & 90.7 & 46.9 & \DASH & 91.8 & 48.2 & \DASH & 93.1 & 49.8 & \DASH & 94.8 & 51.9 & \DASH \\
         & Satellite number $N_l\times N_l$ & $35\times35$ & $71\times71$ & $73\times73$ & $35\times35$ & $69\times69$ & \DASH & $35\times35$ & $67\times67$ & \DASH & $35\times35$ & $65\times65$ & \DASH & $33\times33$ & $61\times61$ & \DASH \\ \hline
    \end{tabular}
    }
    \end{revtable}
\end{minipage}
\begin{minipage}[t]{\textwidth}
    \centering
    \vspace{0.05cm}
    \begin{minipage}[t]{0.30\textwidth}
        \centering
        \subfloat[Antenna diameter ($P_t=0.2$ W)]{\includegraphics[width=\linewidth]{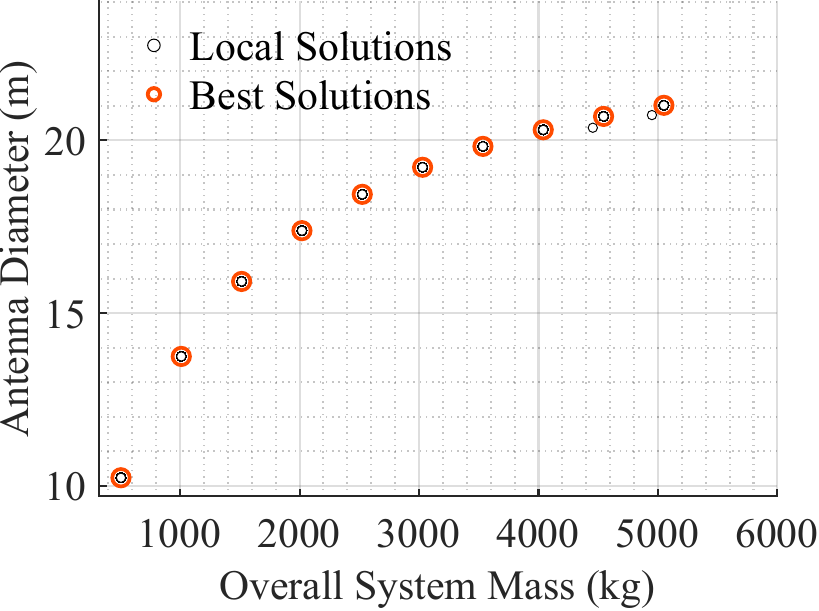}\label{fig:R_g_Pt=200mW_case3}}
    \end{minipage}
    \begin{minipage}[t]{0.30\textwidth}
        \centering
        \subfloat[Antenna diameter]{\includegraphics[width=\linewidth]{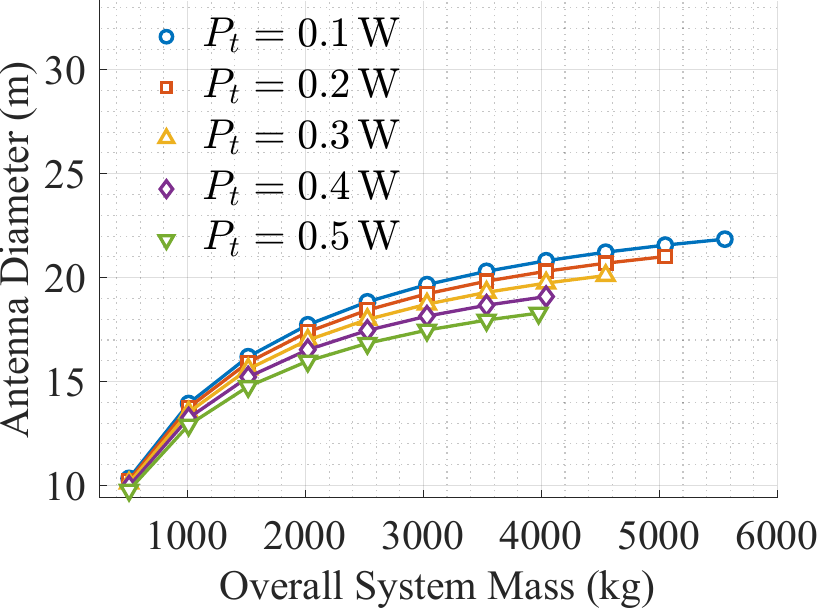}\label{fig:case3_lambda_R}}
    \end{minipage}
    \begin{minipage}[t]{0.30\textwidth}
    \centering
    \subfloat[EIRP]{\includegraphics[width=\linewidth]{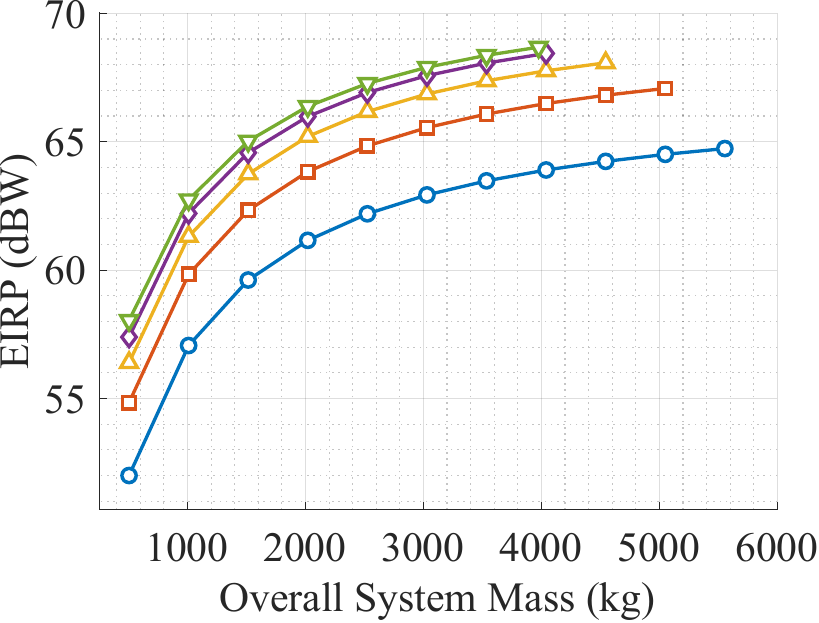}\label{fig:case3_lambda_eirp}}
    \end{minipage}\\
    \begin{minipage}[t]{0.30\textwidth}
        \centering
        \subfloat[Satellite number]{\includegraphics[width=\linewidth]{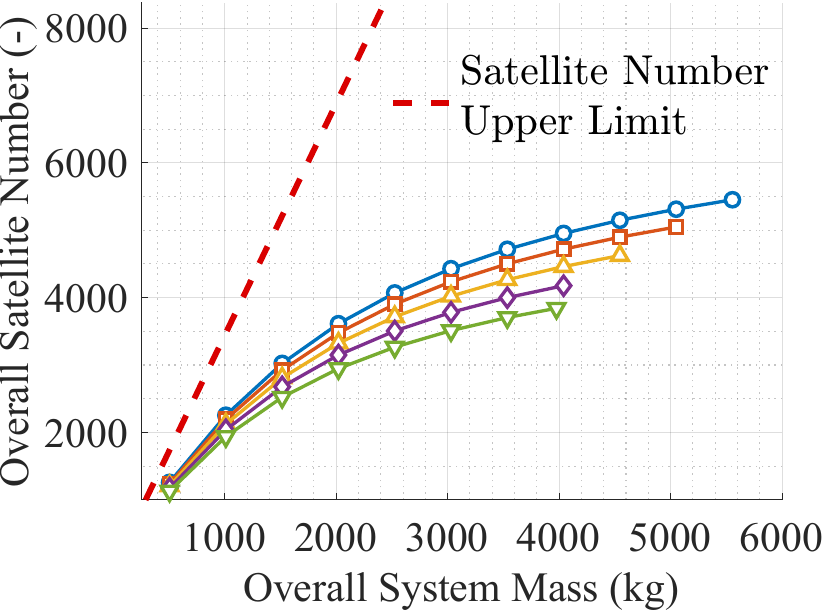}\label{fig:case3_lambda_Nall}}
    \end{minipage}
    \begin{minipage}[t]{0.30\textwidth}
        \centering
        \subfloat[Satellite mass]{\includegraphics[width=\linewidth]{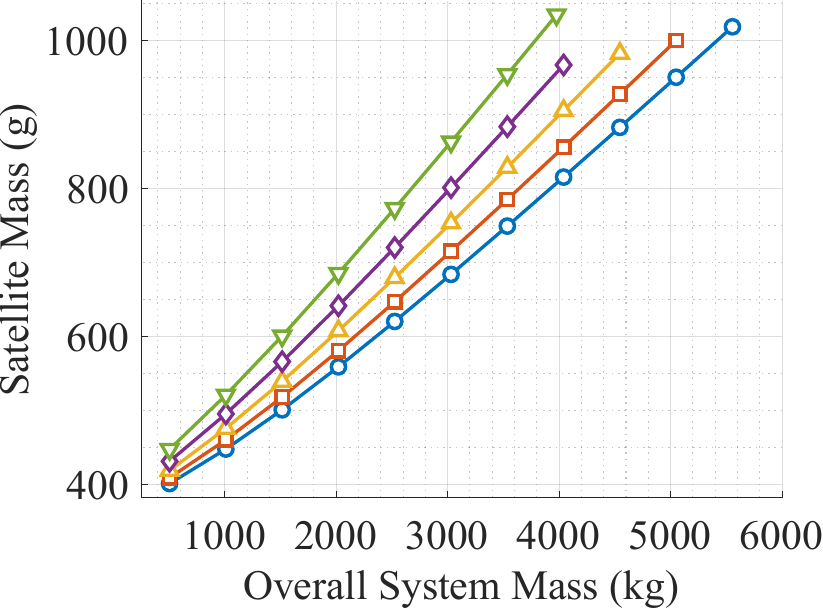}\label{fig:sat_mass_case3}}
    \end{minipage}
    \begin{minipage}[t]{0.30\textwidth}
        \centering
        \subfloat[Coil mass]{\includegraphics[width=\linewidth]{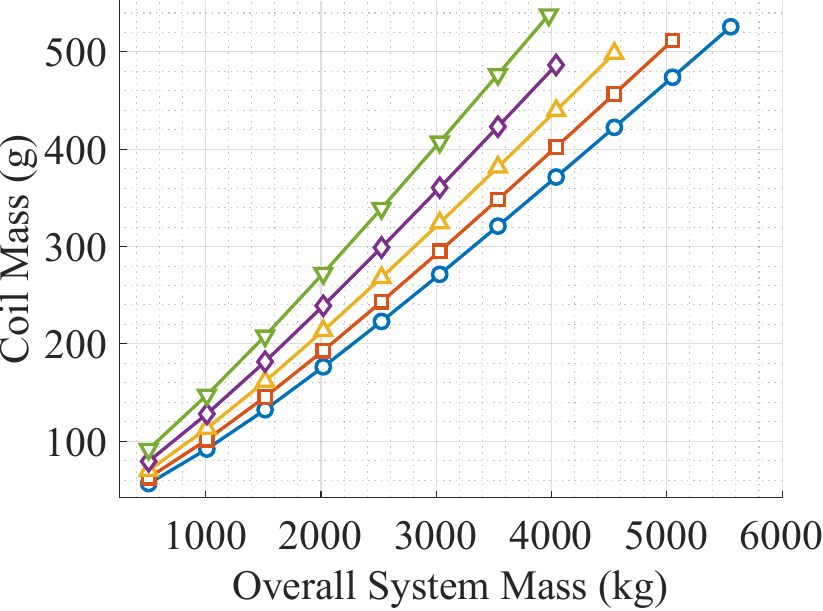}\label{fig:coil_mass_case3}}
    \end{minipage}
    \begin{minipage}[t]{0.30\textwidth}
        \centering
        \subfloat[Satellite size]{\includegraphics[width=\linewidth]{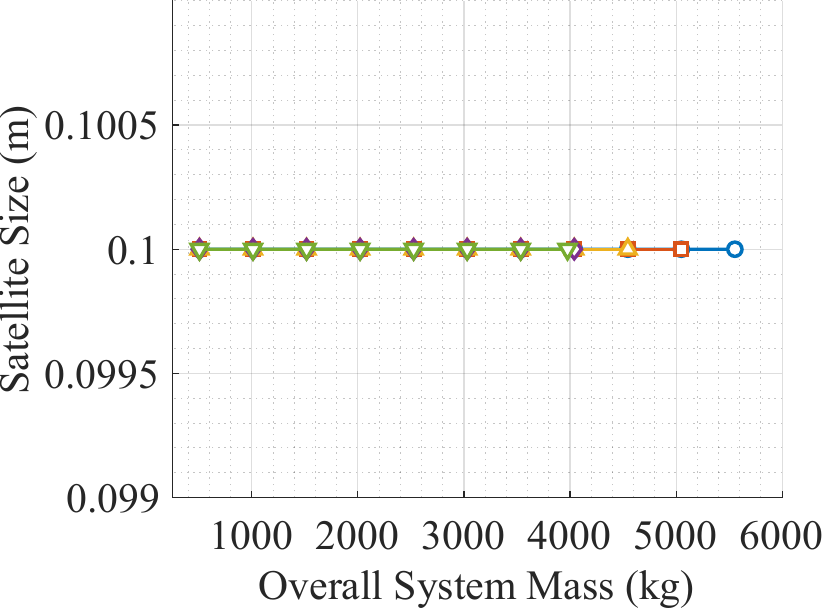}\label{fig:satellite_size_case3}}
    \end{minipage}
    \begin{minipage}[t]{0.30\textwidth}
        \centering
        \subfloat[Coil diameter]{\includegraphics[width=\linewidth]{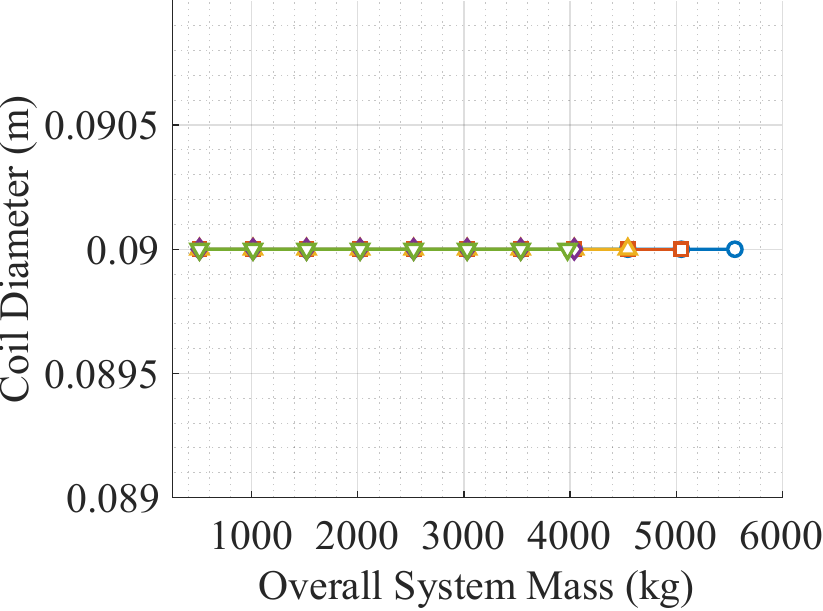}\label{fig:coil_size_case3}}
    \end{minipage}
    \begin{minipage}[t]{0.30\textwidth}
        \centering
        \subfloat[Coil parameter]{\includegraphics[width=\linewidth]{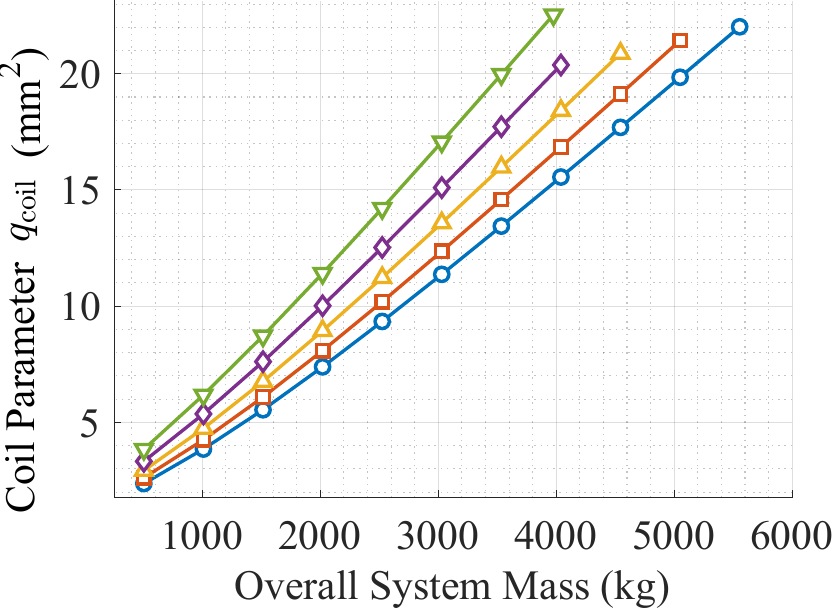}\label{fig:Coil_parameter_case3}}
    \end{minipage}
    \caption{Representative design solutions for Case~3, where the transmit power is swept under large inter-satellite spacing at $\lambda=\rev{0.60}~\mathrm{m}$, $d_{\mathrm{sat}}=\rev{0.30}~\mathrm{m}$, and $\bar{\mu}_{\mathrm{mar}}=0.25~\mathrm{Am^2}$. 
The larger spacing increases the geometric aperture but makes EMFF control more demanding. 
Because the satellite-size margin is nearly exhausted, the coil diameter cannot be enlarged further, and the optimizer increases $q_{\mathrm{coil}}$ to reduce coil-related power; this increases the coil and satellite mass, producing infeasible points when the satellite-level mass, size, and power capacities can no longer accommodate the required coil.}
    \label{fig:optimization_results_case_3}
\end{minipage}
\end{figure*}

\begin{figure*}[!tb]
\centering
\vspace{-2.5mm}
\begin{minipage}[t]{\textwidth}
\centering
    \begin{minipage}[t]{0.30\textwidth}
        \centering
        \subfloat[Case 1: Satellite mass \rev{upper} margin]{\includegraphics[width=\linewidth]{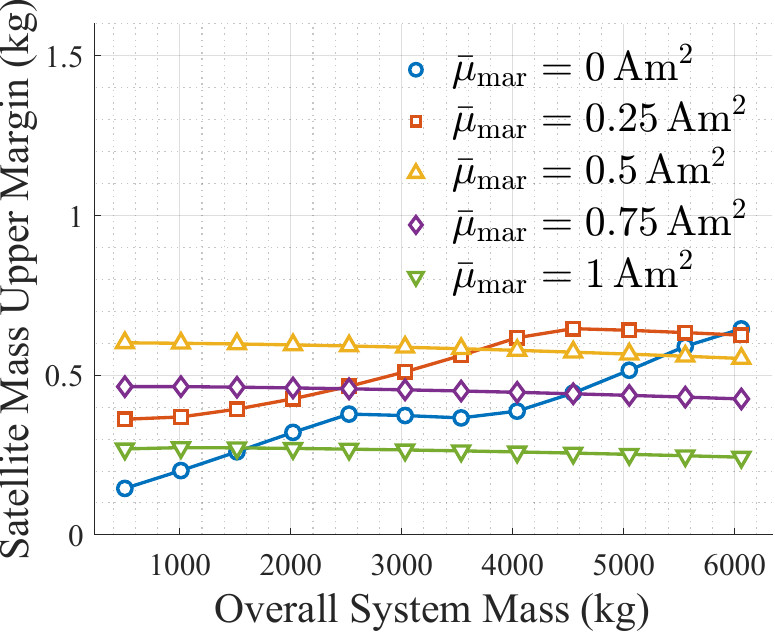}\label{fig:sat_mass_constraint_case1}}
    \end{minipage}
    \begin{minipage}[t]{0.30\textwidth}
        \centering
        \subfloat[Case 1: Satellite size upper margin.]{\includegraphics[width=\linewidth]{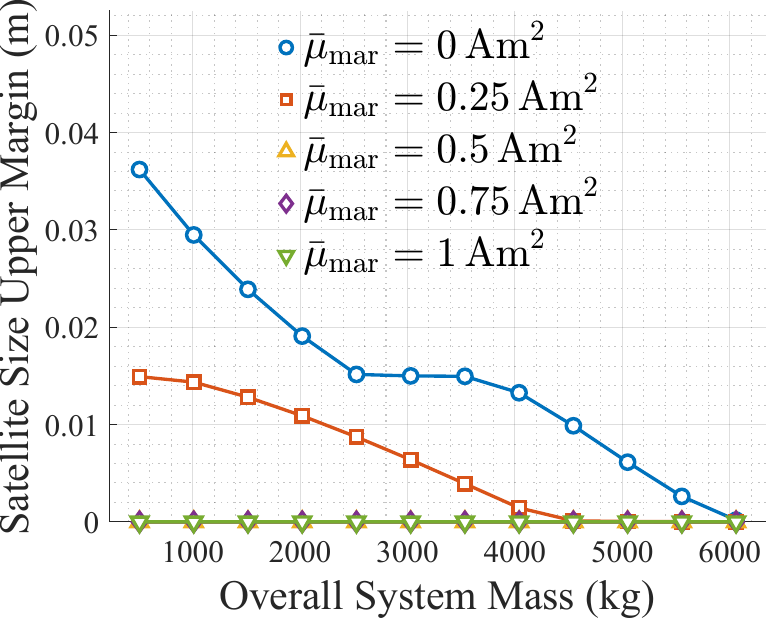}\label{fig:sat_size_constraint_case1}}
    \end{minipage}
    \begin{minipage}[t]{0.30\textwidth}
        \centering
        \subfloat[Case 1: Power margin.]{\includegraphics[width=\linewidth]{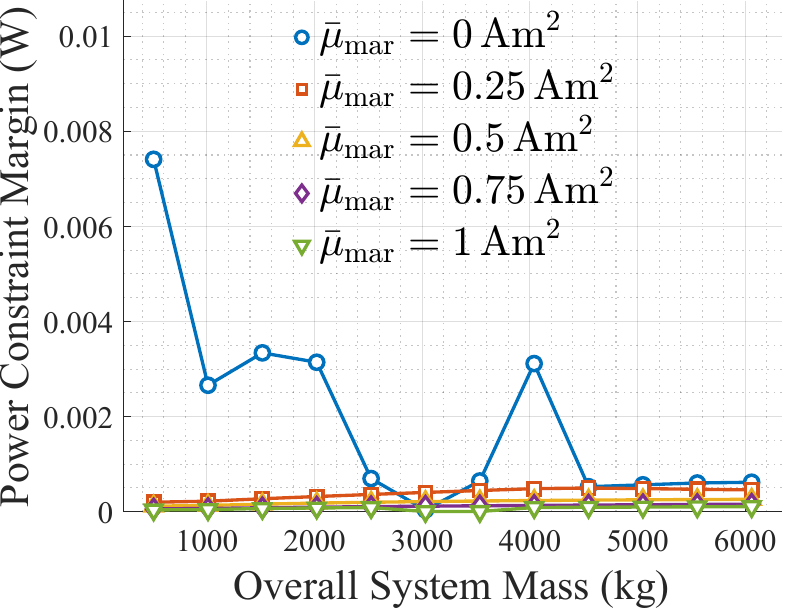}\label{fig:power_constraint_case1}}
    \end{minipage}\\
    \begin{minipage}[t]{0.30\textwidth}
        \centering
        \subfloat[Case 2: Satellite mass upper margin]{\includegraphics[width=\linewidth]{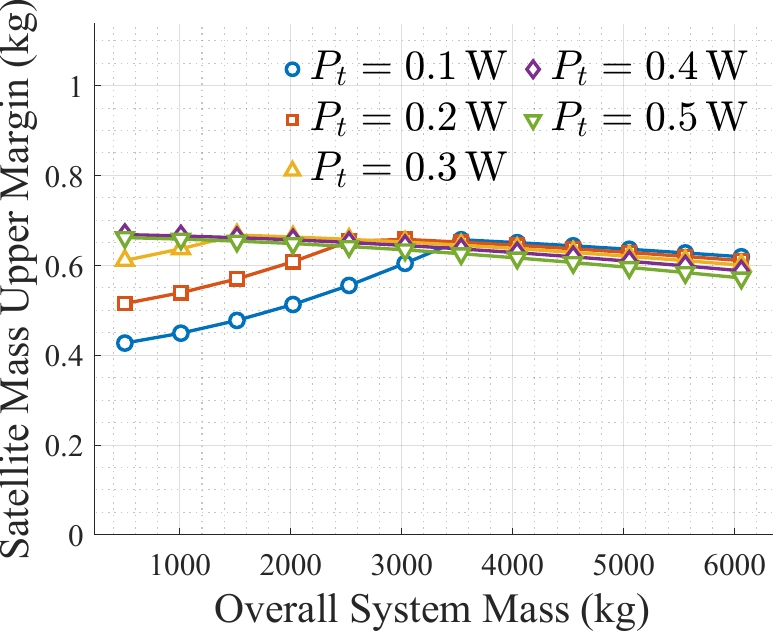}\label{fig:sat_mass_constraint_case2}}
    \end{minipage}
    \begin{minipage}[t]{0.30\textwidth}
        \centering
        \subfloat[Case 2: Satellite size upper margin]{\includegraphics[width=\linewidth]{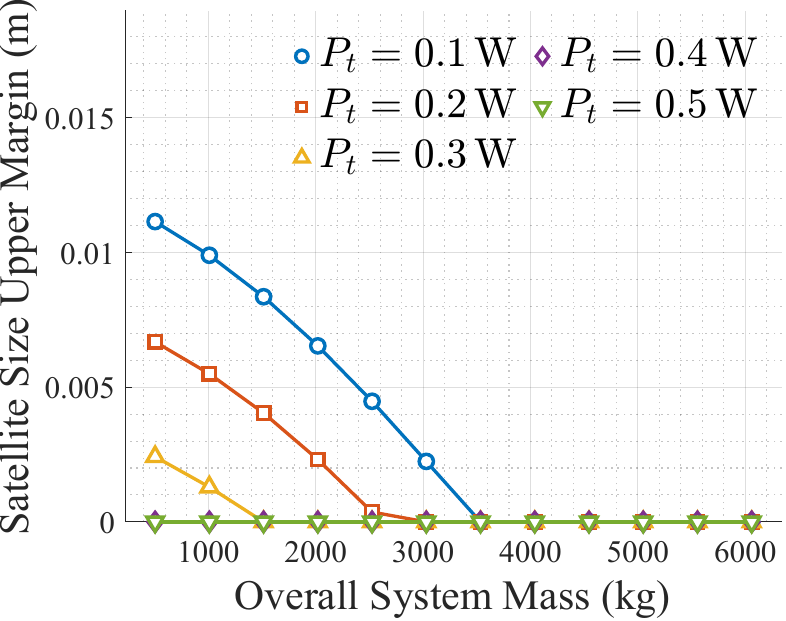}\label{fig:sat_size_constraint_case2}}
    \end{minipage}
    \begin{minipage}[t]{0.30\textwidth}
        \centering
        \subfloat[Case 2: Power margin]{\includegraphics[width=\linewidth]{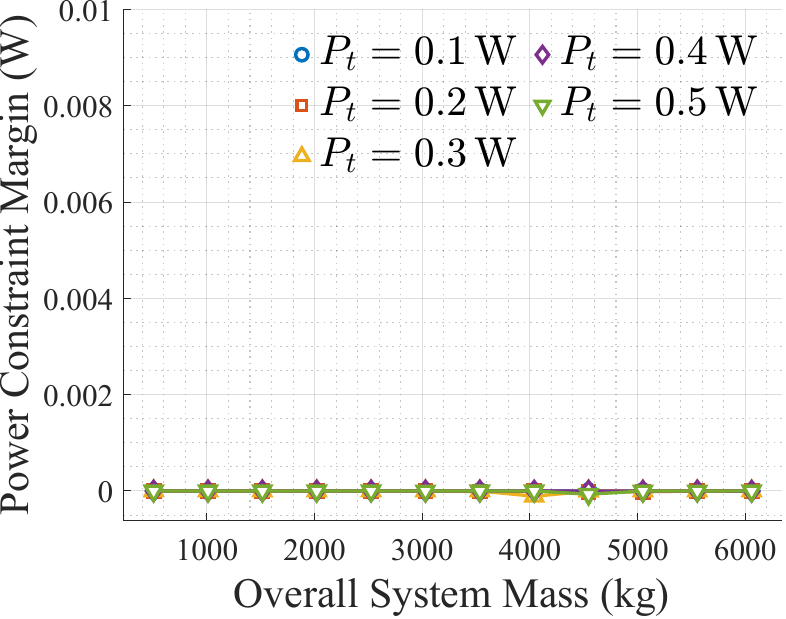}\label{fig:power_constraint_case2}}
    \end{minipage}\\
    \begin{minipage}[t]{0.30\textwidth}
        \centering
        \subfloat[Case 3: Satellite mass upper margin]{\includegraphics[width=\linewidth]{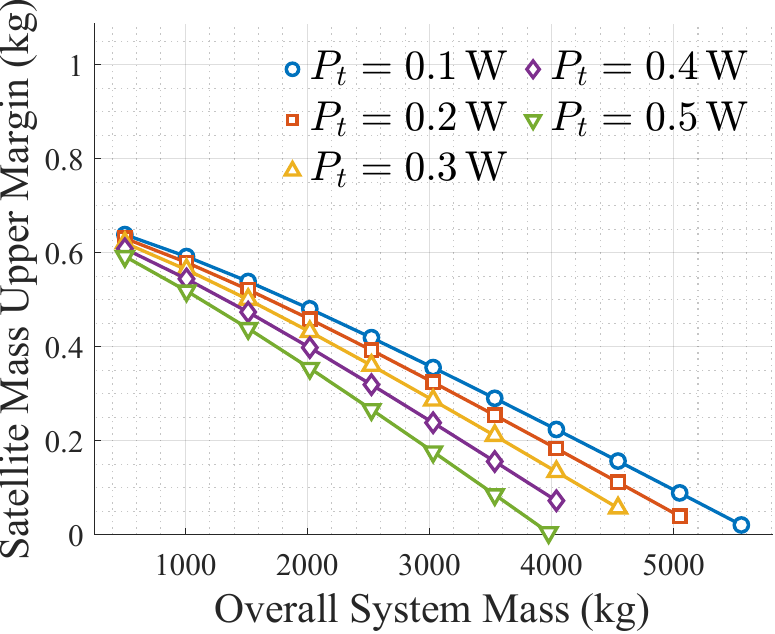}\label{fig:sat_mass_constraint_case3}}
    \end{minipage}
    \begin{minipage}[t]{0.30\textwidth}
        \centering
        \subfloat[Case 3: Satellite size margin]{\includegraphics[width=\linewidth]{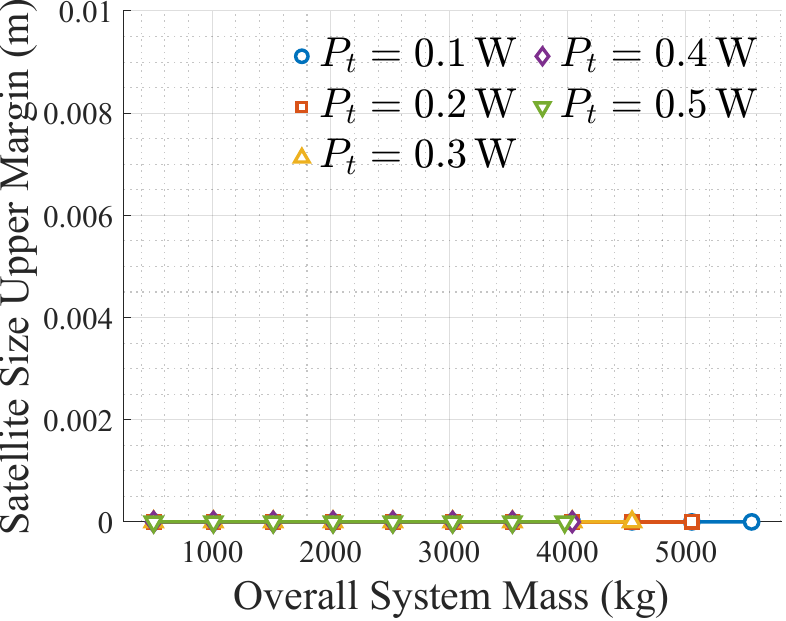}\label{fig:sat_size_constraint_case3}}
    \end{minipage}
    \begin{minipage}[t]{0.33\textwidth}
        \centering
        \subfloat[Case 3: Power margin]{\includegraphics[width=\linewidth]{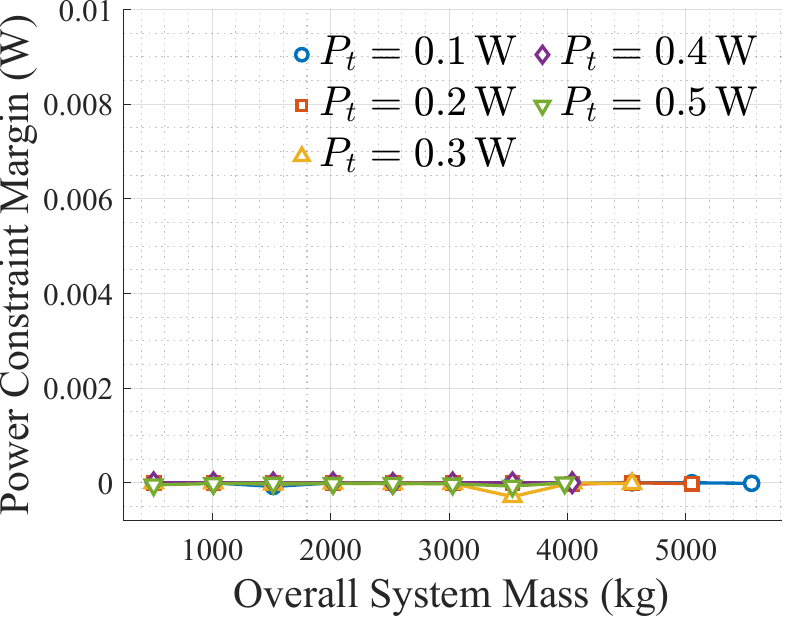}\label{fig:power_constraint_case3}}
    \end{minipage}
\caption{Constraint margins of satellite mass, satellite size, and power in each case. 
For satellite mass and satellite size, the active margin with respect to either the upper or lower bound is shown. 
The power-constraint margin is defined as the generated power from the solar panels minus the total consumed power. 
The margins indicate how the active bottleneck changes from generated power and satellite size in the baseline-spacing cases to reduced power and size margins with a decreasing satellite-mass upper margin in the large-spacing case.}
\label{fig:margin_figure_case123}
\end{minipage}
\vspace{-0.3cm}
\end{figure*}

Case~3 evaluates the feasibility limit caused by increasing the inter-satellite spacing to $d_{\mathrm{sat}}=\rev{0.30}~\mathrm{m}$ with $\lambda=\rev{0.60}~\mathrm{m}$. 
The margin magnetic moment is fixed at $\bar{\mu}_{\mathrm{mar}}=0.25~\mathrm{Am^2}$, and $P_t$ is swept as in Case~2 to compare how the larger spacing increases the coil burden required for EMFF formation maintenance. 
The numerical setup, representative design values, optimized trends, and constraint margins are shown in Table~\ref{table:case3}, Table~\ref{tab:sat_spec_case3}, Fig.~\ref{fig:optimization_results_case_3}, and Figs.~\ref{fig:sat_mass_constraint_case3}--\ref{fig:power_constraint_case3}, respectively.

\label{Trade_off_Study}
\begin{figure*}[!tb]
\centering
\begin{minipage}[t]{\textwidth}
    \centering
    \vspace{-0.25cm}
    
    \captionof{table}{Numerical setup and input parameters for Case 1.}
    \vspace{-0.3cm}
    \label{table:case1}
    \resizebox{\textwidth}{!}{ \begin{tabular}{c|c|c|c|c}
            \hline
            Frequency [GHz] & Wavelength $\lambda$ [m] & Inter-satellite distance $d_{\mathrm{sat}}$ [m] & Surplus magnetic moment $\bar{\mu}_{\mathrm{mar}}$ [Am$^2$] & Transmitting power from an element $P_t$ [W] \\ \hline
            1 & 0.30 & 0.15 & [0, 0.25, 0.50, 0.75, 1.0] & (\ref{3-3:: P_T maximum sidelobe constraint_d2d}) (Calculated for $P_r = -87.2$~dBm and $G_R = 0$~dBi.)\\ \hline
        \end{tabular}
    }

    \vspace{0.15cm} 
    \captionof{table}{Case 1: Breakdown of size, mass, and power with overall system masses from 500 kg to 6000 kg. In terms of power, only the solar panel generates power, while all other components consume power. The bus mass and bus power are set as constants at 200 g and 200 mW, respectively. The SLL denotes the normalized peak sidelobe-envelope level relative to the main-beam peak.}
    \vspace{-0.25cm}
    \begin{revtable}
    \label{tab:sat_spec_case1}
    \resizebox{\textwidth}{!}{
        \begin{tabular}{@{}ll|c|c|c|c|c|c|c|c|c|c|c|c|c|c|c@{}}
        \hline
            \multicolumn{2}{c|}{$\bar{\mu}_{\mathrm{mar}}$ (Am$^2$)} & \multicolumn{3}{c|}{\textbf{0}} & \multicolumn{3}{c|}{\textbf{0.25}} & \multicolumn{3}{c|}{\textbf{0.50}} & \multicolumn{3}{c|}{\textbf{0.75}} & \multicolumn{3}{c}{\textbf{1.0}} \\ \hline
            \multicolumn{2}{l|}{\textbf{Overall system mass $\overline{m}_{\mathrm{sys}}$ (kg)}}  & \textbf{500} & \textbf{4000} & \textbf{6000} & \textbf{500} & \textbf{4000} & \textbf{6000} & \textbf{500} & \textbf{4000} & \textbf{6000} & \textbf{500} & \textbf{4000} & \textbf{6000} & \textbf{500} & \textbf{4000} & \textbf{6000} \\ \hline
            \multirow{3}{*}{Size} & Satellite $2a_{\mathrm{sat}}$ (mm) & 63.8 & 86.7 & 99.8 & 85.1 & 98.6 & 100.0 & 100.0 & 100.0 & 100.0 & 100.0 & 100.0 & 100.0 & 100.0 & 100.0 & 100.0 \\
             & Coil $2a_{\mathrm{coil}}$ (mm) & 53.7 & 75.0 & 75.0 & 75.0 & 75.0 & 75.0 & 75.0 & 75.0 & 75.0 & 75.0 & 75.0 & 75.0 & 75.0 & 75.0 & 75.0 \\
             & Coil parameter $q_{\mathrm{coil}}$ (mm$^2$) & 0.590 & 1.67 & 2.41 & 1.58 & 2.26 & 3.32 & 4.25 & 5.15 & 6.12 & 9.49 & 10.2 & 11.0 & 16.9 & 17.3 & 17.9 \\ \hline
            \multirow{5}{*}{Mass (g)} & Satellite & 300 & 355 & 390 & 351 & 384 & 414 & 438 & 462 & 487 & 576 & 593 & 614 & 771 & 780 & 797 \\
             & 3-axis coil & 8.40 & 33.2 & 48.0 & 31.5 & 45.0 & 66.0 & 84.5 & 102 & 122 & 189 & 202 & 218 & 337 & 344 & 357 \\
             & Battery & 10.0 & 18.5 & 24.5 & 17.8 & 23.9 & 24.6 & 24.6 & 24.6 & 24.6 & 24.6 & 24.6 & 24.6 & 24.6 & 24.6 & 24.6 \\
             & Body & 75.1 & 88.7 & 97.5 & 87.7 & 96.1 & 103 & 110 & 115 & 122 & 144 & 148 & 154 & 193 & 195 & 199 \\
             & 4 Solar panels & 9.77 & 18.1 & 23.9 & 17.4 & 23.3 & 24.0 & 24.0 & 24.0 & 24.0 & 24.0 & 24.0 & 24.0 & 24.0 & 24.0 & 24.0 \\ \hline
            \multirow{4}{*}{Power (mW)} & 4 Solar panels ($\times 10^{3}$) & 1.67 & 3.08 & 4.09 & 2.97 & 3.98 & 4.10 & 4.10 & 4.10 & 4.10 & 4.10 & 4.10 & 4.10 & 4.10 & 4.10 & 4.10 \\
             & Disturbance & 1450 & 2880 & 3890 & 203 & 2000 & 2680 & 79.0 & 766 & 1260 & 38.0 & 330 & 588 & 23.3 & 169 & 301 \\
             & Transmitter & 10.4 & 0.228 & 0.122 & 14.2 & 0.267 & 0.138 & 22.1 & 0.386 & 0.191 & 38.1 & 0.636 & 0.303 & 68.0 & 1.10 & 0.510 \\
             & Margin ($\times 10^{3}$) & 0 & 0 & 0 & 2.55 & 1.79 & 1.22 & 3.80 & 3.13 & 2.64 & 3.82 & 3.57 & 3.31 & 3.81 & 3.73 & 3.60 \\ \hline
            \multirow{5}{*}{Performance} & Main-beam EIRP (dBW) & 39.5 & 39.5 & 39.5 & 39.5 & 39.5 & 39.5 & 39.4 & 39.5 & 39.5 & 39.4 & 39.5 & 39.5 & 39.4 & 39.5 & 39.5 \\
             & Peak SLL envelope (dB) & -13.2 & -13.3 & -13.3 & -13.2 & -13.3 & -13.3 & -13.2 & -13.3 & -13.3 & -13.2 & -13.3 & -13.3 & -13.2 & -13.3 & -13.3 \\
             & Maximum antenna gain (dBi) & 32.3 & 40.6 & 41.9 & 31.6 & 40.2 & 41.7 & 30.6 & 39.4 & 40.9 & 29.4 & 38.3 & 39.9 & 28.2 & 37.1 & 38.8 \\
             & Footprint diameter (km) & 78.0 & 30.4 & 26.0 & 84.1 & 31.6 & 26.8 & 93.8 & 34.6 & 29.0 & 107 & 39.1 & 32.6 & 124 & 44.8 & 37.1 \\
             & Satellite number $N_l\times N_l$ & $41\times41$ & $107\times107$ & $125\times125$ & $37\times37$ & $103\times103$ & $121\times121$ & $33\times33$ & $93\times93$ & $111\times111$ & $29\times29$ & $83\times83$ & $99\times99$ & $25\times25$ & $71\times71$ & $87\times87$ \\ \hline
        \end{tabular}
    }
    \end{revtable}
\end{minipage}

\begin{minipage}[t]{\textwidth}
    \centering
    \vspace{0.1cm}
    \begin{minipage}[t]{0.295\textwidth}
        \centering
        \subfloat[Antenna diameter ($\bar{\mu}_{\mathrm{mar}}=0.25$ Am$^2$)]{\includegraphics[width=\linewidth]{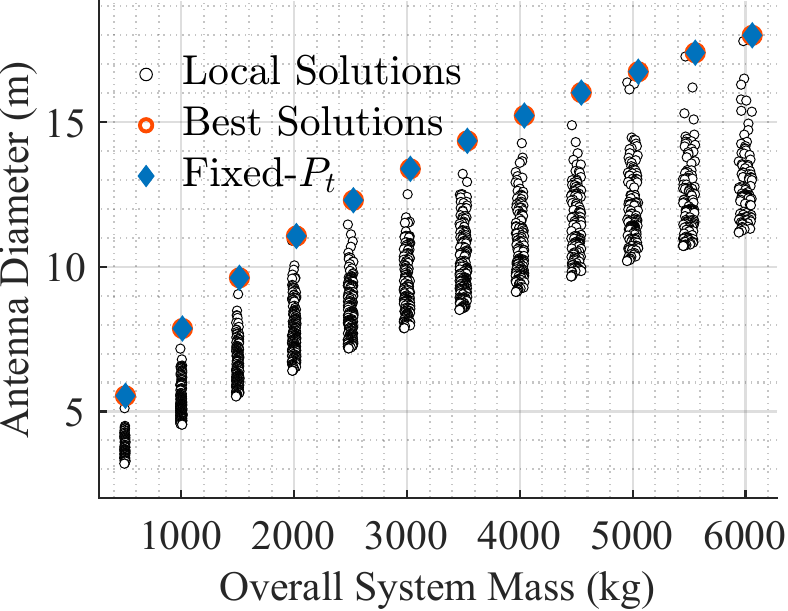}\label{fig:antenna_diameter_case1_mu=025}}
    \end{minipage}
    \begin{minipage}[t]{0.295\textwidth}
        \centering
        \subfloat[Antenna diameter]{\includegraphics[width=\linewidth]{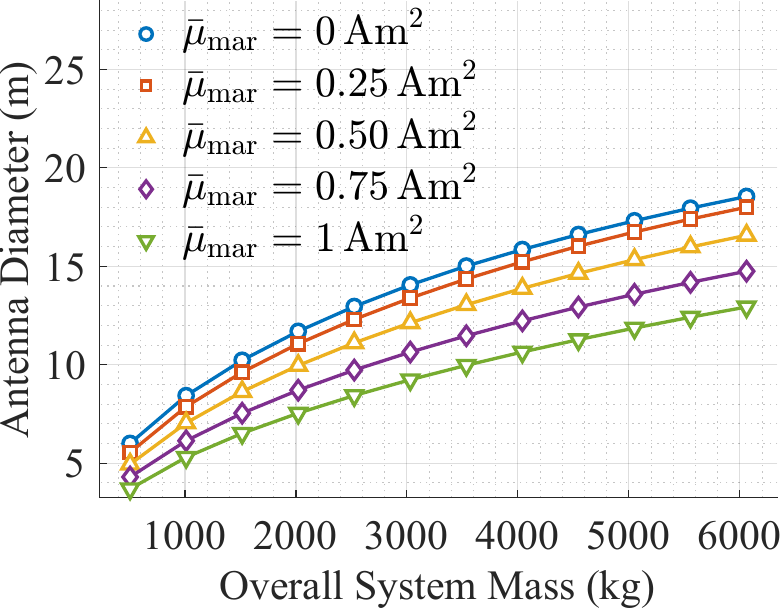}\label{fig:R_g_case1}}
    \end{minipage}
    \begin{minipage}[t]{0.295\textwidth}
        \centering
        \subfloat[Footprint diameter]{\includegraphics[width=\linewidth]{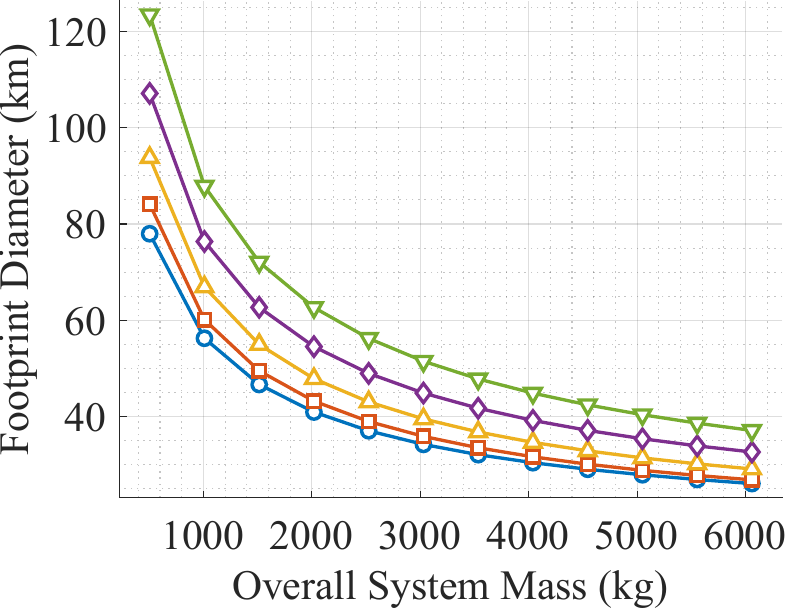}\label{fig:main_lobe_case1}}
    \end{minipage}\\
    \begin{minipage}[t]{0.295\textwidth}
        \centering
        \subfloat[Satellite number]{\includegraphics[width=\linewidth]{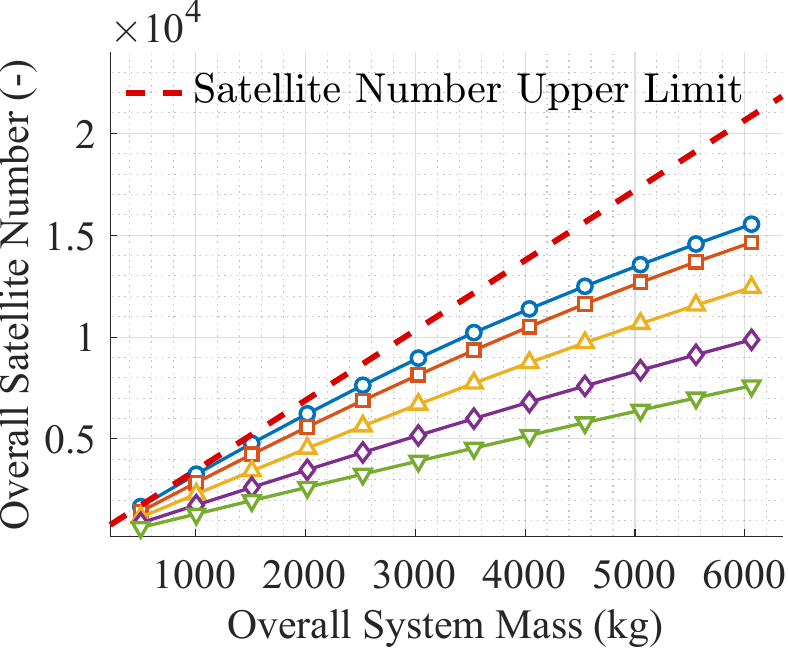}\label{fig:N_sat_case1}}
    \end{minipage}
    \begin{minipage}[t]{0.295\textwidth}
        \centering
        \subfloat[Satellite mass]{\includegraphics[width=\linewidth]{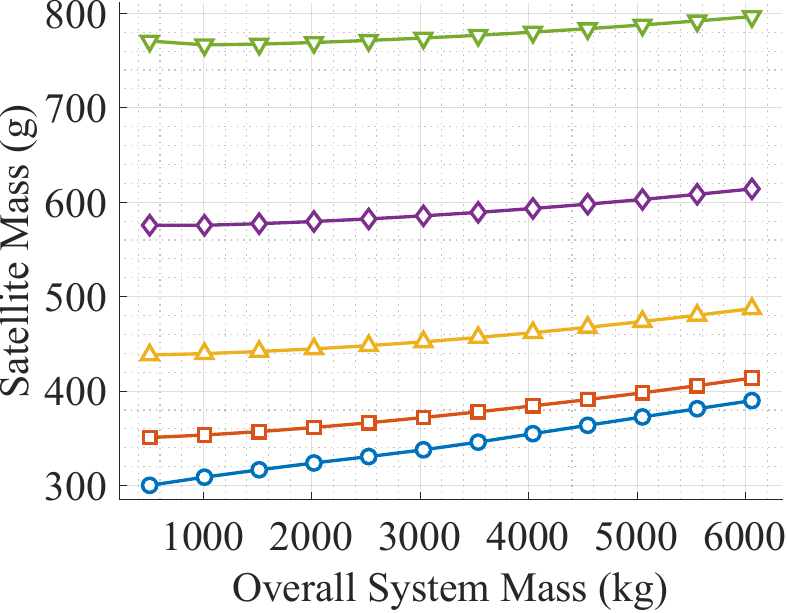}\label{fig:sat_mass_case1}}
    \end{minipage}
    \begin{minipage}[t]{0.295\textwidth}
        \centering
        \subfloat[Coil mass]{\includegraphics[width=\linewidth]{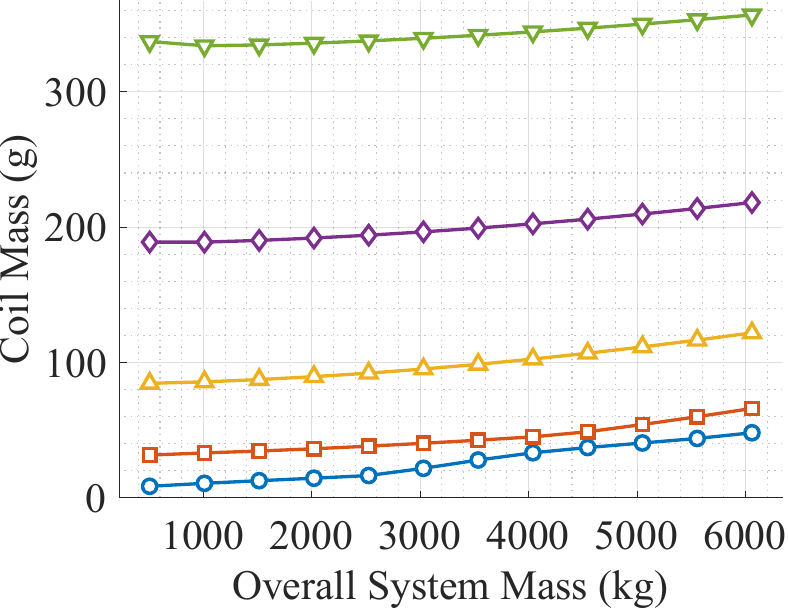}\label{fig:coil_mass_case1}}
    \end{minipage}\\
    \begin{minipage}[t]{0.295\textwidth}
        \centering
        \subfloat[Satellite size]{\includegraphics[width=\linewidth]{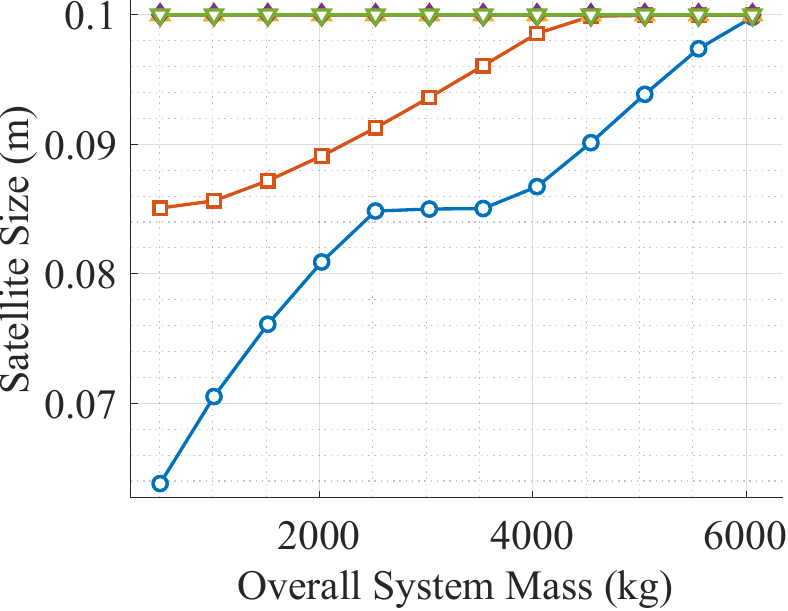}\label{fig:sat_size_case1}}
    \end{minipage}
    \begin{minipage}[t]{0.295\textwidth}
        \centering
        \subfloat[Coil diameter]{\includegraphics[width=\linewidth]{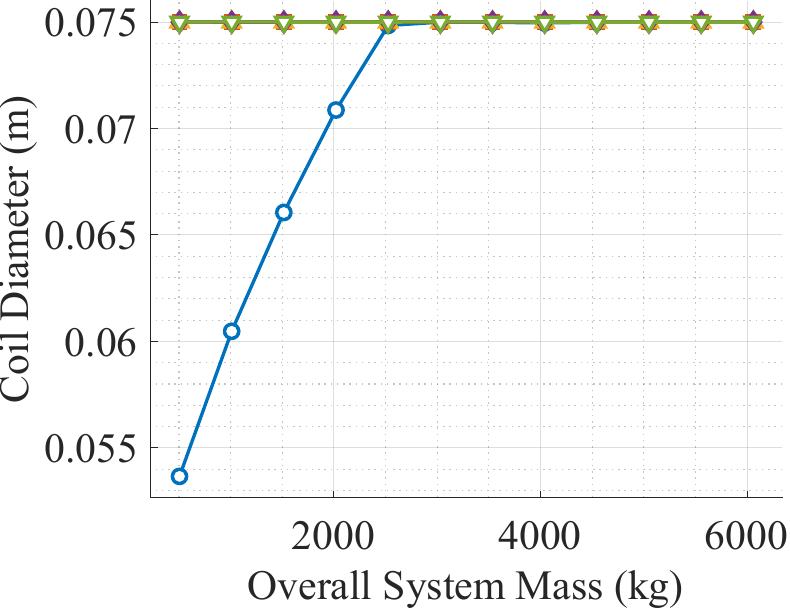}\label{fig:coil_diameter_case1}}
    \end{minipage}
    \begin{minipage}[t]{0.295\textwidth}
        \centering
        \subfloat[Coil parameter]{\includegraphics[width=\linewidth]{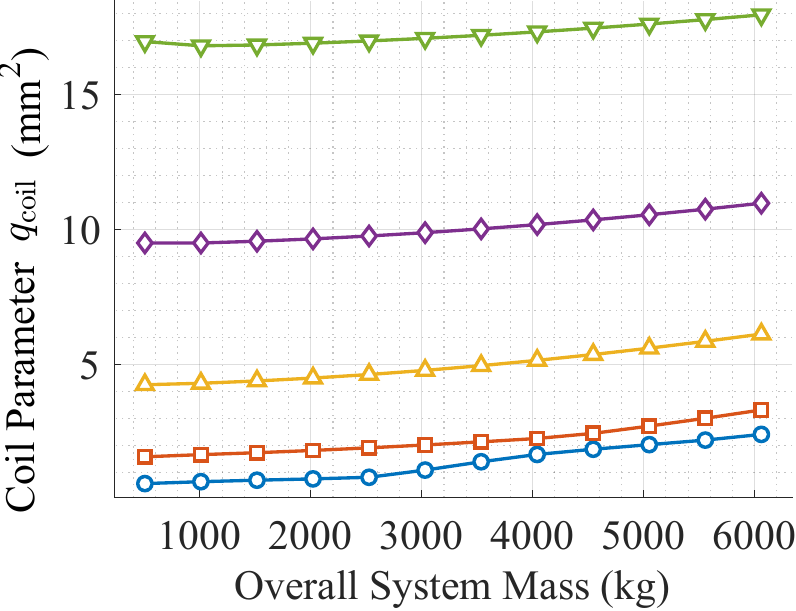}\label{fig:coil_parameter_case1}}
    \end{minipage}
    \caption{Representative design solutions for Case~1, where the margin magnetic moment is swept at $\lambda=0.30~\mathrm{m}$ and $d_{\mathrm{sat}}=0.15~\mathrm{m}$. 
    Panel~(a) compares the original formulation with the fixed-$P_t$ formulation at $\bar{\mu}_{\mathrm{mar}}=0.25~\mathrm{Am^2}$, showing the scatter of local solutions caused by the peak-sidelobe-related variable. 
    Panels~(b)--(i) show the optimal design trends for $\bar{\mu}_{\mathrm{mar}}=0$--$1.0~\mathrm{Am^2}$; the zero-margin case approaches the satellite-number upper bound with near-minimum satellite mass, whereas nonzero margin magnetic moment reduces the feasible aperture and shifts the coil adjustment toward $q_{\mathrm{coil}}$ as the generated-power and satellite-size margins become restrictive.}
    \label{fig:optimization_results_case_1}
\end{minipage}
\end{figure*}

\subsection{\rev{System-Level Design Analysis}}

\subsubsection{System-Level Trade-Off}
The three case studies show that aperture enlargement improves antenna performance only while the satellite-level design retains sufficient headroom for EMFF control. Figure~\ref{fig:optimization_results_case_1} and Fig.~\ref{fig:optimization_results_case_2} show that increasing the overall system mass generally enlarges the optimized aperture in the baseline-spacing cases, leading to footprint reduction in Case~1 and EIRP improvement in Case~2. Figure~\ref{fig:optimization_results_case_3}, however, shows that the large-spacing case reaches infeasible design points even when the overall system mass is increased. Figure~\ref{fig:margin_figure_case123} supports this interpretation by showing that, in Case~3, the satellite-size and power margins are already nearly exhausted while the satellite-mass margin decreases toward the upper-bound limit. Tables~\rev{\ref{tab:sat_spec_case1}}, \rev{\ref{tab:sat_spec_case2}}, and \rev{\ref{tab:sat_spec_case3}} summarize the corresponding representative satellite specifications and antenna-performance values. The following subsections discuss the mechanisms that determine this design headroom: fixed-transmit-power simplification, margin magnetic moment, transmit power, and large inter-satellite spacing.

\subsubsection{Trends in Optimized Design Solutions}
\rev{Case~1 demonstrates that increasing the margin magnetic moment shifts the dominant design adjustment from geometric sizing to \(q_{\mathrm{coil}}\). Since \(P_{\mathrm{cont}}\) and \(P_{\mathrm{mar}}\) scale as \(1/(q_{\mathrm{coil}}a_{\mathrm{coil}}^3)\), the coil diameter initially increases until reaching its distance-dependent upper bound. For \(\bar{\mu}_{\mathrm{mar}}=0\), the resulting satellite-size bound produces an intermediate plateau, during which \(q_{\mathrm{coil}}\) increases and the satellite-mass upper margin decreases slightly (Figs.~\ref{fig:sat_size_case1}, \ref{fig:coil_diameter_case1}, \ref{fig:coil_parameter_case1}, and~\ref{fig:sat_mass_constraint_case1}). For \(\bar{\mu}_{\mathrm{mar}}=0.25~\mathrm{A\,m^2}\), the satellite size increases after the coil-diameter bound becomes active, followed by \(q_{\mathrm{coil}}\) adjustment after the satellite-size bound is reached. For \(\bar{\mu}_{\mathrm{mar}}\geq0.5~\mathrm{A\,m^2}\), both size bounds are active from \(500~\mathrm{kg}\). Thus, higher magnetic-moment requirements activate the geometric constraints earlier and make \(q_{\mathrm{coil}}\) the primary remaining design variable.}

\rev{Cases~2 and 3 show that the same \(q_{\mathrm{coil}}\)-driven trade-off becomes dominant as the power or spacing requirement increases. In Case~2, the additional demand for \(P_t=0.1\)--\(0.3~\mathrm{W}\) is initially accommodated by satellite-size growth and then by increasing \(q_{\mathrm{coil}}\) after the size bound becomes active (Figs.~\ref{fig:satellite_size_case2}, \ref{fig:coil_parameter_case2}, and~\ref{fig:sat_mass_constraint_case2}). For \(P_t=0.4\) and \(0.5~\mathrm{W}\), this bound is active from \(500~\mathrm{kg}\), and the \(q_{\mathrm{coil}}\)-driven trend appears over the entire mass range. In Case~3, the size and power margins are nearly exhausted, and, thus, increasing \(q_{\mathrm{coil}}\) reduces the EMFF power requirement at the expense of coil mass (Figs.~\ref{fig:coil_mass_case3}, \ref{fig:Coil_parameter_case3}, \ref{fig:sat_size_constraint_case3}, and~\ref{fig:power_constraint_case3}). The resulting loss of satellite-mass margin eventually eliminates feasible solutions (Fig.~\ref{fig:sat_mass_constraint_case3}). Accordingly, increasingly restrictive requirements shift the active bottleneck from geometric sizing to the \(q_{\mathrm{coil}}\)-dependent power--mass trade-off, which determines the feasibility boundary.}

\subsubsection{Convex Relaxation of the Fixed-$P_t$ Formulation}
The fixed-$P_t$ formulation explains why the local solutions become concentrated in Case~2. In the original Case~1 formulation, $P_t$ is derived through $u_{\mathrm{psl}}$, which introduces a strongly nonlinear peak-sidelobe constraint. When the $P_t$ values obtained from Case~1 are fixed and the problem is recalculated, the solutions nearly follow the selected optimal trend of the original formulation, as shown in Fig.~\ref{fig:antenna_diameter_case1_mu=025}. Case~2 has the same simplified structure because $P_t$ is prescribed, and Fig.~\ref{fig:R_g_Pt200mW_case2} shows a concentrated local-solution trend. \rev{This structural simplification is also reflected in the total wall-clock times reported in Table~\ref{tab:computation_time}.} This behavior indicates that removing $u_{\mathrm{psl}}$ leaves a mostly monotonic active-constraint structure, but it does not prove global convexity of the original formulation.
\begin{table}[!t]
    \begin{revtable}
    \caption{Total elapsed wall-clock times for Cases~1--3.
    Each case comprised 60 optimization problems
    (five parameter settings $\times$ twelve system-mass points).}
    \label{tab:computation_time}
    \centering
    \renewcommand{\arraystretch}{1.2}
    \begin{tabularx}{\columnwidth}{
        >{\centering\arraybackslash}X
        |>{\centering\arraybackslash}X}
        \hline
        Case & Total wall-clock time \\
        \hline
        Case 1 & $17.9~\mathrm{h}$ \\
        \hline
        Case 2 & $47.3~\mathrm{min}$ \\
        \hline
        Case 3 & $49.3~\mathrm{min}$ \\
        \hline
    \end{tabularx}
    \end{revtable}
    \vspace{-0.3cm}
\end{table}

\begin{remark}
The original problem is not convex because it includes $u_{\mathrm{psl}}$, integer-like satellite-number sizing, empirical satellite-mass bounds, and a numerically fitted $J_d^*(N_l)$ with negative polynomial coefficients.
\rev{A relaxed fixed-$P_t$ problem can be written as a geometric program under the following conditions.
The variable $u_{\mathrm{psl}}$ is removed, $N_l$ is relaxed as a positive continuous variable, and $m_{\mathrm{sys}}$ is fixed.
The satellite mass $m_{\mathrm{sat}}$ is eliminated as a dependent posynomial, and the total-mass constraint is relaxed to $N_l^2m_{\mathrm{sat}}\leq m_{\mathrm{sys}}$.
The satellite-mass bound is replaced by a monomial envelope, and $J_d^*(N_l)$ is replaced by a positive-coefficient posynomial upper envelope.}
Under these relaxations, the problem becomes convex after \rev{a} logarithmic variable transformation~\cite{boyd2004convex}.
This convexified problem is a conservative relaxation, not a proof of convexity of the original formulation.
\end{remark}

\subsection{\rev{Higher-Fidelity Verification of Designs}}
\rev{This subsection verifies representative design solutions obtained using the proposed framework from antenna-performance and formation-control perspectives. Closed-loop simulations under nonlinear $J_2$ dynamics assess the control-power budget, while electromagnetic-field simulations evaluate the antenna gain considering a non-ideal element model and mutual coupling.}

\subsubsection{\rev{Antenna-Performance Verification With Mutual Coupling}}
\begin{figure}[!t]
    \centering
    \subfloat[Low-gain antenna element approximating an omnidirectional element]{\parbox{\linewidth}{\centering
            \includegraphics[width=0.45\linewidth]{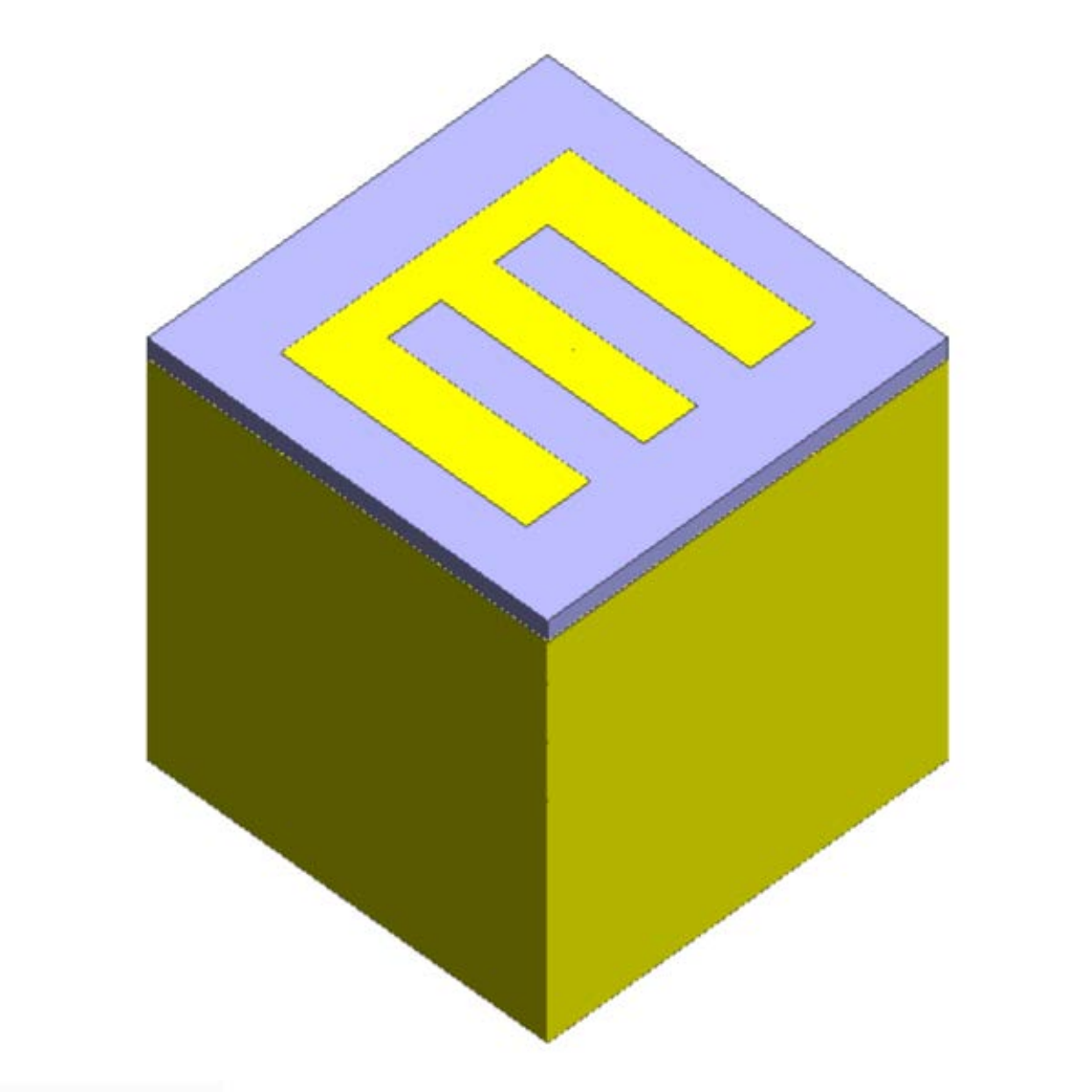}\label{fig:antenna_array_element}
        }}
    \vspace{1mm}
    \subfloat[Simulated array gain pattern including mutual coupling for Case~1 with $\bar{\mu}_{\mathrm{mar}}=\rev{0.50}~\mathrm{A\,m^2}$ and $\bar{m}_{\mathrm{sys}}=\rev{4000}~\mathrm{kg}$]{\parbox{\linewidth}{\centering
            \includegraphics[width=0.85\linewidth]{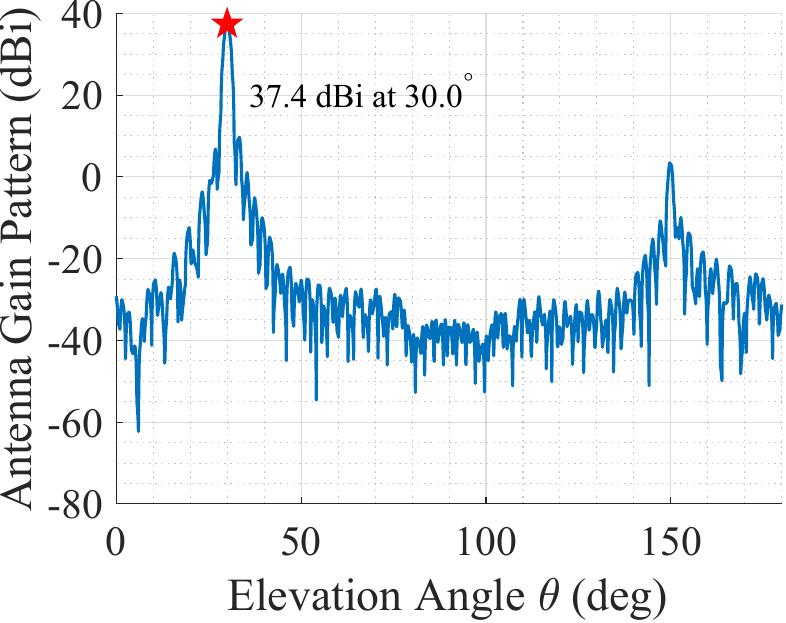}\label{fig:3dimage}
        }}
    \caption{Electromagnetic-field simulation used to assess the combined effects of the non-ideal antenna-element model and mutual coupling. Panel (a) shows the low-gain element used to approximate the omnidirectional element assumed in the system-level model. Panel (b) shows the radiated gain pattern of the representative Case~1 design with $\bar{\mu}_{\mathrm{mar}}=\rev{0.50}~\mathrm{A\,m^2}$ and $\bar{m}_{\mathrm{sys}}=\rev{4000}~\mathrm{kg}$.}
    \label{fig:mutual coupling discussion}
\end{figure}

\rev{Figure~\ref{fig:mutual coupling discussion} evaluates a representative Case~1 design using a non-ideal antenna-element model with mutual coupling. The resulting gain is $37.4~\mathrm{dBi}$, compared with $39.4~\mathrm{dBi}$ for the simplified model used in the system-level design. Maintaining the same EIRP under this approximately $2~\mathrm{dB}$ gain reduction would require additional RF transmit power, consistent with the trade-off in Case~2. This additional power demand may reduce the feasible aperture under a fixed mass budget.}

\subsubsection{\rev{Verification for Drift Compensation Design}}
\begin{figure*}[!bt]
    \centering
    \subfloat[Minimum neighboring-pair distance.\label{fig:validation_distance}]{\includegraphics[width=0.485\textwidth]{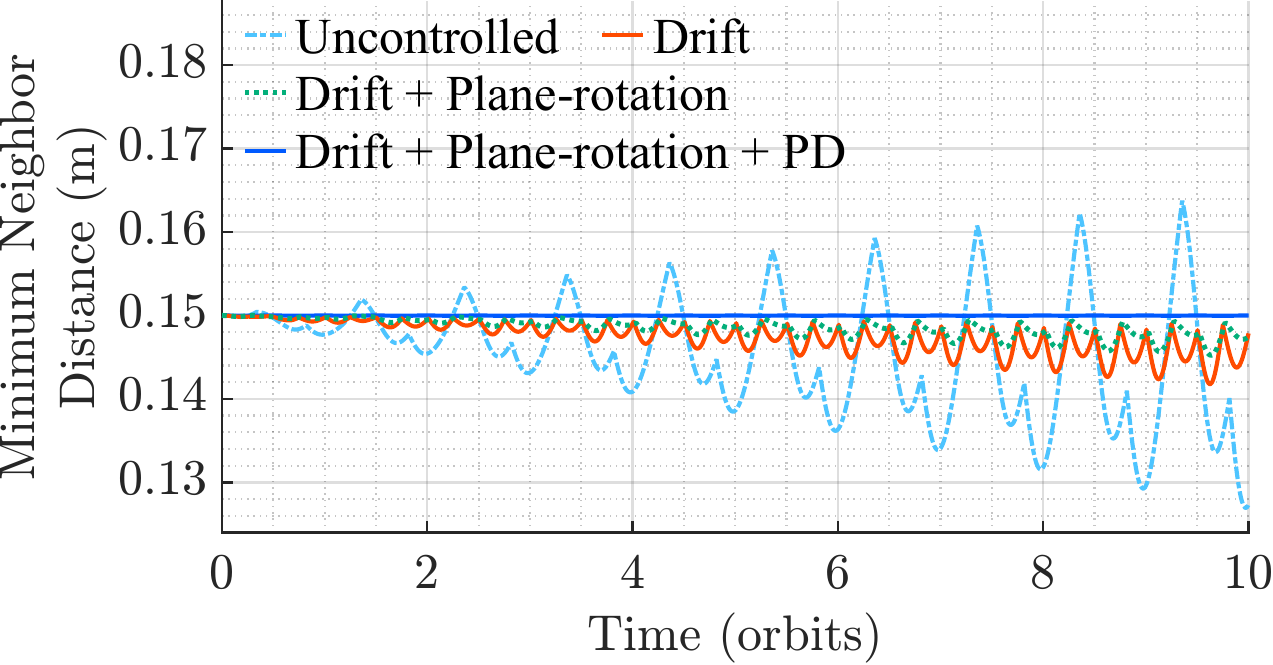}
    }
    \subfloat[Maximum residual drift vector.\label{fig:validation_drift}]{\includegraphics[width=0.485\textwidth]{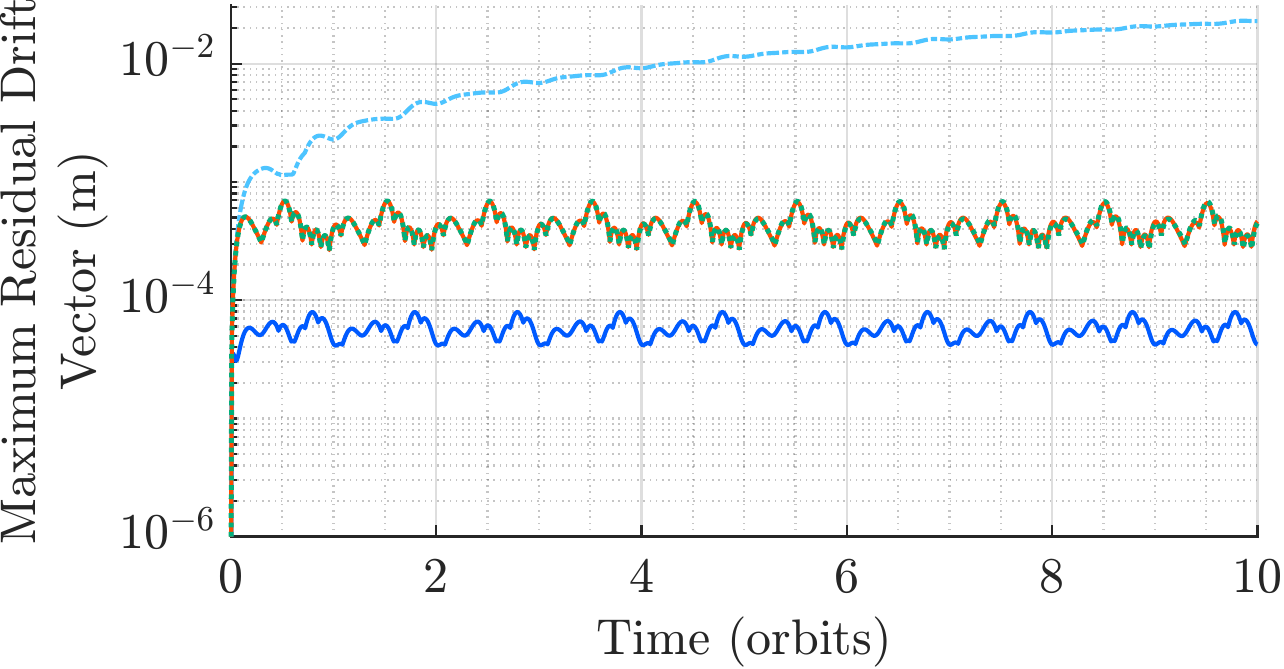}
    }
    
    \vspace{-0.3cm}
    
    \subfloat[Maximum intersatellite control force.\label{fig:validation_force}]{\includegraphics[width=0.485\textwidth]{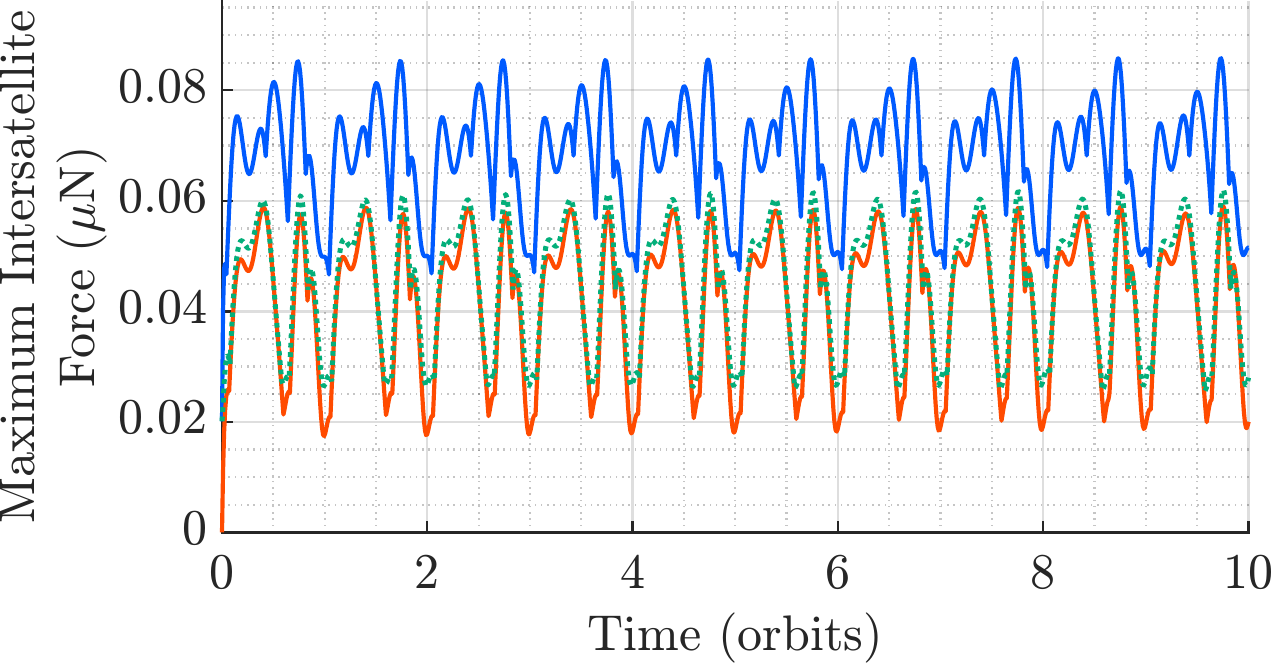}
    }
    \subfloat[Maximum per-satellite control power in each orbit.\label{fig:validation_power}]{\includegraphics[width=0.485\textwidth]{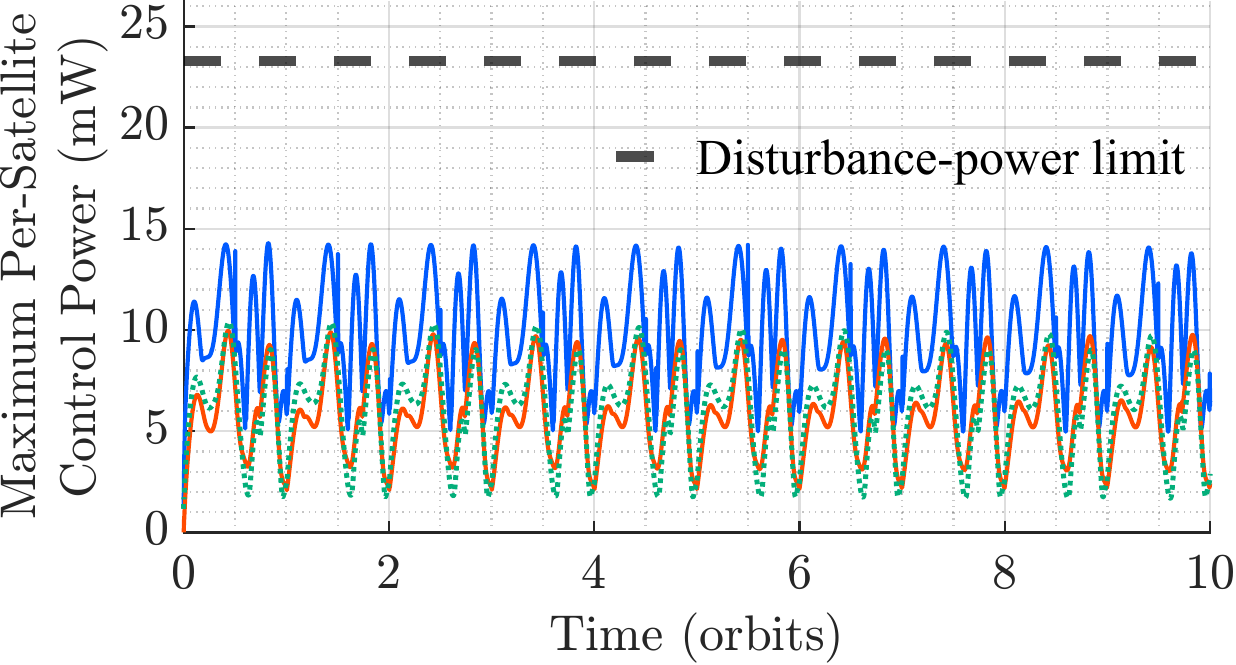}
    }
    
    \vspace{0.15cm}
    
    \par\noindent
    \makebox[\textwidth][s]{%
\begin{minipage}[t]{0.24\textwidth}\vspace{0pt}\centering
\subfloat[Uncontrolled.\label{fig:validation_formation_uncontrolled}]{\includegraphics[width=\linewidth]{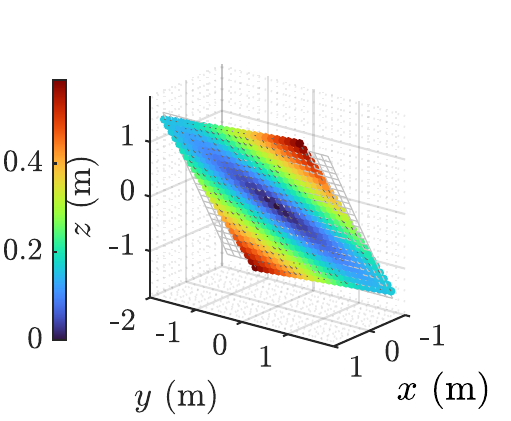}}%
\end{minipage}%
\hfill%
\begin{minipage}[t]{0.24\textwidth}\vspace{0pt}\centering
\subfloat[Drift.\label{fig:validation_formation_drift}]{\includegraphics[width=\linewidth]{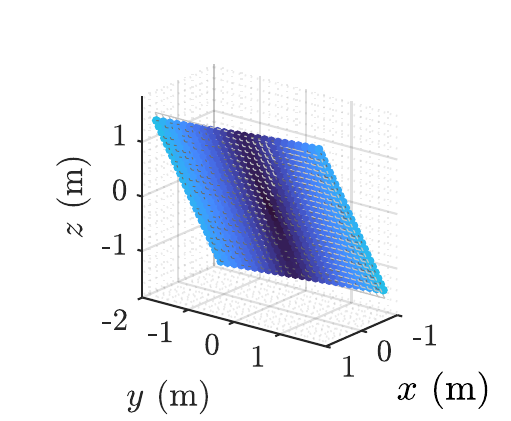}}%
\end{minipage}%
\hfill%
\begin{minipage}[t]{0.24\textwidth}\vspace{0pt}\centering
\subfloat[Drift and plane-rotation.\label{fig:validation_formation_rotation}]{\includegraphics[width=\linewidth]{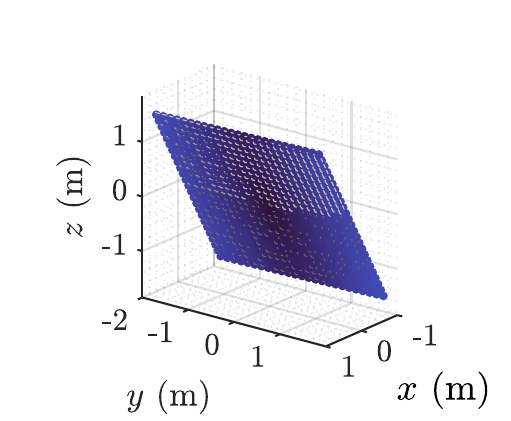}}%
\end{minipage}%
\hfill%
\begin{minipage}[t]{0.24\textwidth}\vspace{0pt}\centering
\subfloat[Drift, plane-rotation, and PD.\label{fig:validation_formation_pd}]{\includegraphics[width=\linewidth]{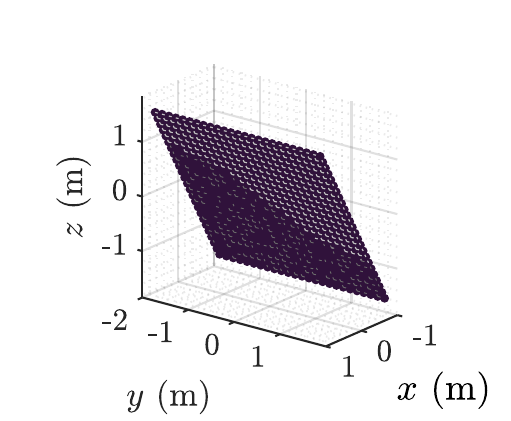}}%
\end{minipage}%
    }%
    \par
\caption{\rev{Ten-orbit validation under nonlinear $J_2$ dynamics for the selected Case~1 design. The formation consisted of $25\times25$ satellites with an initial intersatellite spacing of $0.15~\mathrm{m}$, a surplus magnetic moment of $1.0~\mathrm{A\,m^2}$, and a total-system mass budget of $500~\mathrm{kg}$. A circular reference orbit at an altitude of $500~\mathrm{km}$ and an inclination of $45^\circ$ was used. The fourth-order Runge--Kutta integration step and control-update interval were $0.125~\mathrm{s}$ and $0.25~\mathrm{s}$, respectively, and all controlled cases used $k_A=0.199~\mathrm{s^{-1}}$. The relative-position PD controller used $K_p=1.0\times10^{-4}~\mathrm{s^{-2}}$ and $K_d=2.0\times10^{-2}~\mathrm{s^{-1}}$. The control power was evaluated at $10~\mathrm{s}$ intervals. The panels show (a) minimum neighboring-pair distance, (b) maximum residual-drift vector, (c) maximum intersatellite control force, (d) maximum per-satellite control power, and (e)--(h) formation configurations at the end of the tenth orbit. The light-blue dash-dotted, red solid, green dotted, and blue solid curves denote no control, drift control, drift and plane-rotation control, and drift, plane-rotation, and PD control, respectively. Among the time-history plots, the uncontrolled trajectory is included only in (a). The black dashed line in (d) denotes the $23.3~\mathrm{mW}$ disturbance-control power budget. In (e)--(h), the gray grids denote the corresponding reference configurations, while the marker colors denote the centered position errors using the common color scale shown in (e). The intersatellite force in (c) is the force component of the AC-cycle-averaged pairwise electromagnetic input defined in \eqref{Feasibility_bilinear_polynomial_model}; the control power in (d) is derived by \eqref{2-2::opt1}.}
}
\label{fig:nonlinear_j2_validation}
\vspace{-0.3cm}
\end{figure*}

\rev{The numerical closed-loop simulations validated the preliminary control-power estimate derived from the mass-normalized control index \(j_d^*(N_l)\) in Fig.~\ref{fig:for_Jd_calculation}. As shown in Fig.~\ref{fig:nonlinear_j2_validation}, the drift controller in \eqref{closed_loop_ro} suppresses the $C_1$ and $C_4$ components responsible for secular in-plane drift with a maximum evaluated power of $9.96~\mathrm{mW}$ per satellite. This confirms that drift cancellation can be achieved within the designed control-power budget since this value is below the preliminary design value of $23.3~\mathrm{mW}$.}

\rev{We show that the additional controllers can remain within the available power range. To improve formation accuracy, we add a plane-rotation controller that uses the feedforward term in \eqref{d_fz_disturbance} to compensate for orbital plane rotation.
We also add the PD controller $\boldsymbol{f}^{\mathrm{cmd}}_{\mathrm{PD},i}=-m_{\mathrm{sat}}(K_p\boldsymbol{e}_{p,i}+K_d\boldsymbol{e}_{v,i})$.
Here, $\boldsymbol{e}_{p,i}$ and $\boldsymbol{e}_{v,i}$ denote the position and velocity errors relative to the reference formation in \eqref{desired_position}. The maximum per-satellite control powers were \(10.4~\mathrm{mW}\) for combined drift and plane-rotation control and \(14.3~\mathrm{mW}\) for combined drift, plane-rotation, and PD control. Both values remained within the original design-power budget.}

\subsection{\rev{Design Implications and Future Work}}
\subsubsection{\rev{Static-Grid Design Implications}}
The proposed framework should be regarded as a design-space analysis framework based on a static grid reference, rather than as a final antenna shape-optimization method. The sizing results in Figs.~\ref{fig:optimization_results_case_1}--\ref{fig:margin_figure_case123} show how the feasible antenna aperture is governed by coupled constraints on satellite mass, size, power, and coil design. Under the assumptions of a uniform square grid, simplified radiation modeling, fixed beam direction, and no mutual-coupling-aware optimization, the framework clarifies how antenna requirements are translated into satellite designs. These results indicate that the static grid-based EMFF antenna is a useful reference configuration for identifying the dominant bottlenecks in distributed space antenna design under a fixed launch mass.

\subsubsection{\rev{Future Work on Practical Implementation Effects}}
\rev{Future work should quantify how practical implementation effects alter antenna performance and EMFF design margins. State-estimation errors and communication delays can provide inaccurate or outdated information for control and beamforming \cite{ran2017relative}. Coil-generated electromagnetic interference can contaminate onboard magnetic measurements \cite{olsen2021magnetometer}. Timing, carrier-frequency, and phase-synchronization errors can degrade coherent transmission \cite{tan2022transmit,ouassal2021decentralized}. Beamforming-weight updates can mitigate geometric phase errors \cite{tuzi2025control}. Residual errors should be evaluated using error-perturbed array-factor calculations and closed-loop simulations \cite{debruijn2014influence,ran2017relative}. Formation-control margins should also be examined at larger spacing. The inverse-fourth-power force scaling increases magnetic-moment and power requirements at larger spacing. Cases~1 and 3 provide a qualitative reference for this trend. Component failures should be assessed together with excitation reoptimization \cite{acharya2014limits} and redundancy trade-offs \cite{erlank2018reliability}. The associated gain loss and resource costs may reduce the feasible aperture. These studies will clarify whether the nominal design remains feasible under realistic operating conditions.}

\subsubsection{\rev{Future Work on Array and Formation Extensions}}
\rev{The static grid reference should be extended to non-uniform and time-varying formations. Future studies should also consider relaxed beam-direction constraints and mutual-coupling-aware optimization. Dynamic reconfiguration requires trajectory optimization. This optimization should jointly consider orbital motion, design headroom, antenna performance, and control time.}

\section{Conclusion}
\label{conclusion}
This paper proposed a system-level design-space analysis framework for distributed space antennas using electromagnetic formation flight. By linking antenna requirements with satellite-level mass, power, and coil-design constraints, the framework provides a static grid-based reference for designing feasible distributed apertures under a fixed system mass. Unlike our previous bucket-brigade model \cite{shim2025feasibility}, the formation-maintenance requirement was incorporated through a control index derived from distributed-control simulations. The case studies showed that antenna performance improves with system mass only while satellite-level design headroom remains. In the direct-to-device-oriented case with \(d_{\mathrm{sat}}=0.15~\mathrm{m}\), generated-power and coil-geometry constraints mainly govern the feasible aperture, whereas the \(d_{\mathrm{sat}}=\rev{0.30}~\mathrm{m}\) large-spacing case can become infeasible because the required coil burden exceeds satellite-level mass, size, and power capacities. Future work will extend the framework to non-uniform and time-varying formations, mutual-coupling-aware radiation models, and dynamic reconfiguration with trajectory optimization.
\appendix[Nomenclature]
\label{app:nomenclature}
\rev{Table~\ref{tab:nomenclature} summarizes the principal variables and parameters used in the proposed design framework.}
\begin{table*}[t]
\begin{revtable}
    \caption{Nomenclature of the principal variables and parameters used in the proposed design framework. }
    \vspace{-0.3cm}
    \label{tab:nomenclature}
    \centering
    \scriptsize
    \setlength{\tabcolsep}{4pt}
    \renewcommand{\arraystretch}{1.08}
    \begin{tabularx}{\textwidth}{
        >{\raggedright\arraybackslash}p{0.27\textwidth}
        >{\raggedright\arraybackslash}X
        >{\centering\arraybackslash}p{0.10\textwidth}}
        \hline
        Symbol & Description & Unit \\
        \hline

        \multicolumn{3}{l}{\textit{System and formation parameters}}\\
        $\bar m_{\mathrm{sys}},\ m_{\mathrm{sys}}$
        & Prescribed total system mass and mass evaluated within its tolerance
        & kg \\

        $h,\ i_0$
        & Orbital altitude and inclination of the reference orbit
        & km, rad \\

        $d_{\mathrm{sat}}$
        & Nominal nearest-neighbor distance in the square grid
        & m \\

        $\Theta_P,\ \Theta_{z-xy}$
        & Swarm-plane angle and relative phase between cross-track and in-plane motions
        & rad \\

        $n,\ N_l=2n+1,\allowbreak\ N_{\mathrm{all}}=N_l^2$
        & Array half-width index, satellites per side, and total satellite count
        & -- \\

        \hline
        \multicolumn{3}{l}{\textit{Optimization quantities}}\\
        $\boldsymbol{X}$
        & Design-variable vector
        $\{2a_{\mathrm{coil}},2a_{\mathrm{sat}},q_{\mathrm{coil}},n,u_{\mathrm{psl}}\}$
        & -- \\

        $\chi_{\mathrm{feas}}$
        & Feasible region of the design variables
        & -- \\

        $\chi_{\mathrm{inv}}$
        & Set of investigation parameters, such as $\bar\mu_{\mathrm{mar}}$ or $P_t$
        & -- \\

        $\mathcal{P}_{\mathrm{DSA}}$
        & Nonlinear system-level design optimization problem
        & -- \\

        $J(\boldsymbol{X})$
        & User-defined evaluation or objective function
        & -- \\

        $N_{\mathrm{GS}}$
        & Number of GlobalSearch or multistart trials
        & -- \\

        \hline
        \multicolumn{3}{l}{\textit{Satellite and coil design quantities}}\\
        $a_{\mathrm{sat}}$
        & Half side length of the cubic satellite
        & m \\

        $a_{\mathrm{coil}}$
        & Radius of a single-axis electromagnetic coil
        & m \\

        $r_{\mathrm{coil}},\ N_t,\allowbreak\
         q_{\mathrm{coil}}=N_t r_{\mathrm{coil}}^2$
        & Coil-wire radius, turn count, and combined coil design parameter
        & m, --, m$^2$ \\

        $r_{\mathrm{mar}},\ k_F$
        & Coil-size margin and coil-separation coefficient
        & m, -- \\

        $m_{\mathrm{sat}},\ \bar m_{\mathrm{sat}}(a_{\mathrm{sat}})$
        & Satellite mass and its size-dependent allowable maximum
        & kg \\

        $m_{\mathrm{3coil}},\allowbreak\
         m_{\mathrm{sap}},\allowbreak\
         m_{\mathrm{bat}},\allowbreak\
         m_{\mathrm{str}},\allowbreak\
         m_{\mathrm{bus}}$
        & Masses of the three-axis coils, four solar-panel sheets, battery, structure, and bus
        & kg \\

        $P_{\mathrm{sap}}$
        & Available solar-array power during sunlit operation
        & W \\

        \hline
        \multicolumn{3}{l}{\textit{EMFF and power quantities}}\\
        $\boldsymbol{\mu},\ \mu_{\mathrm{1d}},\allowbreak\
         \bar\mu_{\mathrm{mar}}$
        & Generated magnetic-moment vector, peak nominal requirement, and prescribed additional margin
        & A\,m$^2$ \\

        $R_{\mathrm{coil}},\ c_{\mathrm{coil}}$
        & Electrical resistance and current of a single-axis coil
        & $\Omega$, A \\

        \rev{\(J_d^*(n)\)}
        & \rev{Satellite-level control index used for magnetic-moment sizing}
        & \rev{\(\mathrm{A^2\,m^4}\)} \\
        
        \rev{\(j_d^*(N_l),\,\widehat{j}_d^*(N_l;d_{\mathrm{sat}})\)}
        & \rev{Mass-normalized control index and its polynomial upper-envelope approximation}
        & \rev{\(\mathrm{A^2\,m^4\,kg^{-1}}\)} \\

        $P_{\mathrm{cont}},\ P_{\mathrm{mar}}$
        & Formation-maintenance power and power reserved for the magnetic-moment margin
        & W \\

        $P_{\mathrm{mis}},\ P_{\mathrm{bus}},\ P_{\mathrm{tot}}$
        & Mission, bus, and total steady-state satellite power consumption
        & W \\

        $k_A$
        & Feedback gain used to construct the distributed-control requirement
        & $\mathrm{s}^{-1}$ \\

        $L_e$
        & Graph Laplacian of the nearest-neighbor formation-control network
        & -- \\

        \hline
        \multicolumn{3}{l}{\textit{Antenna and communication quantities}}\\
        $\lambda$
        & RF wavelength
        & m \\

        $(\theta_{\mathrm{a}},\phi_{\mathrm{a}}),\allowbreak\
         (\theta_0,\phi_0)$
        & Pattern-evaluation angles and commanded main-beam direction
        & rad \\

        $P_t,\ P_T=N_{\mathrm{all}}P_t$
        & RF transmit power per element and total array transmit power
        & W \\

        $G_T$
        & Maximum transmitting-array gain
        & -- or dBi \\

        $\mathrm{EIRP}$
        & Effective isotropic radiated power of the main beam
        & W or dBW \\

        $B_{\mathrm{env}}(u),\ u_{\mathrm{psl}}$
        & Normalized sidelobe envelope and spatial-phase coordinate of its first peak
        & -- \\

        $D_{\mathrm{fp}}$
        & Approximated beam-footprint diameter on the ground
        & m or km \\

        $P_R,\ G_R$
        & Required received power and receiving-antenna gain
        & W or dBm, -- or dBi \\

        $L_f,\ \zeta$
        & Free-space path loss and attenuation or normalized threshold factor
        & -- \\

        $I_R=\zeta P_R L_f/G_R$
        & Link-budget-derived upper limit on sidelobe EIRP
        & W \\
        \hline
    \end{tabularx}
\end{revtable}
\end{table*}

\section*{Acknowledgment}
The authors used ChatGPT (OpenAI) only to improve language clarity. The tool did not generate the scientific content, results, figures, equations, or references. The authors reviewed and verified all edited text.

\bibliographystyle{IEEEtran}
\bibliography{references_takahashi}
\begin{IEEEbiography}[{\includegraphics[width=1in,height=1.25in,clip,keepaspectratio]{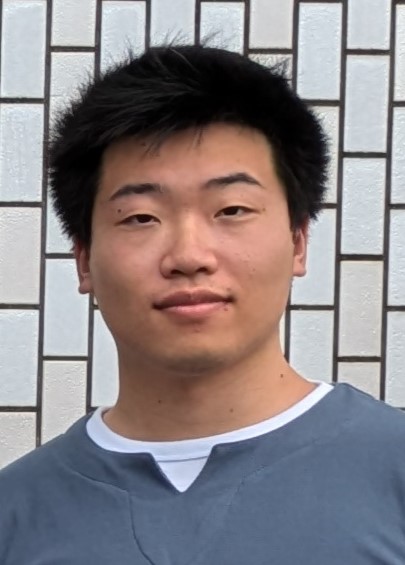}}]{Seang Shim} is currently enrolled in a five-year doctoral program in the Department of Space and Astronautical Science at the Graduate University for Advanced Studies (SOKENDAI). He graduated with a B.S. degree in aerospace engineering from Osaka Prefecture University in 2022. During his undergraduate studies, he researched vibration suppression for flexible structures. He led the ground station team for ``HIROGARI,'' a satellite that demonstrated the on-orbit measurement system of thick deployable structures. His recent research interests include the design of distributed space systems and control methods for multi-systems.
\end{IEEEbiography}
\begin{IEEEbiography}[{\includegraphics[width=1in,height=1.25in,clip,keepaspectratio]{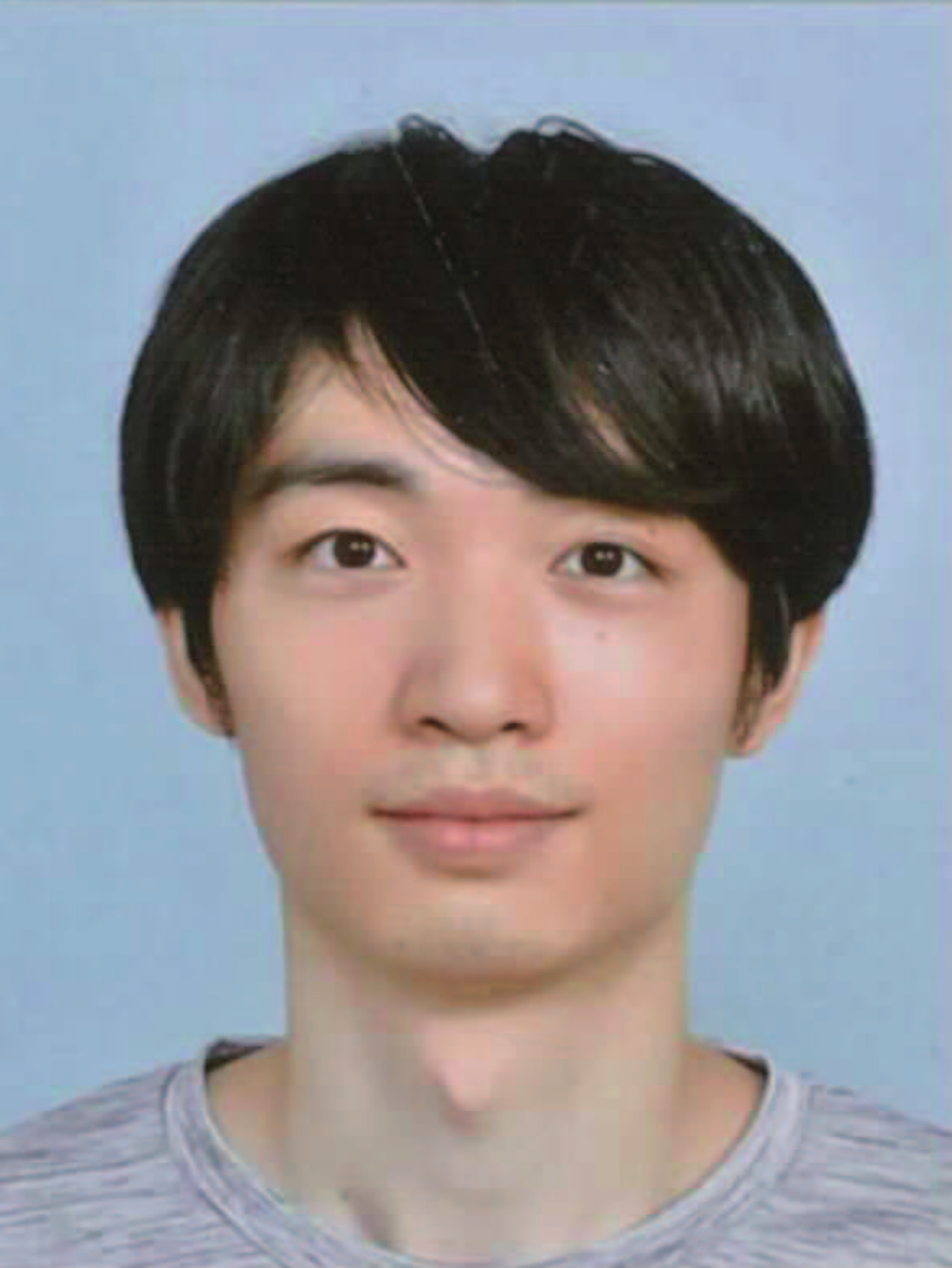}}]{Yuta Takahashi}(Student Member, IEEE) received his B.S. and M.S. degrees in mechanical engineering (ME) in 2019 and 2021 (early graduations), respectively, from the Tokyo Institute of Technology, Japan. He was the recipient of the Best Student Awards 2017 and 2018 (school class rank 1 of 7/149) and 2019--2021 (school class rank 1/140) in ME at Tokyo Tech. He was a Visiting Student Researcher at the California Institute of Technology in 2022 and 2023. 

He is currently pursuing a Ph.D. degree in ME at the Institute of Science Tokyo, Japan, under the JAXA/ISAS research program. His research fields include multi-agent and learning-based nonlinear control theory, with primary applications in distributed space and under-actuated systems. He led the avionics team for the space demonstration mission of the membrane-array antenna device ``5G/HELIOS'' from 2019 to 2022, with launches in 2023 and 2026. From 2023, he worked as a Researcher in the satellite R\&D division at Interstellar Technologies Inc., contributing to a constellation project based on distributed space antenna technology.

Mr. Takahashi was the recipient of the Best Student Paper Finalist Award of the GNC Graduate Student Paper Competition at the AIAA SciTech Forum 2025, Best Paper Nominee of the 2nd Space Robotics Workshop at IEEE SMC-IT/SCC 2025, First Place Award at the MTT-Sat Challenge by the IEEE MTT-S in 2023, Miura prize from The Japan Society of Mechanical Engineers in 2021 (1 of 221/31000), and Best Group Presentation Award of the AOTULE Conference, which he joined for his B.S. work in 2019.
\end{IEEEbiography}
\begin{IEEEbiography}[{\includegraphics[width=1in,height=1.25in,clip,keepaspectratio]{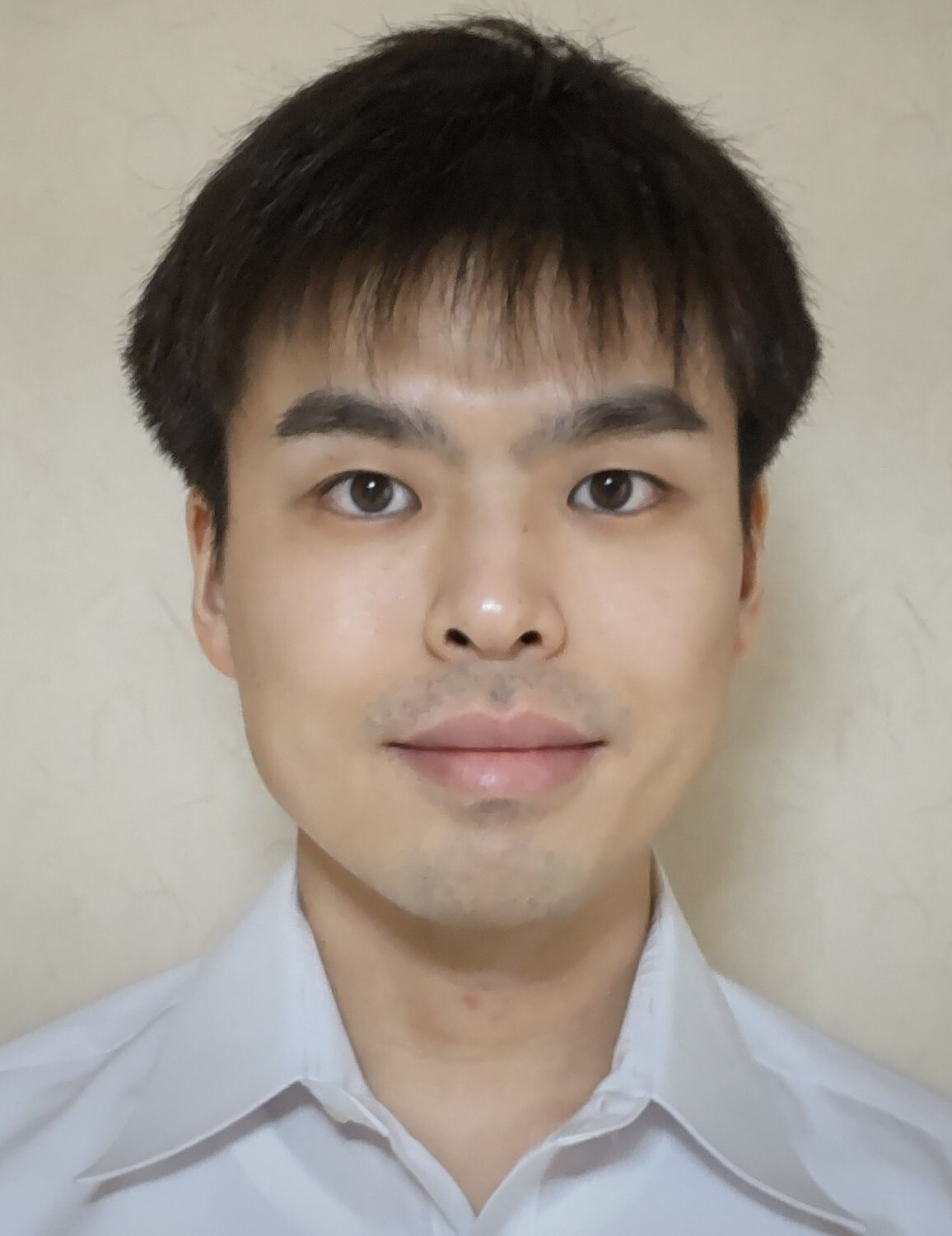}}]{Naoto Usami} (Member, IEEE) received his B.E., M.E., and Ph.D. degrees in electrical engineering from The University of Tokyo in 2014, 2016, and 2019, respectively. In 2014, he served as a Project Manager for the realization of the deep-space sculpture ``DESPATCH,'' which was part of the ARTSAT Project. He is an Assistant Professor of the Institute of Space and Astronautical Science, Japan Aerospace Exploration Agency, since 2021. His current interests include robotic and wireless systems for planetary exploration and micro- and nanofabrication.
\end{IEEEbiography}
\begin{IEEEbiography}[{\includegraphics[width=1in,height=1.25in,clip,keepaspectratio]{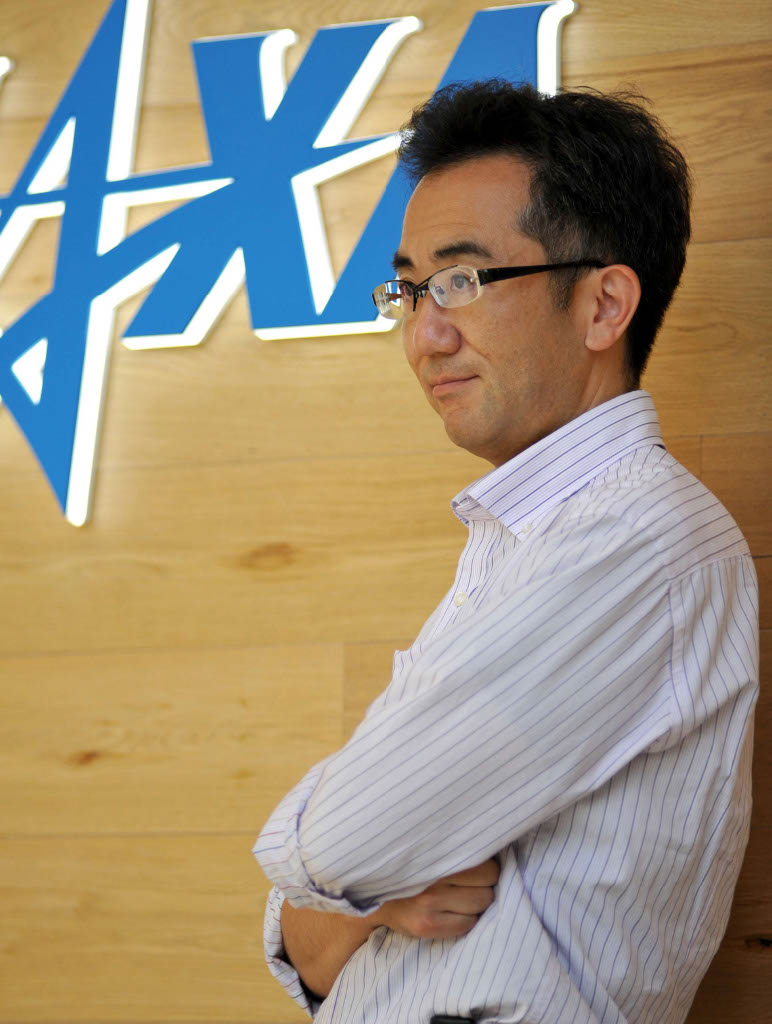}}]{Shin-ichiro Sakai} is a professor at the ISAS, which is part of the JAXA. He received his Ph.D. in electrical engineering from the University of Tokyo in 2000. In 2001, he joined ISAS as an Assistant Professor. He became an Associate Professor in 2005 and has been a Professor since 2019. He has developed attitude control systems for scientific satellites such as Reimei (INDEX), ASTRO-G, Hisaki (SPRINT-A), and Arase (ERG). His research fields include satellite attitude control, electromagnetic formation flight, and spacecraft landing guidance, navigation, and control. He served as the project manager for JAXA's lunar lander, ``Smart Lander for Investigating Moon (SLIM),'' which successfully demonstrated a pinpoint moon landing. Mr. Sakai and his SLIM team received the "Minister of Education, Culture, Sports, Science and Technology Award'' at the 54th Japan Industrial Technology Awards for the SLIM mission.
\end{IEEEbiography}

\end{document}